\documentclass{ptephy_v1}

\usepackage{graphics} 

\usepackage{dcolumn}
\usepackage{bm}
\usepackage{amsfonts}
\usepackage{latexsym}
\usepackage{amssymb}
\usepackage{verbatim}
\usepackage{amsthm}
\usepackage{amsmath}
\usepackage{mathrsfs}
\usepackage{wrapft}

\newcommand{\FF}{{{\mathbb F}}}

\newcommand{\RF}{{{\mathbb R}}}

\newcommand{\ScrF}{{{\mathscr F}}}
\newcommand{\ScrG}{{{\mathscr G}}}
\newcommand{\ScrH}{{{\mathscr H}}}

\newcommand{\ScrM}{{{\mathscr M}}}

\newcommand{\ScrQ}{{{\mathscr Q}}}

\newcommand{\ScrT}{{{\mathscr T}}}

\newcommand{\ScrX}{{{\mathscr X}}}
\newcommand{\ScrY}{{{\mathscr Y}}}
\newcommand{\ScrZ}{{{\mathscr Z}}}

\newtheorem{theorem}{Theorem}[section]

\newtheorem{proposal}{Proposal}[section]
\newtheorem{conjecture}{Conjecture}[section]

\begin{document}


\numberwithin{equation}{section}

\title{
  Gauge-invariant perturbation theory on the Schwarzschild
  background spacetime Part III:\\
  --- Realization of exact solutions ---
}
\author{
  Kouji Nakamura
  \footnote{E-mail address: dr.kouji.nakamura@gmail.com}
}
\address{
  Gravitational-Wave Science Project,
  National Astronomical Observatory of Japan,\\
  2-21-1, Osawa, Mitaka, Tokyo 181-8588, Japan
}
\date{\today}
\begin{abstract}
  This is the Part III paper of our series of papers on a
  gauge-invariant perturbation theory on the Schwarzschild background
  spacetime.
  After reviewing our general framework of the gauge-invariant
  perturbation theory and the proposal on the gauge-invariant
  treatments for $l=0,1$ mode perturbations on the Schwarzschild
  background spacetime in [K.~Nakamura, arXiv:2110.13508 [gr-qc]], we
  examine the problem whether the $l=0,1$ even-mode solutions derived
  in the Part II paper [K.~Nakamura, arXiv:2110.13512 [gr-qc]] are
  physically reasonable, or not.
  We consider the linearized versions of the Lema\^itre-Tolman-Bondi
  solution and the non-rotating C-metric.
  As the result, we show that our derived even-mode solutions to the
  linearized Einstein equations actually realize above two linearized
  solutions.
  This fact supports that our derived solutions are physically
  reasonable, which implies that our proposal on the gauge-invariant
  treatments for $l=0,1$ mode perturbations are also physically
  reasonable.
  We also briefly summarize our conclusions of our series of papers.
\end{abstract}

\maketitle


\section{Introduction}
\label{sec:introduction}


Gravitational-wave observations are now on the stage where we can
directly measure many events through the ground-based
gravitational-wave
detectors~\cite{LIGO-home-page,Virgo-home-page,KAGRA-home-page,LIGO-INDIA-home-page}.
One of the future directions of gravitational-wave astronomy will be a
precise science through the statistics of many events.
Toward further development, the projects of future ground-based
gravitational-wave
detectors~\cite{ET-home-page,CosmicExplorer-home-page} are also
progressing to achieve more sensitive detectors and some projects of
space gravitational-wave antenna are also
progressing~\cite{LISA-home-page,DECIGO-PTEP-2021,TianQin-PTEP-2021,Taiji-PTEP-2021}.
Although there are many targets of these detectors, the
Extreme-Mass-Ratio-Inspiral (EMRI), which is a source of gravitational
waves from the motion of a stellar mass object around a supermassive
black hole, is a promising target of the Laser Interferometer Space
Antenna~\cite{LISA-home-page}.
Since the mass ratio of this EMRI is very small, we can describe the
gravitational waves from EMRIs through black hole
perturbations~\cite{L.Barack-A.Pound-2019}.
Furthermore, the sophistication of higher-order black hole
perturbation theories is required to support these gravitational-wave
physics as a precise science.
The motivation of our series of papers,
Refs.~\cite{K.Nakamura-2021a,K.Nakamura-2021b,K.Nakamura-2021c,K.Nakamura-2021d}
and this paper, is in this theoretical sophistication of black hole
perturbation theories toward higher-order perturbations.


In black hole perturbation theories, further sophistications are
possible even in perturbation theories on the Schwarzschild background
spacetime.
From the works by Regge and Wheeler~\cite{T.Regge-J.A.Wheeler-1957}
and Zerilli~\cite{F.Zerilli-1970-PRL,F.Zerilli-1970-PRD}, many studies
on the perturbations in the Schwarzschild background spacetime are
carried out~\cite{H.Nakano-2019,V.Moncrief-1974a,V.Moncrief-1974b,C.T.Cunningham-R.H.Price-V.Moncrief-1978,Chandrasekhar-1983,Gerlach_Sengupta-1979a,Gerlach_Sengupta-1979b,Gerlach_Sengupta-1979c,Gerlach_Sengupta-1980,T.Nakamura-K.Oohara-Y.Kojima-1987,Gundlach-Martine-Garcia-2000,Gundlach-Martine-Garcia-2001,A.Nagar-L.Rezzolla-2005-2006,K.Martel-E.Poisson-2005}.
In these works, perturbations are decomposed through the spherical
harmonics $Y_{lm}$ because of the spherical symmetry of the background
spacetime, and $l=0,1$ modes should be separately treated.
Due to this separate treatments, ``{\it gauge-invariant}'' treatments
for $l=0$ and $l=1$ modes were unclear.


Owing to this situation,  in the previous
papers~\cite{K.Nakamura-2021a,K.Nakamura-2021c}, we proposed the
strategy of the gauge-invariant treatments of these $l=0,1$ mode
perturbations, which is declared as
Proposal~\ref{proposal:treatment-proposal-on-pert-on-spherical-BG}
in this paper below.
We have been developing the general formulation of a higher-order
gauge-invariant perturbation theory on a generic background spacetime
toward unambiguous sophisticated nonlinear general-relativistic
perturbation theories~\cite{K.Nakamura-2003,K.Nakamura-2005,K.Nakamura-2011,K.Nakamura-IJMPD-2012,K.Nakamura-2013,K.Nakamura-2014}.
Although we have been applied this general framework to cosmological
perturbations~\cite{K.Nakamura-2006,K.Nakamura-2007,K.Nakamura-LTVII-2008,K.Nakamura-2009a,K.Nakamura-2009b,K.Nakamura-2010,A.J.Christopherson-K.A.Malik-D.R.Matravers-K.Nakamura-2011,K.Nakamura-2020},
we applied it to black hole perturbations in the series of papers,
i.e.,
Refs.~\cite{K.Nakamura-2021a,K.Nakamura-2021b,K.Nakamura-2021c,K.Nakamura-2021d}
and this paper.


In the Part I paper~\cite{K.Nakamura-2021c}, we also derived the
linearized Einstein equations in a gauge-invariant manner following
Proposal~\ref{proposal:treatment-proposal-on-pert-on-spherical-BG}.
Perturbations on the spherically symmetric background spacetime are
classified into even- and odd-mode perturbations.
In the same paper~\cite{K.Nakamura-2021c}, we also gave the strategy
to solve the odd-mode perturbations including $l=0,1$ modes.
Furthermore, we also derived the explicit solutions for the $l=0,1$
odd-mode perturbations to the linearized Einstein equations following
Proposal~\ref{proposal:treatment-proposal-on-pert-on-spherical-BG}.
In the Part II paper~\cite{K.Nakamura-2021d}, we gave the strategy to
solve the even-mode perturbations including $l=0,1$ modes and we also
derive the explicit solutions for the $l=0,1$ even-mode perturbations
following
Proposal~\ref{proposal:treatment-proposal-on-pert-on-spherical-BG}.
This series of papers is the full paper version of our short
paper~\cite{K.Nakamura-2021a}.


In this paper, we check whether the solutions for even-mode
perturbations derived in the Part II paper~\cite{K.Nakamura-2021d} are
physically reasonable, or not.
We consider the correspondence between our linearized solutions and
two exact solutions.
One of exact solutions discussed in this paper is the
Lema\^itre-Tolman-Bondi (LTB) solution and the other is the
non-rotating
C-metric~\cite{W.Kinnersley-M.Walker-1970,J.B.Griffiths-P.Krtous-J.Podolsky-2006}.


The organization of this Part III paper is as follows.
In Sec.~\ref{sec:review-of-general-framework-GI-perturbation-theroy},
after briefly review the framework of the gauge-invariant perturbation
theory, we summarize our proposal in
Refs.~\cite{K.Nakamura-2021a,K.Nakamura-2021c}.
Then, we also summarize the linearized even-mode Einstein equations on
the Schwarzschild background spacetime derived in
Ref.~\cite{K.Nakamura-2021c}.
These are derived based on
Proposal~\ref{proposal:treatment-proposal-on-pert-on-spherical-BG}.
In
Sec.~\ref{sec:LTB_solution_as_a_perturbation_of_Schwarzschild_spacetime},
we discuss the realization of the LTB solution by our derived
solutions for $l=0,1$-mode even-mode perturbations to the linearized
Einstein equation following
Proposal~\ref{proposal:treatment-proposal-on-pert-on-spherical-BG}.
In Sec.~\ref{sec:Reconsideration_of_the_linearized_C-metric_again}, we
discuss the realization of the C-metric from our derived solutions in
the Part II paper~\cite{K.Nakamura-2021d}.
The final section (Sec.~\ref{sec:Summary_and_Discussions}) devoted to
our summary and discussion.
This final section also includes a brief conclusion of our series of
papers on a gauge-invariant perturbation theory on the Schwarzschild
background spacetime.


We use the notation used in the previous
papers~\cite{K.Nakamura-2021a,K.Nakamura-2021b,K.Nakamura-2021c} and
the unit $G=c=1$, where $G$ is Newton's constant of gravitation and
$c$ is the velocity of light.


\section{Brief review of the general-relativistic gauge-invariant
  perturbation theory}
\label{sec:review-of-general-framework-GI-perturbation-theroy}


In this section, we review the premises of our series of
papers~\cite{K.Nakamura-2021a,K.Nakamura-2021c,K.Nakamura-2021d} which
are necessary to understand the ingredients of this paper.
In Sec.~\ref{sec:general-framework-GI-perturbation-theroy}, we briefly
review our framework of the gauge-invariant perturbation
theory~\cite{K.Nakamura-2003,K.Nakamura-2005}.
This is an important premise of the series of our
papers~\cite{K.Nakamura-2021a,K.Nakamura-2021c,K.Nakamura-2021d} and
this paper.
In Sec.~\ref{sec:spherical_background_case}, we review the
gauge-invariant perturbation theory on spherically symmetric
spacetimes which includes our proposal in
Refs.~\cite{K.Nakamura-2021a,K.Nakamura-2021c}.
In Sec.~\ref{sec:Even_Einstein_equations}, we summarize the even-mode
linearized Einstein equations on the Schwarzschild background
spacetime and their explicit solutions for $l=0,1$ modes, which are
necessary for the arguments in this paper.


\subsection{General framework of gauge-invariant perturbation theory}
\label{sec:general-framework-GI-perturbation-theroy}


In any perturbation theory, we always treat two spacetime manifolds.
One is the physical spacetime $(\ScrM_{{\rm ph}},\bar{g}_{ab})$,
which is identified with our nature itself, and we want to describe
this spacetime $(\ScrM_{{\rm ph}},\bar{g}_{ab})$ by perturbations.
The other is the background spacetime $(\ScrM,g_{ab})$,
which is prepared as a reference by hand.
Note that these two spacetimes are distinct.
Furthermore, in any perturbation theory, we always write equations
for the perturbation of the variable $Q$ as follows:
\begin{equation}
  \label{eq:variable-symbolic-perturbation}
  Q(``p\mbox{''}) = Q_{0}(p) + \delta Q(p).
\end{equation}
Equation (\ref{eq:variable-symbolic-perturbation}) gives a
relation between variables on different manifolds.
Actually, $Q(``p\mbox{''})$ in
Eq.~(\ref{eq:variable-symbolic-perturbation}) is a variable on
$\ScrM_{\epsilon}=\ScrM_{\rm ph}$, whereas $Q_{0}(p)$ and
$\delta Q(p)$ are variables on $\ScrM$.
Because we regard Eq.~(\ref{eq:variable-symbolic-perturbation}) as
a field equation, Eq.~(\ref{eq:variable-symbolic-perturbation})
includes an implicit assumption of the existence of a point
identification map $\ScrX_{\epsilon}$ $:$
$\ScrM\rightarrow\ScrM_{\epsilon}$ $:$
$p\in\ScrM\mapsto ``p\mbox{''}\in\ScrM_{\epsilon}$.
This identification map is a {\it gauge choice} in
general-relativistic perturbation theories.
This is the notion of the {\it second-kind gauge} pointed out by
Sachs~\cite{R.K.Sachs-1964}.
Note that this second-kind gauge is a different notion from the
degree of freedom of the coordinate transformation on a single
manifold, which is called the {\it first-kind
  gauge}~\cite{K.Nakamura-2010,K.Nakamura-2020,K.Nakamura-2021c}.
This distinction of the first- and the second-kind of gauges
extensively explained in the Part I paper~\cite{K.Nakamura-2021c} and
is also important to understand the results in
Secs.~\ref{sec:LTB_solution_as_a_perturbation_of_Schwarzschild_spacetime}
and~\ref{sec:Reconsideration_of_the_linearized_C-metric_again} in this
paper.


To compare the variable $Q$ on $\ScrM_{\epsilon}$ with its
background value $Q_{0}$ on $\ScrM$, we use the pull-back
$\ScrX_{\epsilon}^{*}$ of the identification map
$\ScrX_{\epsilon}$ $:$ $\ScrM\rightarrow\ScrM_{\epsilon}$ and
we evaluate the pulled-back variable $\ScrX_{\epsilon}^{*}Q$ on the
background spacetime $\ScrM$.
Furthermore, in perturbation theories, we expand the pull-back
operation $\ScrX_{\epsilon}^{*}$ to the variable $Q$ with respect
to the infinitesimal parameter $\epsilon$ for the perturbation as
\begin{eqnarray}
  \ScrX_{\epsilon}^{*}Q
  =
  Q_{0}
  + \epsilon {}^{(1)}_{\;\ScrX}Q
  + O(\epsilon^{2})
  .
  \label{eq:perturbative-expansion-of-Q-def}
\end{eqnarray}
Equation~(\ref{eq:perturbative-expansion-of-Q-def}) are evaluated on
the background spacetime $\ScrM$.
When we have two different gauge choices $\ScrX_{\epsilon}$ and
$\ScrY_{\epsilon}$, we can consider the {\it gauge-transformation},
which is the change of the point-identification
$\ScrX_{\epsilon}\rightarrow\ScrY_{\epsilon}$.
This gauge-transformation is given by the diffeomorphism
$\Phi_{\epsilon}$ $:=$
$\left(\ScrX_{\epsilon}\right)^{-1}\circ\ScrY_{\epsilon}$
$:$ $\ScrM$ $\rightarrow$ $\ScrM$.
Actually, the diffeomorphism $\Phi_{\epsilon}$ induces a pull-back from
the representation $\ScrX_{\epsilon}^{*}\!Q_{\epsilon}$ to the
representation $\ScrY_{\epsilon}^{*}\!Q_{\epsilon}$ as
$\ScrY_{\epsilon}^{*}\!Q_{\epsilon}=\Phi_{\epsilon}^{*}\ScrX_{\epsilon}^{*}\!Q_{\epsilon}$.
From general arguments of the Taylor
expansion~\cite{M.Bruni-S.Matarrese-S.Mollerach-S.Sonego-1997}, the
pull-back $\Phi_{\epsilon}^{*}$ is expanded as
\begin{eqnarray}
  \ScrY_{\epsilon}^{*}\!Q_{\epsilon}
  &=&
  \ScrX_{\epsilon}^{*}\!Q_{\epsilon}
  + \epsilon {\pounds}_{\xi_{(1)}} \ScrX_{\epsilon}^{*}\!Q_{\epsilon}
  + O(\epsilon^{2}),
  \label{eq:Bruni-46-one}
\end{eqnarray}
where $\xi_{(1)}^{a}$ is the generator of $\Phi_{\epsilon}$.
From Eqs.~(\ref{eq:perturbative-expansion-of-Q-def}) and
(\ref{eq:Bruni-46-one}), the gauge-transformation for the first-order
perturbation ${}^{(1)}Q$ is given by
\begin{eqnarray}
  \label{eq:Bruni-47-one}
  {}^{(1)}_{\;\ScrY}\!Q - {}^{(1)}_{\;\ScrX}\!Q &=&
  {\pounds}_{\xi_{(1)}}Q_{0}.
\end{eqnarray}
We also employ the {\it order by order gauge invariance} as a
concept of gauge invariance~\cite{K.Nakamura-2009a}.
We call the $k$th-order perturbation ${}^{(k)}_{\ScrX}\!Q$ as
gauge invariant if and only if
\begin{eqnarray}
  \label{eq:orderbyorder-gauge-inv-def}
{}^{(k)}_{\;\ScrX}\!Q = {}^{(k)}_{\;\ScrY}\!Q
\end{eqnarray}
for any gauge choice $\ScrX_{\epsilon}$ and $\ScrY_{\epsilon}$.


Based on the above setup, we proposed a formulation to construct
gauge-invariant variables of higher-order
perturbations~\cite{K.Nakamura-2003,K.Nakamura-2005}.
First, we expand the metric on the physical spacetime
$\ScrM_{\epsilon}$, which was pulled back to the background
spacetime $\ScrM$ through a gauge choice $\ScrX_{\epsilon}$ as
\begin{eqnarray}
  \ScrX^{*}_{\epsilon}\bar{g}_{ab}
  &=&
  g_{ab} + \epsilon {}_{\ScrX}\!h_{ab}
  + O(\epsilon^{2}).
  \label{eq:metric-expansion}
\end{eqnarray}
Although the expression (\ref{eq:metric-expansion}) depends
entirely on the gauge choice $\ScrX_{\epsilon}$, henceforth,
we do not explicitly express the index of the gauge choice
$\ScrX_{\epsilon}$ in the expression if there is no
possibility of confusion.
The important premise of our formulation of higher-order
gauge-invariant perturbation theory was the following
conjecture~\cite{K.Nakamura-2003,K.Nakamura-2005} for the linear
metric perturbation $h_{ab}$:
\begin{conjecture}
  \label{conjecture:decomposition-conjecture}
  If the gauge-transformation rule for a perturbative pulled-back
  tensor field $h_{ab}$ to the background spacetime $\ScrM$ is
  given by ${}_{\ScrY}\!h_{ab}$ $-$ ${}_{\ScrX}\!h_{ab}$ $=$
  ${\pounds}_{\xi_{(1)}}g_{ab}$ with the background metric $g_{ab}$,
  there then exist a tensor field $\ScrF_{ab}$ and a vector field
  $Y^{a}$ such that $h_{ab}$ is decomposed as $h_{ab}$ $=:$
  $\ScrF_{ab}$ $+$ ${\pounds}_{Y}g_{ab}$, where $\ScrF_{ab}$ and
  $Y^{a}$ are transformed as ${}_{\ScrY}\!\ScrF_{ab}$ $-$
  ${}_{\ScrX}\!\ScrF_{ab}$ $=$ $0$ and ${}_{\ScrY}\!Y^{a}$
  $-$ ${}_{\ScrX}\!Y^{a}$ $=$ $\xi^{a}_{(1)}$ under the gauge
  transformation, respectively.
\end{conjecture}
We call $\ScrF_{ab}$ and $Y^{a}$ as the
{\it gauge-invariant} and {\it gauge-variant} parts
of $h_{ab}$, respectively.


The proof of Conjecture~\ref{conjecture:decomposition-conjecture} is
highly nontrivial~\cite{K.Nakamura-2011,K.Nakamura-2013}, and it was
found that the gauge-invariant variables are essentially non-local.
Despite this non-triviality, once we accept
Conjecture~\ref{conjecture:decomposition-conjecture},
we can decompose the linear perturbation of an arbitrary tensor field
${}_{\ScrX}^{(1)}\!Q$, whose gauge-transformation is given by
Eq.~(\ref{eq:Bruni-47-one}), through the gauge-variant part $Y_{a}$ of
the metric perturbation in
Conjecture~\ref{conjecture:decomposition-conjecture} as
\begin{eqnarray}
  \label{eq:arbitrary-Q-decomp}
  {}_{\ScrX}^{(1)}\!Q = {}^{(1)}\!\ScrQ + {\pounds}_{{}_{\ScrX}\!Y}Q_{0},
\end{eqnarray}
where ${}^{(1)}\!\ScrQ$ is the gauge invariant part of the perturbation
${}_{\ScrX}^{(1)}\!Q$.
As examples, the linearized Einstein tensor
${}_{\ScrX}^{(1)}G_{a}^{\;\;b}$ and the linear perturbation of the
energy-momentum tensor ${}_{\ScrX}^{(1)}T_{a}^{\;\;b}$ are also
decomposed as
\begin{eqnarray}
  \label{eq:Gab-Tab-decomp}
  {}_{\ScrX}^{(1)}\!G_{a}^{\;\;b}
  =
  {}^{(1)}\!\ScrG_{a}^{\;\;b}\left[\ScrF\right] + {\pounds}_{{}_{\ScrX}\!Y}G_{a}^{\;\;b}
  ,
  \quad
  {}_{\ScrX}^{(1)}\!T_{a}^{\;\;b}
  =
  {}^{(1)}\!\ScrT_{a}^{\;\;b} + {\pounds}_{{}_{\ScrX}\!Y}T_{a}^{\;\;b}
  ,
\end{eqnarray}
where $G_{ab}$ and $T_{ab}$ are the background values of the Einstein
tensor and the energy-momentum tensor, respectively.
The explicit form of the gauge-invariant part
${}^{(1)}\!\ScrG_{a}^{\;\;b}$ of the linear-order perturbation of
the Einstein tensor is not important within this paper.
Using the background Einstein equation
$G_{a}^{\;\;b}=8\pi T_{a}^{\;\;b}$, the linearized Einstein equation
${}_{\ScrX}^{(1)}\!G_{ab}=8\pi{}_{\ScrX}^{(1)}\!T_{ab}$ is
automatically given in the gauge-invariant form
\begin{eqnarray}
  \label{eq:einstein-equation-gauge-inv}
  {}^{(1)}\!\ScrG_{a}^{\;\;b}\left[\ScrF\right]
  =
  8 \pi
  {}^{(1)}\!\ScrT_{a}^{\;\;b}\left[\ScrF,\phi\right]
\end{eqnarray}
even if the background Einstein equation is nontrivial.
Here, ``$\phi$'' in Eq.~(\ref{eq:einstein-equation-gauge-inv})
symbolically represents the matter degree of freedom.


For the ingredients of this paper, it is important to note that the
decomposition of the metric perturbation $h_{ab}$ into its
gauge-invariant part $\ScrF_{ab}$ and into its gauge-variant part
$Y^{a}$ is not unique as noted in
Refs.~\cite{K.Nakamura-2009a,K.Nakamura-2010,K.Nakamura-2021c}.
Actually, the decomposition of the metric perturbation $h_{ab}$ is
also given by
\begin{eqnarray}
  \label{eq:gauge-inv-nonunique}
  h_{ab}
  =
  \ScrF_{ab} - {\pounds}_{Z}g_{ab}
  + {\pounds}_{Z+Y}g_{ab}
  =:
  \ScrH_{ab} + {\pounds}_{X}g_{ab}
  ,
\end{eqnarray}
where $Z^{a}$ is gauge-invariant in the second kind, i.e.,
${}_{\ScrY}\!Z^{a}$ $-$ ${}_{\ScrX}\!Z^{a}$ $=$ $0$.
The tensor field $\ScrH_{ab}:=\ScrF_{ab} - {\pounds}_{Z}g_{ab}$
is also regarded as the gauge-invariant part (in the sense of the
second-kind) of the perturbation $h_{ab}$ because
${}_{\ScrY}\!\ScrH_{ab}$ $-$ ${}_{\ScrX}\!\ScrH_{ab}$ $=$
$0$.
Similarly, the vector field $X^{a}$ $:=$ $Z^{a}$ $+$ $Y^{a}$ is also
regarded as the gauge-variant part of the perturbation $h_{ab}$
because ${}_{\ScrY}\!X^{a}$ $-$ ${}_{\ScrX}\!X^{a}$ $=$
$\xi^{a}_{(1)}$.
The difference between the variables $\ScrH_{ab}$ and
$\ScrF_{ab}$ is given by ${\pounds}_{-Z}g_{ab}$.
Here, we note that if the gauge-invariant variable $\ScrF_{ab}$ is
a solution to the linearized Einstein equation, $\ScrH_{ab}$ $=$
$\ScrF_{ab}$ $-$ ${\pounds}_{Z}g_{ab}$ is also a solution to the
linearized Einstein equation (\ref{eq:einstein-equation-gauge-inv})
due to a symmetry of the linearized Einstein equation
(\ref{eq:einstein-equation-gauge-inv}) as explained in Part I
paper~\cite{K.Nakamura-2021c}.
This implies that the terms in the form ${\pounds}_{Z}g_{ab}$ may
always appear in the solutions to the linearized Einstein equation due
to the above symmetry of the linearized Einstein equation.
Since our formulation already exclude the second-kind gauge
completely, we should regard that the gauge-invariant term
${\pounds}_{-Z}g_{ab}$ as the first-kind gauge of the background
spacetime induced by the infinitesimal coordinate transformations on
the physical spacetime $\ScrM_{\epsilon}$ as discussed in the Part I
paper~\cite{K.Nakamura-2021c}.


\subsection{Linear perturbations on spherically symmetric background}
\label{sec:spherical_background_case}


Here, we consider the 2+2 formulation of the perturbation of a
spherically symmetric background spacetime, which originally proposed
by Gerlach and Sengupta~\cite{Gerlach_Sengupta-1979a,Gerlach_Sengupta-1979b,Gerlach_Sengupta-1979c,Gerlach_Sengupta-1980}.
Spherically symmetric spacetimes are characterized by the direct
product $\ScrM=\ScrM_{1}\times S^{2}$ and their metric is
\begin{eqnarray}
  \label{eq:background-metric-2+2}
  g_{ab}
  &=&
  y_{ab} + r^{2}\gamma_{ab}
  , \\
  y_{ab} &=& y_{AB} (dx^{A})_{a}(dx^{B})_{b}, \quad
             \gamma_{ab} = \gamma_{pq} (dx^{p})_{a} (dx^{q})_{b},
\end{eqnarray}
where $x^{A} = (t,r)$, $x^{p}=(\theta,\phi)$, and $\gamma_{pq}$ is the
metric on the unit sphere.
In the Schwarzschild spacetime, the metric
(\ref{eq:background-metric-2+2}) is given by
\begin{eqnarray}
  \label{eq:background-metric-2+2-y-comp-Schwarzschild}
  y_{ab}
  &=&
      - f (dt)_{a}(dt)_{b}
      +
      f^{-1} (dr)_{a}(dr)_{b}
      ,
      \quad
      f = 1 - \frac{2M}{r}
  ,\\
  \label{eq:background-metric-2+2-gamma-comp-Schwarzschild}
  \gamma_{ab}
  &=&
      (d\theta)_{a}(d\theta)_{b}
      +
      \sin^{2}\theta(d\phi)_{a}(d\phi)_{b}
      =
      \theta_{a}\theta_{b} + \phi_{a}\phi_{b}
      ,
  \\
  \label{eq:S2-unit-basis-def}
  \theta_{a}
  &=&
      (d\theta)_{a}, \quad
      \phi_{a}
      =
      \sin\theta (d\phi)_{a}
      .
\end{eqnarray}


On this background spacetime $(\ScrM,g_{ab})$, the components of
the metric perturbation is given by
\begin{eqnarray}
  \label{eq:metric-perturbation-components}
  h_{ab}
  =
  h_{AB} (dx^{A})_{a}(dx^{B})_{b}
  +
  2 h_{Ap} (dx^{A})_{(a}(dx^{p})_{b)}
  +
  h_{pq} (dx^{p})_{a}(dx^{q})_{b}
  .
\end{eqnarray}
Here, we note that the components $h_{AB}$, $h_{Ap}$, and
$h_{pq}$ are regarded as components of scalar, vector, and
tensor on $S^{2}$, respectively.
In the Part I paper~\cite{K.Nakamura-2021c}, we showed the
linear-independence of the set of harmonic functions
\begin{eqnarray}
  \label{eq:harmonic-fucntions-set}
  \left\{
  S_{\delta},
  \hat{D}_{p}S_{\delta},
  \epsilon_{pq}\hat{D}^{q}S_{\delta},
  \frac{1}{2}\gamma_{pq}S_{\delta},
  \left(\hat{D}_{p}\hat{D}_{q}-\frac{1}{2}\gamma_{pq}\right)S_{\delta},
  2\epsilon_{r(p}\hat{D}_{q)}\hat{D}^{r}S_{\delta}
  \right\}
  ,
\end{eqnarray}
where $\hat{D}_{p}$ is the covariant derivative associated with
the metric $\gamma_{pq}$ on $S^{2}$,
$\hat{D}^{p}=\gamma^{pq}\hat{D}_{q}$,
$\epsilon_{pq}=\epsilon_{[pq]}=2\theta_{[p}\phi_{q]}$ is the totally
antisymmetric tensor on $S^{2}$.
In the set of harmonic function (\ref{eq:harmonic-fucntions-set}), the
scalar harmonic function $S_{\delta}$ is given by
\begin{eqnarray}
  \label{eq:harmonics-extended-choice-sum}
  S_{\delta}
  =
  \left\{
  \begin{array}{ccccc}
    Y_{lm} & \quad & \mbox{for} & \quad & l\geq 2; \\
    k_{(\hat{\Delta}+2)m} & \quad & \mbox{for} & \quad & l=1; \\
    k_{(\hat{\Delta})} & \quad & \mbox{for} & \quad & l=0.
  \end{array}
  \right.
\end{eqnarray}
Here, functions $k_{(\hat{\Delta})}$ and $k_{(\hat{\Delta}+2)m}$ are
the kernel modes of the derivative operator $\hat{\Delta}$ and
$[\hat{\Delta}+2]$, respectively, and we employ the explicit form of
these functions as
\begin{eqnarray}
  k_{(\hat{\Delta})}
  &=&
      1 + \delta \ln\left(\frac{1-\cos\theta}{1+\cos\theta}\right)^{1/2},
      \quad \delta \in\RF
      ,
      \label{eq:l=0-general-mode-func-specific}
  \\
  k_{(\hat{\Delta}+2,m=0)}
  &=&
      \cos\theta
      +
      \delta \left(\frac{1}{2} \cos\theta \ln\frac{1+\cos\theta}{1-\cos\theta} -1\right)
      ,
     \quad \delta \in \RF
     ,
     \label{eq:l=1-m=0-mode-func-explicit}
     \\
  k_{(\hat{\Delta}+2,m=\pm1)}
  &=&
      \left[
      \sin\theta
      +
      \delta \left(
      + \frac{1}{2} \sin\theta \ln\frac{1+\cos\theta}{1-\cos\theta}
      + \cot\theta
      \right)
      \right]
      e^{\pm i\phi}
      .
      \label{eq:l=1-m=pm1-mode-func-explicit}
\end{eqnarray}


Then, we consider the mode decomposition of the components
$\{h_{AB},h_{Ap},h_{pq}\}$ as follows:
\begin{eqnarray}
  \label{eq:hAB-fourier}
  \!\!\!\!\!\!
  h_{AB}
  \!\!\!\!\!&=&\!\!\!\!\!
      \sum_{l,m} \tilde{h}_{AB} S_{\delta}
      ,
  \\
  \label{eq:hAp-fourier}
  \!\!\!\!\!\!
  h_{Ap}
  \!\!\!\!\!&=&\!\!\!\!\!
      r \sum_{l,m} \left[
      \tilde{h}_{(e1)A} \hat{D}_{p}S_{\delta}
      +
      \tilde{h}_{(o1)A} \epsilon_{pq} \hat{D}^{q}S_{\delta}
      \right]
      ,
  \\
  \label{eq:hpq-fourier}
  \!\!\!\!\!\!
  h_{pq}
  \!\!\!\!\!&=&\!\!\!\!\!
      r^{2} \sum_{l,m} \left[
      \frac{1}{2} \gamma_{pq} \tilde{h}_{(e0)} S_{\delta}
      +
      \tilde{h}_{(e2)} \left(
      \hat{D}_{p}\hat{D}_{q} - \frac{1}{2} \gamma_{pq} \hat{D}^{r}\hat{D}_{r}
      \right) S_{\delta}
      +
      2 \tilde{h}_{(o2)} \epsilon_{r(p} \hat{D}_{q)}\hat{D}^{r} S_{\delta}
      \right]
      .
\end{eqnarray}
Since the linear-independence of each element of the set of harmonic
function (\ref{eq:harmonic-fucntions-set}) is guaranteed, the
one-to-one correspondence between the components $\{h_{AB},$ $h_{Ap},$
$h_{pq}\}$ and the mode coefficients $\{\tilde{h}_{AB},$
$\tilde{h}_{(e1)A},$ $\tilde{h}_{(o1)A},$ $\tilde{h}_{(e0)},$
$\tilde{h}_{(e2)},$ $\tilde{h}_{(o2)}\}$ with the decomposition
formulae (\ref{eq:hAB-fourier})-(\ref{eq:hpq-fourier}) is guaranteed
including $l=0,1$ mode if $\delta\neq 0$.
Then, the mode-by-mode analysis including $l=0,1$ is possible when
$\delta\neq 0$.
However, the mode functions
(\ref{eq:l=0-general-mode-func-specific})--(\ref{eq:l=1-m=pm1-mode-func-explicit})
are singular if $\delta\neq 0$.
When $\delta=0$, we have $k_{(\hat{\Delta})}\propto Y_{00}$ and
$k_{(\hat{\Delta}+2)m}\propto Y_{1m}$.
Because of this situation, we proposed the following strategy:
\begin{proposal}
  \label{proposal:treatment-proposal-on-pert-on-spherical-BG}
  We decompose the metric perturbation $h_{ab}$ on the background
  spacetime with the metric
  (\ref{eq:background-metric-2+2})--(\ref{eq:background-metric-2+2-gamma-comp-Schwarzschild})
  through Eqs.~(\ref{eq:hAB-fourier})--(\ref{eq:hpq-fourier}) with the
  harmonic function $S_{\delta}$ given by
  Eq.~(\ref{eq:harmonics-extended-choice-sum}).
  Then, Eqs.~(\ref{eq:hAB-fourier})--(\ref{eq:hpq-fourier}) become
  invertible including $l=0,1$ modes.
  After deriving the mode-by-mode field equations such as linearized
  Einstein equations by using the harmonic functions $S_{\delta}$, we
  choose $\delta=0$ as regular boundary condition for solutions when
  we solve these field equations.
\end{proposal}


As shown in the Part I paper~\cite{K.Nakamura-2021c}, once we accept
Proposal~\ref{proposal:treatment-proposal-on-pert-on-spherical-BG},
the Conjecture~\ref{conjecture:decomposition-conjecture} becomes the
following statement:
\begin{theorem}
  \label{theorem:decomposition-theorem-with-spherical-symmetry}
  If the gauge-transformation rule for a perturbative pulled-back
  tensor field $h_{ab}$ to the background spacetime $\ScrM$ is
  given by ${}_{\ScrY}\!h_{ab}$ $-$ ${}_{\ScrX}\!h_{ab}$ $=$
  ${\pounds}_{\xi_{(1)}}g_{ab}$ with the background metric $g_{ab}$
  with spherically symmetry, there then exist a tensor field
  $\ScrF_{ab}$ and a vector  field $Y^{a}$ such that $h_{ab}$ is
  decomposed as $h_{ab}$ $=:$ $\ScrF_{ab}$ $+$
  ${\pounds}_{Y}g_{ab}$, where $\ScrF_{ab}$ and $Y^{a}$ are
  transformed into ${}_{\ScrY}\!\ScrF_{ab}$ $-$
  ${}_{\ScrX}\!\ScrF_{ab}$ $=$ $0$ and ${}_{\ScrY}\!Y^{a}$
  $-$ ${}_{\ScrX}\!Y^{a}$ $=$ $\xi^{a}_{(1)}$ under the gauge
  transformation, respectively.
\end{theorem}
Actually, the gauge-variant variable $Y_{a}$ is given by
\begin{eqnarray}
  Y_{a}
  &:=&
       \sum_{l,m} \tilde{Y}_{A} S_{\delta} (dx^{A})_{a}
       +
       \sum_{l,m} \left(
       \tilde{Y}_{(e1)} \hat{D}_{p}S_{\delta}
       +
       \tilde{Y}_{(o1)} \epsilon_{pq}\hat{D}^{q}S_{\delta}
       \right)
       (dx^{p})_{a}
       ,
  \label{eq:2+2-Ya-def}
\end{eqnarray}
where
\begin{eqnarray}
  \tilde{Y}_{A}
  &:=&
       r \tilde{h}_{(e1)A}
       - \frac{r^{2}}{2} \bar{D}_{A}\tilde{h}_{(e2)}
       \label{eq:2+2-gauge-trans-tildeYA-def-sum}
       ,
  \\
  \label{eq:tildeYe-def}
  \tilde{Y}_{(e1)}
  &:=&
       \frac{r^{2}}{2} \tilde{h}_{(e2)}
       ,
  \\
  \label{eq:tildeYo-def}
  \tilde{Y}_{(o1)}
  &:=&
       - r^{2} \tilde{h}_{(o2)}
       .
\end{eqnarray}
Furthermore, including $l=0,1$ modes, the components of the
gauge-invariant part $\ScrF_{ab}$ of the metric perturbation
$h_{ab}$ is given by
\begin{eqnarray}
  \label{eq:2+2-gauge-invariant-variables-calFAB}
  \ScrF_{AB}
  &=&
      \sum_{l,m} \tilde{F}_{AB} S_{\delta}
      ,
  \\
  \label{eq:2+2-gauge-invariant-variables-calFAp}
  \ScrF_{Ap}
  &=&
      r \sum_{l,m} \tilde{F}_{A} \epsilon_{pq}
      \hat{D}^{q}S_{\delta}, \quad
      \hat{D}^{p}\ScrF_{Ap} = 0
      ,
  \\
  \label{eq:2+2-gauge-invariant-variables-calFpq}
  \ScrF_{pq}
  &=&
      \frac{1}{2} \gamma_{pq} r^{2} \sum_{l,m} \tilde{F} S_{\delta}
      ,
\end{eqnarray}
where $\tilde{F}_{AB}$, $\tilde{F}_{A}$, and $\tilde{F}$ are given by
\begin{eqnarray}
  \label{eq:gauge-inv-tildeFAB-def-sum}
  \tilde{F}_{AB}
  &:=&
       \tilde{h}_{AB}
       - 2 \bar{D}_{(A}\tilde{Y}_{B)}
       ,
  \\
  \label{eq:2+2-gauge-inv-def-tildeFA-sum}
  \tilde{F}_{A}
  &:=&
       \tilde{h}_{(o1)A}
       + r \bar{D}_{A}\tilde{h}_{(o2)}
       ,
  \\
  \label{eq:2+2-gauge-inv-tildeF-def-sum}
  \tilde{F}
  &:=&
       \tilde{h}_{(e0)}
       - \frac{4}{r} \tilde{Y}_{A} \bar{D}^{A}r
       + \tilde{h}_{(e2)} l(l+1)
  .
\end{eqnarray}
Thus, we have constructed gauge-invariant metric perturbations on
the Schwarzschild background spacetime including $l=0,1$ modes..


To discuss the linearized Einstein equation
(\ref{eq:einstein-equation-gauge-inv}) and the linear perturbation of
the continuity equation
\begin{eqnarray}
  \label{eq:continuity-linearized-energy-momentum-general}
  \nabla_{a}{}^{(1)}\!\ScrT_{b}^{\;\;a} = 0
\end{eqnarray}
of the gauge-invariant energy-momentum tensor
${}^{(1)}\!\ScrT_{b}^{\;\;a}:=g^{ac}{}^{(1)}\!\ScrT_{bc}$ on a
vacuum background spacetime, we consider the mode-decomposition of the
gauge-invariant part ${}^{(1)}\!\ScrT_{bc}$ of the linear perturbation
of the energy-momentum tensor through the set
(\ref{eq:harmonic-fucntions-set}) of the harmonics as follows:
\begin{eqnarray}
  {}^{(1)}\!\ScrT_{ab}
  &=&
      \sum_{l,m}
      \tilde{T}_{AB}
      S_{\delta}
      (dx^{A})_{a}(dx^{B})_{b}
      +
      r
      \sum_{l,m} \left\{
      \tilde{T}_{(e1)A} \hat{D}_{p}S_{\delta}
      +
      \tilde{T}_{(o1)A} \epsilon_{pr}\hat{D}^{r}S_{\delta}
      \right\}
      2 (dx^{A})_{(a}(dx^{p})_{b)}
      \nonumber\\
  &&
     +
     \sum_{l,m} \left\{
     \tilde{T}_{(e0)} \frac{1}{2} \gamma_{pq} S_{\delta}
     +
     \tilde{T}_{(e2)} \left(
     \hat{D}_{p}\hat{D}_{q}S_{\delta}
     -
     \frac{1}{2} \gamma_{pq} \hat{D}_{r}\hat{D}^{r}S_{\delta}
     \right)
     \right.
     \nonumber\\
  && \quad\quad\quad
     \left.
     +
     \tilde{T}_{(o2)}
     2 \epsilon_{r(p}\hat{D}_{q)}\hat{D}^{r}S_{\delta}
     \right\}
     (dx^{p})_{a}(dx^{q})_{b}
     .
     \label{eq:1st-pert-calTab-dd-decomp}
\end{eqnarray}
In terms of these mode coefficients, the components of the continuity
equation (\ref{eq:continuity-linearized-energy-momentum-general}) for
the gauge-invariant part of the linearized energy-momentum tensor are
summarized as follows:
\begin{eqnarray}
  &&
     \bar{D}^{C}\tilde{T}_{C}^{\;\;B}
     + \frac{2}{r} (\bar{D}^{D}r)\tilde{T}_{D}^{\;\;\;B}
     -  \frac{1}{r} l(l+1) \tilde{T}_{(e1)}^{B}
     -  \frac{1}{r} (\bar{D}^{B}r) \tilde{T}_{(e0)}
     =
     0
     ,
     \label{eq:div-barTab-linear-A}
  \\
  &&
     \bar{D}^{C}\tilde{T}_{(e1)C}
     + \frac{3}{r} (\bar{D}^{C}r) \tilde{T}_{(e1)C}
     + \frac{1}{2r} \tilde{T}_{(e0)}
     -  \frac{1}{2r} (l-1)(l+2) \tilde{T}_{(e2)}
     =
     0
     ,
     \label{eq:div-barTab-linear-p-even}
  \\
  &&
     \bar{D}^{C}\tilde{T}_{(o1)C}
     + \frac{3}{r} (\bar{D}^{D}r) \tilde{T}_{(o1)D}
     + \frac{1}{r} (l-1)(l+2) \tilde{T}_{(o2)}
     =
     0
     .
     \label{eq:div-barTab-linear-p-odd}
\end{eqnarray}


In the Part I paper~\cite{K.Nakamura-2021c}, we derived the linearized
Einstein equations, discussed the odd-mode perturbation $\tilde{F}_{Ap}$ in
Eq.~(\ref{eq:2+2-gauge-invariant-variables-calFAp}), and derived the
$l=1$ odd-mode solutions to these equations.
The Einstein equation for even mode $\tilde{F}_{AB}$ and $\tilde{F}$ in
Eqs.~(\ref{eq:2+2-gauge-invariant-variables-calFAB}) and
(\ref{eq:2+2-gauge-invariant-variables-calFpq}) also derived in the
Part I paper~\cite{K.Nakamura-2021c}, and derived $l=0,1$ even-mode
solutions are derived in the Part II paper~\cite{K.Nakamura-2021d}.
These solutions include the Kerr parameter perturbation and the
Schwarzschild mass parameter perturbation of the linear order in the
vacuum case and are physically reasonable.
Then, we conclude that our proposal is also physically reasonable.
The purpose of this paper is to check that our derived solutions
include the linearized LTB solution and non-rotating C-metric with the
Schwarzschild background.
For this purpose, the even-mode solutions are necessary.
Therefore, we review the strategy to derive the even-mode solutions,
below.


\subsection{Even-mode linearized Einstein equations}
\label{sec:Even_Einstein_equations}


The even-mode part of the linearized Einstein equation
(\ref{eq:einstein-equation-gauge-inv}) is summarized as follows:
\begin{eqnarray}
  \label{eq:linearized-Einstein-pq-traceless-even}
  \tilde{F}_{D}^{\;\;\;D}
  &=&
      -
      16 \pi r^{2}
      \tilde{T}_{(e2)}
      ,
  \\
  \bar{D}^{D}\tilde{\FF}_{AD}
  - \frac{1}{2} \bar{D}_{A}\tilde{F}
  &=&
      16 \pi
      \left[
      r \tilde{T}_{(e1)A}
      - \frac{1}{2} r^{2} \bar{D}_{A}\tilde{T}_{(e2)}
      \right]
      =:
      16 \pi S_{(ec)A}
      ,
      \label{eq:even-FAB-divergence-3}
\end{eqnarray}
where the variable $\tilde{\FF}_{AB}$ is the traceless part of the
variable $\tilde{F}_{AB}$ defined by
\begin{eqnarray}
  \label{eq:FF-def}
  \tilde{\FF}_{AB} := \tilde{F}_{AB} - \frac{1}{2} y_{AB} \tilde{F}_{C}^{\;\;C}.
\end{eqnarray}
We also have the evolution equations
\begin{eqnarray}
  &&
     \!\!\!\!\!\!\!\!\!\!\!\!\!\!\!\!\!\!\!\!\!\!\!\!
     \left(
     \bar{D}_{D}\bar{D}^{D}
     + \frac{2}{r} (\bar{D}^{D}r) \bar{D}_{D}
     -  \frac{(l-1)(l+2)}{r^{2}}
     \right) \tilde{F}
     - \frac{4}{r^{2}} (\bar{D}_{C}r) (\bar{D}_{D}r) \tilde{\FF}^{CD}
  =
      16 \pi S_{(F)}
      ,
     \label{eq:even-mode-tildeF-master-eq-mod-3}
  \\
  S_{(F)}
  \!\!\!\!&:=&\!\!\!\!
       \tilde{T}_{C}^{\;\;\;C}
       + 4 (\bar{D}_{D}r) \tilde{T}_{(e1)}^{D}
       -  2 r (\bar{D}_{D}r) \bar{D}^{D}\tilde{T}_{(e2)}
       -  (l(l+1)+2) \tilde{T}_{(e2)}
               .
               \label{eq:sourcee0-def}
\end{eqnarray}
\begin{eqnarray}
  &&
     \!\!\!\!\!\!\!\!\!\!\!\!\!\!\!\!\!\!\!\!\!\!\!\!
     \left[
     -  \bar{D}_{D}\bar{D}^{D}
     -  \frac{2}{r} (\bar{D}_{D}r) \bar{D}^{D}
     + \frac{4}{r} (\bar{D}^{D}\bar{D}_{D}r)
     + \frac{l(l+1)}{r^{2}}
     \right]
     \tilde{\FF}_{AB}
     \nonumber\\
  &&
     + \frac{4}{r} (\bar{D}^{D}r) \bar{D}_{(A}\tilde{\FF}_{B)D}
     -  \frac{2}{r} (\bar{D}_{(A}r) \bar{D}_{B)}\tilde{F}
     \nonumber\\
  \!\!\!\!&=&\!\!\!\!
  16 \pi S_{(\FF)AB}
              ,
              \label{eq:1st-pert-Einstein-non-vac-AB-traceless-final-3}
\end{eqnarray}
\begin{eqnarray}
  S_{(\FF)AB}
  \!\!\!\!&:=&\!\!\!\!
      \tilde{T}_{AB} - \frac{1}{2} y_{AB} \tilde{T}_{C}^{\;\;\;C}
      - 2 \left( \bar{D}_{(A}(r \tilde{T}_{(e1)B)}) - \frac{1}{2} y_{AB} \bar{D}^{D}(r \tilde{T}_{(e1)D}) \right)
      \nonumber\\
  &&\!\!\!\!
      + 2 \left( (\bar{D}_{(A}r) \bar{D}_{B)} - \frac{1}{2} y_{AB} (\bar{D}^{D}r) \bar{D}_{D}  \right) ( r \tilde{T}_{(e2)} )
      \nonumber\\
  &&\!\!\!\!
      + r \left( \bar{D}_{A}\bar{D}_{B} - \frac{1}{2} y_{AB} \bar{D}^{D}\bar{D}_{D} \right)(r \tilde{T}_{(e2)})
      \nonumber\\
  &&\!\!\!\!
      + 2 \left( (\bar{D}_{A}r) (\bar{D}_{B}r) - \frac{1}{2} y_{AB} (\bar{D}^{C}r) (\bar{D}_{C}r) \right) \tilde{T}_{(e2)}
      \nonumber\\
  &&\!\!\!\!
      + 2 y_{AB} (\bar{D}^{C}r) \tilde{T}_{(e1)C}
      -  r y_{AB} (\bar{D}^{C}r) \bar{D}_{C}\tilde{T}_{(e2)}
     ,
     \nonumber\\
  \label{eq:souce(FF)-def}
\end{eqnarray}
for the variable $\tilde{F}$ and the traceless variable
$\tilde{\FF}_{AB}$.
Of course, we have to take into account of the even-mode part of the
continuity equations (\ref{eq:div-barTab-linear-A}) and
(\ref{eq:div-barTab-linear-p-even}) of the
linearized energy-momentum tensor.
We note that these equations are valid not only for $l\geq 2$ modes
but also $l=0,1$ modes in our formulation.


To evaluate
Eqs.~(\ref{eq:even-mode-tildeF-master-eq-mod-3})--(\ref{eq:souce(FF)-def}),
it is convenient to introduce the component $X_{(e)}$ and $Y_{(e)}$ of
the traceless variable $\tilde{\FF}_{AB}$ by
\begin{eqnarray}
  \label{eq:Xe-Ye-def}
  \tilde{\FF}_{AB}
  =:
  X_{(e)} \left\{ - f (dt)_{A}(dt)_{B} - f^{-1} (dr)_{A}(dr)_{B}\right\}
  +
  2 Y_{(e)} (dt)_{(A}(dr)_{B)}
  ,
\end{eqnarray}
and the Moncrief variable $\Phi_{(e)}$ defined by
\begin{eqnarray}
  \label{eq:Moncrief-master-variable-final}
  \Phi_{(e)}
  :=
  \frac{r}{\Lambda} \left[
  f X_{(e)}
  - \frac{1}{4} \Lambda \tilde{F}
  + \frac{1}{2} r f \partial_{r}\tilde{F}
  \right]
  ,
\end{eqnarray}
where
\begin{eqnarray}
  \label{eq:mu-Lambda-defs}
  \Lambda = \mu + 3(1-f), \quad \mu := (l-1)(l+2).
\end{eqnarray}
From Eqs.~(\ref{eq:even-FAB-divergence-3}) and
(\ref{eq:1st-pert-Einstein-non-vac-AB-traceless-final-3}), we obtain
the initial value constraints for the variable $\tilde{F}$ and
$Y_{(e)}$ as follows:
\begin{eqnarray}
  l(l+1) \Lambda \tilde{F}
  &=&
      -  8 f \Lambda \partial_{r}\Phi_{(e)}
      + \frac{4}{r} \left[ 6 f (1-f) - l(l+1) \Lambda \right] \Phi_{(e)}
      - 64 \pi r^{2} S_{(\Lambda\tilde{F})}
      ,
      \label{eq:-Ein-non-vac-XF-constraint-apha-Phie-with-SLambdaF-sum}
  \\
  l(l+1) Y_{(e)}
  &=&
      r \partial_{t}\left(
      2 X_{(e)}
      + r \partial_{r}\tilde{F}
      \right)
      +  \frac{3f-1}{2f} r \partial_{t}\tilde{F}
      + 16 \pi r^{2} S_{(Y_{(e)})}
      ,
      \label{eq:ll+1fYe-1st-pert-Ein-non-vac-tr-FAB-div-t-r-sum-3}
\end{eqnarray}
where the source term $S_{(\Lambda\tilde{F})}$ and $S_{(Y_{(e)})}$ are
given by
\begin{eqnarray}
  S_{(\Lambda\tilde{F})}
  &:=&
      \tilde{T}_{tt}
      + r f^{2} \partial_{r}\tilde{T}_{(e2)}
      + 2 f (f+1) \tilde{T}_{(e2)}
      +  \frac{1}{2} f (l-1)(l+2) \tilde{T}_{(e2)}
      ,
      \label{eq:SLambdaF-def-explicit}
  \\
  S_{(Y_{(e)})}
  &:=&
       \tilde{T}_{tr}
       + r \partial_{t}\tilde{T}_{(e2)}
       .
       \label{eq:sourceYe-SPsie-explicit-sum}
\end{eqnarray}
Furthermore, we obtain the evolution equations for the variables
$\Phi_{(e)}$ and $\tilde{F}$ as follows:
\begin{eqnarray}
  &&
     -  \frac{1}{f} \partial_{t}^{2}\Phi_{(e)}
     + \partial_{r}\left[ f \partial_{r}\Phi_{(e)} \right]
     -
     V_{even} \Phi_{(e)}
     =
     16 \pi \frac{r}{\Lambda} S_{(\Phi_{(e)})}
     ,
     \label{eq:Zerilli-Moncrief-eq-final-sum}
  \\
  &&
     -  \frac{1}{f} \partial_{t}^{2}\tilde{F}
     + \partial_{r}( f \partial_{r}\tilde{F} )
     + \frac{1}{r^{2}} 3(1-f) \tilde{F}
     + \frac{4\Lambda}{r^{3}} \Phi_{(e)}
     =
     16 \pi S_{(F)}
     ,
     \label{eq:even-mode-tildeF-eq-Phie-sum}
\end{eqnarray}
where the potential function $V_{even}$ in
Eq.~(\ref{eq:Zerilli-Moncrief-eq-final-sum}) is defined by
\begin{eqnarray}
  V_{even}
  &:=&
      \frac{1}{r^{2}\Lambda^{2}}
      \left[
      \Lambda^{3}
      - 2 (2-3f) \Lambda^{2}
      + 6 (1-3f) (1-f) \Lambda
      + 18 f (1-f)^{2}
      \right]
      ,
      \label{eq:Zerilli-Moncrief-master-potential-final-sum}
\end{eqnarray}
and the source terms in Eq.~(\ref{eq:Zerilli-Moncrief-eq-final-sum})
and (\ref{eq:even-mode-tildeF-eq-Phie-sum}) are given by
\begin{eqnarray}
  S_{(\Phi_{(e)})}
  &:=&
      \frac{1}{2} \left(
      \frac{\Lambda}{2f} - 1
      \right) \tilde{T}_{tt}
      + \frac{1}{2} \left(
      (2-f) -  \frac{1}{2} \Lambda
      \right) f \tilde{T}_{rr}
      -  \frac{1}{2} r \partial_{r}\tilde{T}_{tt}
      + \frac{1}{2} f^{2} r \partial_{r}\tilde{T}_{rr}
      \nonumber\\
  &&
      -  \frac{f}{2} \tilde{T}_{(e0)}
      -  l(l+1) f \tilde{T}_{(e1)r}
      \nonumber\\
  &&
      + \frac{1}{2} r^{2} \partial_{t}^{2}\tilde{T}_{(e2)}
      -  \frac{1}{2} f^{2} r^{2} \partial_{r}^{2}\tilde{T}_{(e2)}
      - \frac{1}{2} 3(1+f) r f \partial_{r}\tilde{T}_{(e2)}
      \nonumber\\
  &&
      -  \frac{1}{2} (7-3f) f \tilde{T}_{(e2)}
      + \frac{1}{4} (l(l+1)-1-f) (l(l+1)+2) \tilde{T}_{(e2)}
      \nonumber\\
  &&
      - \frac{3(1-f)}{\Lambda} \left[
      \tilde{T}_{tt}
      + r f^{2} \partial_{r}\tilde{T}_{(e2)}
      + \frac{1}{2} (1+7f) f \tilde{T}_{(e2)}
      \right]
     ,
      \label{eq:SPhie-def-explicit-sum}
  \\
  S_{(F)}
  &:=&
      -  \frac{1}{f} \tilde{T}_{tt}
      + f \tilde{T}_{rr}
      + 4 f \tilde{T}_{(e1)r}
      -  2 r f \partial_{r}\tilde{T}_{(e2)}
      -  (l(l+1)+2) \tilde{T}_{(e2)}
      ,
      \label{eq:source(F)-def-sum-component}
\end{eqnarray}
respectively.
The consistency of evolution equations
(\ref{eq:Zerilli-Moncrief-eq-final-sum}) and
(\ref{eq:even-mode-tildeF-eq-Phie-sum}) with the initial value
constraint
(\ref{eq:-Ein-non-vac-XF-constraint-apha-Phie-with-SLambdaF-sum})
leads the identity
\begin{eqnarray}
  0
  &=&
      r^{2} \Lambda \partial_{t}^{2}S_{(\Lambda\tilde{F})}
      -  \left[
      (5-3f) \Lambda
      + 3 (1-f) (1+f)
      + 18 \frac{1}{\Lambda} f (1-f)^{2}
      \right] f S_{(\Lambda\tilde{F})}
      \nonumber\\
  &&
     - 2 \left[
      3 (1-f) + 2 \Lambda
     \right] f^{2} r \partial_{r}S_{(\Lambda\tilde{F})}
      - \Lambda r^{2} f \partial_{r}\left[ f \partial_{r}S_{(\Lambda\tilde{F})} \right]
      \nonumber\\
  &&
     + \frac{1}{4}  \left[ (1-3f) - \Lambda  \right] \Lambda^{2} f S_{(F)}
      \nonumber\\
  &&
      -  2 r f^{2} \Lambda \partial_{r}S_{(\Phi_{(e)})}
     -  \left[ \Lambda + (1+3f) \right] \Lambda f S_{(\Phi_{(e)})}
      .
      \label{eq:even-mode-tildeF-eq-Phie-remainig-source-tmp}
\end{eqnarray}
Actually, we can confirm
Eq.~(\ref{eq:even-mode-tildeF-eq-Phie-remainig-source-tmp}) from the
definitions (\ref{eq:SLambdaF-def-explicit}),
(\ref{eq:SPhie-def-explicit-sum}), and
(\ref{eq:source(F)-def-sum-component}),
and the continuity equations
(\ref{eq:div-barTab-linear-A}) and
(\ref{eq:div-barTab-linear-p-even}), i.e.,
\begin{eqnarray}
  &&\!\!\!\!\!\!\!\!
     -  \partial_{t}\tilde{T}_{tt}
     + f^{2} \partial_{r}\tilde{T}_{rt}
     + \frac{(1+f)f}{r} \tilde{T}_{rt}
     -  \frac{f}{r} l(l+1) \tilde{T}_{(e1)t}
     =
     0
     ,
     \label{eq:div-barTab-linear-AB-t}
  \\
  &&\!\!\!\!\!\!\!\!
     - \partial_{t}\tilde{T}_{tr}
     + \frac{1-f}{2rf} \tilde{T}_{tt}
     + f^{2} \partial_{r}\tilde{T}_{rr}
     + \frac{(3+f)f}{2r} \tilde{T}_{rr}
     -  \frac{f}{r} l(l+1) \tilde{T}_{(e1)r}
     -  \frac{f}{r} \tilde{T}_{(e0)}
     =
     0
     ,
     \label{eq:div-barTab-linear-AB-r}
  \\
  &&\!\!\!\!\!\!\!\!
     - \partial_{t}\tilde{T}_{(e1)t}
     + f^{2} \partial_{r}\tilde{T}_{(e1)r}
     + \frac{(1+2f)f}{r} \tilde{T}_{(e1)r}
     + \frac{f}{2r} \tilde{T}_{(e0)}
     -  \frac{f}{2r} (l-1)(l+2) \tilde{T}_{(e2)}
     =
     0
     .
     \label{eq:div-barTab-linear-p-even-mode}
\end{eqnarray}


For the mode with $l\neq 0$, the master equation
(\ref{eq:Zerilli-Moncrief-eq-final-sum}) is solved through
appropriate boundary conditions for the Cauchy problem and obtain the
Moncrief variable $\Phi_{(e)}$.
Then, we obtain the variable $\tilde{F}$ through
Eq.~(\ref{eq:-Ein-non-vac-XF-constraint-apha-Phie-with-SLambdaF-sum}).
From the solution $(\Phi_{(e)},\tilde{F})$, we obtain the component
$X_{(e)}$ through the definition
(\ref{eq:Moncrief-master-variable-final}) of the Moncrief variable.
Through the solution $(\Phi_{(e)},\tilde{F},X_{(e)})$, we obtain the
component $Y_{(e)}$ through
Eq.~(\ref{eq:ll+1fYe-1st-pert-Ein-non-vac-tr-FAB-div-t-r-sum-3}).
We can check the evolution equation
(\ref{eq:even-mode-tildeF-eq-Phie-sum}) as a consistency check of
solutions.
Together with Eq.~(\ref{eq:linearized-Einstein-pq-traceless-even}), we
obtain the solution $(\tilde{\FF}_{AB},\tilde{F})$ as a solution to
the linearized Einstein equations when $l\neq 0$.


Actually, from the above strategy, for the $l=1$-mode perturbation, we
can derive the solution to the linearized Einstein equation through
the strategy for $l\neq 0$ mode perturbation described above.
For $m=0$ mode, in the Part II paper~\cite{K.Nakamura-2021d}, we
derived the following solution to the linearized Einstein equation
\begin{eqnarray}
  \ScrF_{ab}
  &=&
      {\pounds}_{V}g_{ab}
      - \frac{16 \pi r^{2}}{3(1-f)}\left[
      f^{2} \left\{
      \frac{1+f}{2} \tilde{T}_{rr}
      + r f \partial_{r}\tilde{T}_{rr}
      -  \tilde{T}_{(e0)}
      -  4 \tilde{T}_{(e1)r}
      \right\}  (dt)_{a}(dt)_{b}
      \right.
      \nonumber\\
  && \quad\quad\quad\quad\quad\quad\quad\quad\quad\quad
     \left.
      + \frac{2r}{f} \left\{
      \partial_{t}\tilde{T}_{tt}
      - \frac{3f(1-f)}{2r} \tilde{T}_{tr}
      \right\} (dt)_{(a}(dr)_{b)}
     \right.
      \nonumber\\
  && \quad\quad\quad\quad\quad\quad\quad\quad\quad\quad
     \left.
      + \frac{r}{f}
      \left\{
      \partial_{r}\tilde{T}_{tt}
      - \frac{3(1-3f)}{2rf} \tilde{T}_{tt}
      \right\} (dr)_{a}(dr)_{b}
     \right.
      \nonumber\\
  && \quad\quad\quad\quad\quad\quad\quad\quad\quad\quad
     \left.
      + r^{2} \tilde{T}_{tt} \gamma_{ab}
     \right] \cos\theta
      ,
      \label{eq:tildeFab-l=1-m=0-nonvacsum-2-cov}
\end{eqnarray}
where the vector field $V_{a}$ is given by
\begin{eqnarray}
  V_{a}
  &:=&
       -  r \partial_{t}\Phi_{(e)} \cos\theta (dt)_{a}
       + \left( \Phi_{(e)} - r \partial_{r}\Phi_{(e)} \right) \cos\theta (dr)_{a}
       -  r \Phi_{(e)} \sin\theta (d\theta)_{a}
       .
       \label{eq:generator-covariant-vacuum-l=1-m=0-result-2}
\end{eqnarray}


On the other hand, for the $l=0$-mode, we may choose
$\tilde{T}_{(e1)A}=0$ and $\tilde{T}_{(e2)}=0$ and we may regard that
the tensor $\tilde{F}_{AB}$ is traceless.
Furthermore,
Eqs.~(\ref{eq:-Ein-non-vac-XF-constraint-apha-Phie-with-SLambdaF-sum})
and (\ref{eq:ll+1fYe-1st-pert-Ein-non-vac-tr-FAB-div-t-r-sum-3})
yield the $r$- and $t$-derivative of the Moncrief variable
$\Phi_{(e)}$, respectively.
The integrability of these equations are guaranteed by the continuity
equation (\ref{eq:div-barTab-linear-A}).
Then, we obtain the Moncrief variable $\Phi_{(e)}$.
In this case, the master equation
(\ref{eq:Zerilli-Moncrief-eq-final-sum}) is trivial and the evolution
equation (\ref{eq:even-mode-tildeF-eq-Phie-sum}) gives the variable
$\tilde{F}$.
Then, we obtain the variable $(\Phi_{(e)},\tilde{F})$.
Through the definition (\ref{eq:Moncrief-master-variable-final}) of
the Moncrief variable $\Phi_{(e)}$, we obtain the component $X_{(e)}$.
To obtain the component $Y_{(e)}$, we regard the constraints
(\ref{eq:even-FAB-divergence-3}) as the equation for the component
$Y_{(e)}$.
Through this strategy, in the Part II paper~\cite{K.Nakamura-2021d},
we derived the $l=0$ mode solution
\begin{eqnarray}
  \label{eq:calFab+poundsVg-l=0-non-vac-final}
  \ScrF_{ab}
  &=&
      \frac{2}{r} \left(M_{1}+4\pi \int dr \frac{r^{2}}{f} T_{tt}\right)
      \left((dt)_{a}(dt)_{a}+ \frac{1}{f^{2}} (dr)_{a}(dr)_{a}\right)
      \nonumber\\
  &&
      + 2 \left[4 \pi r \int dt \left(\frac{1}{f} \tilde{T}_{tt} + f \tilde{T}_{rr} \right)\right] (dt)_{(a}(dr)_{b)}
      + {\pounds}_{V}g_{ab}
      ,
\end{eqnarray}
where
\begin{eqnarray}
  \label{eq:Va-result-non-vac-final}
  V_{a}
  =
  \left(
  \frac{f}{4} \Upsilon
  + \frac{rf}{4} \partial_{r}\Upsilon
  -  \frac{r \Xi(r)}{(1-3f)}
  + f \int dr \frac{2 \Xi(r)}{f(1-3f)^{2}}
  \right) (dt)_{a}
  +
  \frac{1}{4f} r \partial_{t}\Upsilon (dr)_{a}
  .
\end{eqnarray}
Here, the variable $\tilde{F}=:\partial_{t}\Upsilon$ must satisfy
Eq.~(\ref{eq:even-mode-tildeF-eq-Phie-sum}) and $\Xi(r)$ is an
arbitrary function of $r$.


\section{Realization of LTB solution as a perturbation of
  Schwarzschild spacetime}
\label{sec:LTB_solution_as_a_perturbation_of_Schwarzschild_spacetime}


\subsection{Perturbative expression of the LTB solution on
  Schwarzschild background spacetime}
\label{sec:Perturbative_expression_of_the_LTB_sol._on_Schwarzschild_BG_spacetime}


Here, we consider the Lema\^itre-Tolman-Bondi (LTB)
solution~\cite{L.Landau-E.Lifshitz-1962} which is an exact solution to
the Einstein equation with the matter field
\begin{eqnarray}
  \label{eq:dast-energy-momentum-tensor}
  T_{ab} = \rho u_{a}u_{b}, \quad u_{a} = - (d\tau)_{a},
\end{eqnarray}
and the metric
\begin{eqnarray}
  \label{eq:LTB-metric}
  g_{ab}
  &=&
      - (d\tau)_{a} (d\tau)_{b}
      + \frac{(\partial_{R}r)^{2}}{1+f(R)} (dR)_{a}(dR)_{b}
      + r^{2} \gamma_{ab}
      ,
      \quad
      r = r(\tau,R)
     .
\end{eqnarray}
This solution is a spherically symmetric solution to the Einstein equation.
The function $r=r(\tau,R)$ satisfies the differential equation
\begin{eqnarray}
  \label{eq:LTB-Hubble-equation}
  (\partial_{\tau}r)^{2}
  =
  \frac{F(R)}{r}
  +
  f(R)
  .
\end{eqnarray}
Here, we note that $F(R)$ is an arbitrary function of $R$, which
represents initial distribution of the dust matter.
$f(R)$ is also an arbitrary function of $R$ which represents initial
distribution of the energy of dust field in Newtonian sense.
The solution to Eq.~(\ref{eq:LTB-Hubble-equation}) is given in the
three cases
\begin{description}
\item[(i) $f(R)>0$ : ]
  \begin{eqnarray}
    \label{eq:Morita-Nakamura-Kasai-1998-f-positive}
    r = \frac{F(R)}{2f(R)}(\cosh\eta - 1), \quad
    \tau_{0}(R) - \tau  = \frac{F(R)}{2f(R)^{3/2}} (\sinh\eta -\eta)
    ,
  \end{eqnarray}
\item[(ii) $f(R)<0$ : ]
  \begin{eqnarray}
    \label{eq:Morita-Nakamura-Kasai-1998-f-negative}
    r = \frac{F(R)}{-2f(R)}(1-\cos\eta), \quad
    \tau_{0}(R) - \tau = \frac{F(R)}{2(-f(R))^{3/2}} (\eta - \sin\eta)
    ,
  \end{eqnarray}
\item[(iii) $f(R)=0$ : ]
  \begin{eqnarray}
    \label{eq:Morita-Nakamura-Kasai-1998-f=0}
    r = \left(\frac{9F(R)}{4}\right)^{1/3} \left[\tau_{0}(R)-\tau\right]^{2/3}
    .
  \end{eqnarray}
\end{description}
The energy density $\rho$ is given by
\begin{eqnarray}
  \label{eq:LTB-energy-density}
  8\pi \rho = \frac{\partial_{R}F}{(\partial_{R}r)r^{2}}.
\end{eqnarray}
The LTB solution includes the three arbitrary functions $f(R)$,
$F(R)$, and $\tau_{0}(R)$.


Here, we consider the vacuum case $\rho=0$.
In this case, from Eq.~(\ref{eq:LTB-energy-density}), we have
\begin{eqnarray}
  \label{eq:LTB-energy-density-vacuum}
  \partial_{R}F = 0
  .
\end{eqnarray}
Furthermore, we consider the case $f=0$.
Here, we chose $\tau_{0}=R$, i.e., $\partial_{R}\tau_{0}=1$.
In this case, Eq.~(\ref{eq:Morita-Nakamura-Kasai-1998-f=0}) yields
\begin{eqnarray}
  (dr)_{a}
  &=&
      \left(\frac{9F}{4}\right)^{1/3}
      \frac{2}{3}
      \left[R-\tau\right]^{-1/3}
      \left[
      (dR)_{a} - (d\tau)_{a}
      \right]
      \nonumber\\
  &=&
      \left(\frac{F}{r}\right)^{1/2}
      \left[
      (dR)_{a} - (d\tau)_{a}
      \right]
      ,
      \label{eq:derivative-Morita-Nakamura-Kasai-1998-f=0}
  \\
  \label{eq:R-derivative-Morita-Nakamura-Kasai-1998-f=0}
  (\partial_{R}r)
  &=&
      \left(\frac{F}{r}\right)^{1/2}
  ,
\end{eqnarray}
and
\begin{eqnarray}
  (dR)_{a}
  &=&
      (d\tau)_{a}
      +
      \left(\frac{F}{r}\right)^{-1/2}
      (dr)_{a}
      .
      \label{eq:dRa-Morita-Nakamura-Kasai-1998-f=0}
\end{eqnarray}
Then, the metric (\ref{eq:LTB-metric}) is given by
\begin{eqnarray}
  g_{ab}
  &=&
      - (d\tau)_{a} (d\tau)_{b}
      + (\partial_{R}r)^{2} (dR)_{a}(dR)_{b}
      + r^{2} \gamma_{ab}
      \nonumber\\
  &=&
      -
      \left(
      1
      -
      \frac{F}{r}
      \right)
      \left[ (d\tau)_{a} - \left( 1 - \frac{F}{r} \right)^{-1} \left(\frac{F}{r}\right)^{1/2} (dr)_{a} \right]
      \nonumber\\
  && \quad\quad\quad\quad\quad\quad
     \times
      \left[ (d\tau)_{b} - \left( 1 - \frac{F}{r} \right)^{-1} \left(\frac{F}{r}\right)^{1/2} (dr)_{b} \right]
      \nonumber\\
  &&
     +
     \left( 1 - \frac{F}{r} \right)^{-1}
     (dr)_{a}
     (dr)_{b}
     + r^{2} \gamma_{ab}
     .
     \label{eq:dRa-Morita-Nakamura-Kasai-1998-Schwarz}
\end{eqnarray}
Here, we define the time function $t$ by
\begin{eqnarray}
  (dt)_{a}
  :=
  (d\tau)_{a}
  -
  \left(
  1
  -
  \frac{F}{r}
  \right)^{-1}
  \left(\frac{F}{r}\right)^{1/2}
  (dr)_{a}
  .
  \label{eq:Killing-time-function-LTB-1998-f=0}
\end{eqnarray}
Then, we obtain
\begin{eqnarray}
  g_{ab}
  =
  -
  f
  (dt)_{a}
  (dt)_{b}
  +
  f^{-1}
  (dr)_{a}
  (dr)_{b}
  +
  r^{2} \gamma_{ab}
  ,
  \quad
  f = 1 - \frac{2M}{r}
  ,
  \label{eq:Schwarzschild-static-chart-background}
\end{eqnarray}
with the identification $F=2M$.
This is the Schwarzschild metric with the mass parameter $M$.


Now, we consider the perturbation of the Schwarzschild spacetime which
is derived by the exact LTB solution (\ref{eq:LTB-metric}) so that
\begin{eqnarray}
  F(R)
  &=&
      2 \left[
      M
      +
      \epsilon
      m_{1}(R)
      \right]
      + O(\epsilon^{2})
      ,
      \label{eq:FR-perturbation-def}
  \\
  f(R)
  &=&
      0
      +
      \epsilon
      f_{1}(R)
      + O(\epsilon^{2})
      ,
      \label{eq:fR-perturbation-def}
  \\
  \tau_{0}(R)
  &=&
      R
      +
      \epsilon
      \tau_{1}(R)
      + O(\epsilon^{2})
      .
      \label{eq:tau0R-perturbation-def}
\end{eqnarray}
Through these perturbations
(\ref{eq:FR-perturbation-def})--(\ref{eq:tau0R-perturbation-def}), we
consider the perturbative expansion of the function $r$ which is
determined by Eq.~(\ref{eq:LTB-Hubble-equation}):
\begin{eqnarray}
  \label{eq:circum-ference-perturbations}
  r(\tau,R) = r_{s}(\tau,R) + \epsilon r_{1}(\tau,R)
  + O(\epsilon^{2})
  .
\end{eqnarray}
Here, the function $r_{s}(\tau,R)$ is given by
Eq.~(\ref{eq:Morita-Nakamura-Kasai-1998-f=0}), i.e.,
\begin{eqnarray}
  \label{eq:LTB-flat-vacuum-circumference}
  r_{s}(\tau,R)
  =
  r(\tau,R)
  =
  \left(\frac{9M}{2}\right)^{1/3}
  \left[R-\tau\right]^{2/3}
  .
\end{eqnarray}
In Eqs.~(\ref{eq:tau0R-perturbation-def}) and
(\ref{eq:LTB-flat-vacuum-circumference}), we chose the background
value of the function $\tau_{0}(R)$ to be $R$.


Through this perturbative expansion, we evaluate
Eq.~(\ref{eq:LTB-Hubble-equation}) and we obtain
\begin{eqnarray}
  &&
     O(\epsilon^{0})  \quad : \quad
     (\partial_{\tau}r_{s})^{2}
     -
     \frac{2M}{r_{s}}
     =
     0
     \label{eq:LTB-Hubble-equation-background}
  \\
  &&
     O(\epsilon^{1})  \quad : \quad
     (\partial_{\tau}r_{s}) (\partial_{\tau}r_{1})
     -
     \frac{m_{1}(R)}{r_{s}}
     +
     \frac{M}{r_{s}^{2}}
     r_{1}
     -
     \frac{1}{2}
     f_{1}(R)
     =
     0
     .
     \label{eq:LTB-Hubble-equation-linear-pert-pre}
\end{eqnarray}
Using Eq.~(\ref{eq:LTB-Hubble-equation-background}), the linear
perturbation (\ref{eq:LTB-Hubble-equation-linear-pert-pre}) yields
\begin{eqnarray}
     (1-f)^{1/2}
     (\partial_{\tau}r_{1})
     +
     \frac{m_{1}(R)}{r}
     -
     \frac{M}{r^{2}}
     r_{1}
     +
     \frac{1}{2}
     f_{1}(R)
     =
     0
     ,
     \label{eq:LTB-Hubble-equation-linear-pert}
\end{eqnarray}
where we used
\begin{eqnarray}
  \label{eq:partialtaur+sqrt1-f=0}
  \partial_{\tau}r_{s} = - (1-f)^{1/2}
\end{eqnarray}
and the replacement $r_{s}\rightarrow r$.
The solution to Eq.~(\ref{eq:LTB-Hubble-equation-linear-pert}) is
given by
\begin{eqnarray}
  r_{1}
  &=&
      \left(\frac{M}{6}\right)^{1/3}
      \frac{m_{1}(R)}{M}
      \left[R-\tau\right]^{2/3}
      -
      \frac{3}{20}
      \left(\frac{6}{M}\right)^{1/3}
      f_{1}(R)
      \left[R-\tau\right]^{+4/3}
      \nonumber\\
  &&
      +
      B(R)
      \left[R-\tau\right]^{-1/3}
      .
      \label{eq:LTB-Hubble-equation-linear-pert-sol.-sum}
\end{eqnarray}
From the comparison with Eq.~(\ref{eq:LTB-flat-vacuum-circumference}),
$B(R)$ is the perturbation of the $\tau_{1}(R)$ as
$\tau_{0}(R)=R+\tau_{1}(R)$ in the exact solution
(\ref{eq:Morita-Nakamura-Kasai-1998-f-positive})--(\ref{eq:Morita-Nakamura-Kasai-1998-f=0}).
Furthermore, the solution
(\ref{eq:LTB-Hubble-equation-linear-pert-sol.-sum}) can be also derived
from the exact solution
(\ref{eq:Morita-Nakamura-Kasai-1998-f-positive})--(\ref{eq:Morita-Nakamura-Kasai-1998-f=0}).
From Eq.~(\ref{eq:LTB-energy-density}), the perturbative dust energy
density given by
\begin{eqnarray}
  \label{eq:LTB-energy-density-perturbation}
  8\pi \rho = \frac{2\partial_{R}m_{1}(R)}{(\partial_{R}r)r^{2}}.
\end{eqnarray}


Through the perturbative solution
(\ref{eq:LTB-Hubble-equation-linear-pert-sol.-sum}), the metric
(\ref{eq:LTB-metric}) is given by
\begin{eqnarray}
  g_{ab}
  &=&
      - (d\tau)_{a} (d\tau)_{b}
      +
      (\partial_{R}r)^{2}
      (dR)_{a}(dR)_{b}
      +
      r^{2}
      \gamma_{ab}
      \nonumber\\
  &&
     +
     \epsilon
     \left[
     \left(
     2(\partial_{R}r_{1})
     -
     f_{1}
     (\partial_{R}r)
     \right)
     (\partial_{R}r)
     (dR)_{a}(dR)_{b}
     +
     2
     r
     r_{1}
     \gamma_{ab}
     \right]
     + O(\epsilon^{2})
     \nonumber\\
  &=:&
       g_{ab}^{(0)}
       +
       \epsilon
       {}_{\ScrX}\!h_{ab}
       + O(\epsilon^{2})
       .
       \label{eq:LTB-metric-perturbative-expansion}
\end{eqnarray}
As shown in Eq.~(\ref{eq:Schwarzschild-static-chart-background}),
the background metric $g_{ab}^{(0)}$ is given by the Schwarzschild
metric in the static chart.
On the other hand, the linear order perturbation ${}_{\ScrX}\!h_{ab}$
(in the gauge $\ScrX_{\epsilon}$) is given by
\begin{eqnarray}
  {}_{\ScrX}\!h_{ab}
  :=
  \left(
  2
  (\partial_{R}r_{1})
  -
  f_{1}(R)
  (\partial_{R}r)
  \right)
  (\partial_{R}r)
  (dR)_{a}(dR)_{b}
  +
  2
  r
  r_{1}
  \gamma_{ab}
  .
  \label{eq:LTB-linear-pert-with-Schwarzschild-BG}
\end{eqnarray}
Here, we fixed the second-kind gauge so that
\begin{eqnarray}
  \label{eq:second-gauge-fixed-LTB}
  \ScrX_{\epsilon} :
  (\tau,R,\theta,\phi)\in\ScrM_{ph} \mapsto
  (\tau,R,\theta,\phi)\in\ScrM.
\end{eqnarray}
Of course, if we employ the different gauge choice
$\ScrY_{\epsilon}$ from the above gauge-choice
$\ScrX_{\epsilon}$, we obtain the different expression of the
metric perturbation ${}_{\ScrY}\!h_{ab}$ $=$ ${}_{\ScrX}\!h_{ab}$
$+$ ${\pounds}_{\xi}g_{ab}$, where $\xi^{a}$ is the generator of
second-rank gauge transformation
$\ScrX_{\epsilon}\rightarrow\ScrY_{\epsilon}$.


\subsection{Expression of the perturbative LTB solution in static chart}
\label{sec:Schwarzschild_static_chart_LTB_solution}


Here, we consider the expression of the linear perturbation
${}_{\ScrX}\!h_{ab}$ given by
Eq.~(\ref{eq:LTB-linear-pert-with-Schwarzschild-BG}).
Here, the radial coordinate $r$ is related to the coordinates $\tau$
and $R$ through Eq.~(\ref{eq:LTB-flat-vacuum-circumference}) as
\begin{eqnarray}
  \label{eq:LTB-flat-vacuum-circumference-2}
  R
  -
  \tau
  =
  \left(
  \frac{2}{9M}
  \right)^{1/2}
  r^{3/2}
  .
\end{eqnarray}
Then, we obtain
\begin{eqnarray}
  \label{eq:LTB-flat-vacuum-circumference-3}
  R
  =
  \tau
  +
  \frac{4M}{3}
  \left(\frac{r}{2M}\right)^{3/2}
  .
\end{eqnarray}
Furthermore, the relation of the time function $t$ and coordinates
$\tau$ and $r$ is given by
Eq.~(\ref{eq:Killing-time-function-LTB-1998-f=0}) as
\begin{eqnarray}
  t
  =
  \tau
  +
  4M
  \left[
  \left(\frac{r}{2M}\right)^{1/2}
  +
  \ln\left\{
  \frac{\displaystyle
  \left(\frac{\displaystyle r}{\displaystyle 2M}\right)^{1/2}-1
  }{\displaystyle
  \left(\frac{\displaystyle r}{\displaystyle 2M}\right)^{1/2}+1
  }
  \right\}
  \right]
  .
  \label{eq:Killing-time-function-LTB-1998-f=0-2}
\end{eqnarray}
Then, we have obtained
\begin{eqnarray}
  \tau
  &=&
      t
      -
      4M
      \left[
      \left(\frac{r}{2M}\right)^{1/2}
      +
      \ln\left\{
      \frac{\displaystyle
      \left(\frac{\displaystyle r}{\displaystyle 2M}\right)^{1/2}-1
      }{\displaystyle
      \left(\frac{\displaystyle r}{\displaystyle 2M}\right)^{1/2}+1
      }
      \right\}
      \right]
      ,
      \label{eq:Killing-time-function-LTB-1998-f=0-3}
  \\
  R
  &=&
      t
      -
      4M
      \left[
      \frac{1}{3}
      \left(\frac{r}{2M}\right)^{3/2}
      +
      \left(\frac{r}{2M}\right)^{1/2}
      +
      \ln\left\{
      \frac{\displaystyle
      \left(\frac{\displaystyle r}{\displaystyle 2M}\right)^{1/2}-1
      }{\displaystyle
      \left(\frac{\displaystyle r}{\displaystyle 2M}\right)^{1/2}+1
      }
      \right\}
      \right]
      .
      \label{eq:LTB-flat-vacuum-circumference-4}
\end{eqnarray}
Through the coordinates $(t,r)$, we express the linear perturbation
${}_{\ScrX}\!h_{ab}$ in
Eq.~(\ref{eq:LTB-linear-pert-with-Schwarzschild-BG}).
From Eqs.~(\ref{eq:dRa-Morita-Nakamura-Kasai-1998-f=0}) and
(\ref{eq:Killing-time-function-LTB-1998-f=0}) with $F=2M$, we obtain
\begin{eqnarray}
  (dR)_{a}
  &=&
      (dt)_{a}
      +
      f^{-1}
      (1-f)^{-1/2}
      (dr)_{a}
      ,
      \quad
      f = 1 - \frac{2M}{r}
      ,
      \label{eq:dRa-f=0-tau0=R-with-Killing-time-sum-R}
  \\
  (d\tau)_{a}
  &=&
      (dt)_{a}
      +
      f^{-1}
      (1-f)^{1/2}
      (dr)_{a}
      .
      \label{eq:dRa-f=0-tau0=R-with-Killing-time-sum-tau}
\end{eqnarray}


Before evaluating the metric perturbation ${}_{\ScrX}\!h_{ab}$, we
consider the perturbation of the energy momentum tensor of the matter
field.
In the case of the LTB solution, the matter field is characterized by
the dust field whose energy momentum tensor
(\ref{eq:dast-energy-momentum-tensor}) is given by
\begin{eqnarray}
  \label{eq:dust-energy-momentum-tesor-general}
  T_{ab} = \rho u_{a} u_{b}, \quad u_{a} = - (d\tau)_{a},
  \quad u^{a} = \left(\partial_{\tau}\right)^{a}.
\end{eqnarray}
In our case, the linearized Einstein equation gives
Eq.~(\ref{eq:LTB-energy-density-perturbation}), i.e.,
\begin{eqnarray}
  8\pi \rho
  =
  \frac{2\partial_{R}m_{1}(R)}{(\partial_{R}r)r^{2}}
  .
  \label{eq:LTB-energy-density-perturbation-2}
\end{eqnarray}
Since we have
\begin{eqnarray}
  \label{eq:LTB-flat-vacuum-circumference-3-partial_R}
  (\partial_{R}r)
  =
  (1-f)^{1/2}
\end{eqnarray}
from Eq.~(\ref{eq:LTB-flat-vacuum-circumference-2}), we obtain
\begin{eqnarray}
  \rho
  =
  \frac{\partial_{R}m_{1}(R)}{4\pi r^{2}} (1-f)^{-1/2}
  .
  \label{eq:LTB-energy-density-perturbation-3}
\end{eqnarray}
On the other hand, substituting
Eq.~(\ref{eq:dRa-f=0-tau0=R-with-Killing-time-sum-tau}) into
Eq.~(\ref{eq:dust-energy-momentum-tesor-general}), we obtain
\begin{eqnarray}
  T_{ab}
  &=&
      \rho (d\tau)_{a} (d\tau)_{b}
      \nonumber\\
  &=&
      \rho \left(
      (dt)_{a}
      +
      f^{-1}
      (1-f)^{1/2}
      (dr)_{a}
      \right)
      \left(
      (dt)_{b}
      +
      f^{-1}
      (1-f)^{1/2}
      (dr)_{b}
      \right)
      \nonumber\\
  &=&
      \rho
      (dt)_{a}
      (dt)_{b}
      +
      \rho
      \frac{(1-f)^{1/2}}{f}
      2
      (dt)_{(a} (dr)_{b)}
      +
      \rho
      \frac{1-f}{f^{2}}
      (dr)_{a}
      (dr)_{b}
      .
      \label{eq:dust-energy-momentum-tesor-general-static}
\end{eqnarray}
Then, we obtain the components of the energy-momentum tensor for the
static coordinate $(t,r)$ as
\begin{eqnarray}
  \tilde{T}_{tt}
  =
  \rho
  ,
  \quad
  \tilde{T}_{tr}
  =
  \frac{(1-f)^{1/2}}{f}
  \rho
  \quad
  \tilde{T}_{rr}
  =
  \frac{1-f}{f^{2}}
  \rho
  .
  \label{eq:LTB-dust-Ttt-Ttr-Trr-def}
\end{eqnarray}
Here, we note that the function $\rho$ is given by the Einstein
equation (\ref{eq:LTB-energy-density-perturbation-3}) and $R$ is given
by Eq.~(\ref{eq:LTB-flat-vacuum-circumference-4}).


Here, we check the components
(\ref{eq:div-barTab-linear-AB-t}) and
(\ref{eq:div-barTab-linear-AB-r}) with $l=0$ of the divergence of the
energy-momentum tensor in the LTB case.
Using
Eqs.~(\ref{eq:dRa-f=0-tau0=R-with-Killing-time-sum-R}) and
(\ref{eq:LTB-dust-Ttt-Ttr-Trr-def}), we obtain
\begin{eqnarray}
  \partial_{t}\tilde{T}_{tt}
  \!\!\!\!&=&\!\!\!\!
      \partial_{t}\rho
      =
      \frac{\partial_{R}^{2}m_{1}(R)}{4\pi r^{2}} (1-f)^{-1/2}
      \left(\frac{\partial R}{\partial t}\right)
      =
      \frac{\partial_{R}^{2}m_{1}(R)}{4\pi r^{2}} (1-f)^{-1/2}
      ,
      \nonumber
  \\
  -  f^{2} \partial_{r}\tilde{T}_{rt}
  \!\!\!\!&=&\!\!\!\!
      - f^{2} \partial_{r}\left(
      \frac{\partial_{R}m_{1}(R)}{4\pi r^{2} f}
      \right)
      =
      - \frac{\partial_{R}^{2}m_{1}(R)}{4\pi r^{2}} (1-f)^{-1/2}
      + \frac{\partial_{R}m_{1}(R)}{4\pi r^{3}} (1+f)
      \nonumber
      ,
  \\
  -  \frac{(1+f)f}{r} \tilde{T}_{rt}
  \!\!\!\!&=&\!\!\!\!
      - \frac{(1+f)f}{r} \frac{(1-f)^{1/2}}{f} \rho
      =
      -  \frac{\partial_{R}m_{1}(R)}{4\pi r^{3}} (1+f)
      .
      \nonumber
\end{eqnarray}
Then, we can confirm
Eq.~(\ref{eq:div-barTab-linear-AB-t}) with $l=0$.
Next, we check Eq.~(\ref{eq:div-barTab-linear-AB-r}) with $l=0$.
Since we may choose $\tilde{T}_{(e1)A}=0$ and $\tilde{T}_{(e2)}=0$,
Eq.~(\ref{eq:div-barTab-linear-p-even-mode}) yields
$\tilde{T}_{(e0)}=0$.
Furthermore, using
\begin{eqnarray}
  \partial_{t}\tilde{T}_{tr}
  \!\!\!\!&=&\!\!\!\!
      \frac{(1-f)^{1/2}}{f}
      \partial_{t}\rho
      =
      \frac{\partial_{R}^{2}m_{1}(R)}{4\pi r^{2} f}
      ,
      \nonumber
  \\
  -  \frac{1-f}{2rf} \tilde{T}_{tt}
  \!\!\!\!&=&\!\!\!\!
      -  \frac{1-f}{2rf} \rho
      =
      - (1-f)^{1/2} \frac{\partial_{R}m_{1}(R)}{8\pi r^{3} f}
      ,
      \nonumber
  \\
  -  f^{2} \partial_{r}\tilde{T}_{rr}
  \!\!\!\!&=&\!\!\!\!
      \frac{(2-f)(1-f)}{rf} \rho
      -  (1-f) \partial_{r}\rho
      =
      -  \frac{\partial_{R}^{2}m_{1}(R)}{4\pi r^{2} f}
      + (4+f) (1-f)^{1/2} \frac{\partial_{R}m_{1}(R)}{8\pi r^{3}f}
      ,
      \nonumber
  \\
  -  \frac{(3+f)f}{2r} \tilde{T}_{rr}
  \!\!\!\!&=&\!\!\!\!
      -  \frac{(1-f)(3+f)}{2rf} \rho
      =
      - (3+f) (1-f)^{1/2} \frac{\partial_{R}m_{1}(R)}{8\pi r^{3}f}
      ,
      \nonumber
\end{eqnarray}
we can confirm Eq.~(\ref{eq:div-barTab-linear-AB-r}) with $l=0$.
Thus, the definitions (\ref{eq:LTB-dust-Ttt-Ttr-Trr-def}) of the
components $\tilde{T}_{tt}$, $\tilde{T}_{tr}$, $\tilde{T}_{rr}$ and
the result of the Einstein equation
(\ref{eq:LTB-energy-density-perturbation-3}) are justified.
We also note that the continuity equations
(\ref{eq:div-barTab-linear-AB-r}) and
(\ref{eq:div-barTab-linear-AB-t}) with $l=0$ and $\tilde{T}_{(e0)}=0$
are important premises of the solution
(\ref{eq:calFab+poundsVg-l=0-non-vac-final}) for $l=0$ mode
perturbations.


Now, we consider the problem whether the form the perturbation
${}_{\ScrX}\!h_{ab}$ given by
Eq.~(\ref{eq:LTB-linear-pert-with-Schwarzschild-BG}) is described by
the solution (\ref{eq:calFab+poundsVg-l=0-non-vac-final}), or not.
Substituting
Eq.~(\ref{eq:dRa-f=0-tau0=R-with-Killing-time-sum-R})
into Eq.~(\ref{eq:LTB-linear-pert-with-Schwarzschild-BG}), we obtain
\begin{eqnarray}
  {}_{\ScrX}\!h_{ab}
  &=&
      \left(
      2 (\partial_{R}r_{1}) (1-f)^{1/2}
      -  f_{1}(R) (1-f)
      \right)
      (dt)_{a} (dt)_{b}
      \nonumber\\
  &&
      +
      \frac{1}{f}
      \left(
      2 (\partial_{R}r_{1})
      -  f_{1}(R) (1-f)^{1/2}
      \right)
      2 (dt)_{(a} (dr)_{b)}
      \nonumber\\
  &&
      +
      \left(
      2 (\partial_{R}r_{1}) (1-f)^{-1/2}
      -  f_{1}(R)
      \right)
      f^{-2} (dr)_{a} (dr)_{b}
      \nonumber\\
  &&
     +
     2
     r
     r_{1}
     \gamma_{ab}
  .
  \label{eq:LTB-linear-pert-with-Schwarzschild-BG-static-chart}
\end{eqnarray}
Here, we used
Eq.~(\ref{eq:LTB-flat-vacuum-circumference-3-partial_R}).
Comparing
Eq.~(\ref{eq:LTB-linear-pert-with-Schwarzschild-BG-static-chart}) with
Eq.~(\ref{eq:calFab+poundsVg-l=0-non-vac-final}), we easily see that
the last term $2rr_{1}\gamma_{ab}$ should be included in the term of
the Lie derivative of the background metric $g_{ab}$.
Then, we consider the components of ${\pounds}_{V_{(1)}}g_{ab}$ with the
generator $V_{(1)a}=V_{(1)r}(dr)_{a}$.
The components of ${\pounds}_{V_{(1)}}g_{ab}$ are summarized as
\begin{eqnarray}
  &&
  {\pounds}_{V_{(1)}}g_{tt}
  =
  - f f' V_{(1)r}
  ,
  \quad
  {\pounds}_{V_{(1)}}g_{tr}
  =
  \partial_{t}V_{(1)r}
  ,
  \quad
  {\pounds}_{V_{(1)}}g_{rr}
  =
  2\partial_{r}V_{(1)r} + \frac{f'}{f} V_{(1)r}
  ,
  \label{eq:poundsVgab-components-tt-tr-rr-even-l=0-LTB-2}
  \\
  &&
  {\pounds}_{V_{(1)}}g_{\theta\theta}
  =
  2 rf V_{(1)r}
  ,
  \quad
  {\pounds}_{V_{(1)}}g_{\phi\phi}
  =
  2 rf \sin^{2}\theta V_{r_{(1)}}
  .
  \label{eq:poundsVgab-components-thetatheta-phiphi-even-l=0-LTB-2}
\end{eqnarray}
If the last term $2rr_{1}\gamma_{ab}$ in
Eq.~(\ref{eq:LTB-linear-pert-with-Schwarzschild-BG-static-chart}) is included in
the term ${\pounds}_{V_{(1)}}g_{ab}$, we should choose the component
$V_{(1)r}$ as
\begin{eqnarray}
  \label{eq:Vr-choice-LTB}
  V_{(1)r} = \frac{r_{1}}{f}.
\end{eqnarray}
Substituting Eq.~(\ref{eq:Vr-choice-LTB}) into
Eq.~(\ref{eq:poundsVgab-components-tt-tr-rr-even-l=0-LTB-2}), we
obtain
\begin{eqnarray}
  {\pounds}_{V_{(1)}}g_{tt} = - \frac{1-f}{r} r_{1},
  \quad
  {\pounds}_{V_{(1)}}g_{tr} = \frac{1}{f} \partial_{t}r_{1},
  \quad
  {\pounds}_{V_{(1)}}g_{rr}
  =
  - \frac{1-f}{rf^{2}} r_{1} + \frac{2}{f} (\partial_{r}r_{1})
  .
  \label{eq:poundsVgab-components-tt-tr-rr-even-l=0-LTB-3}
\end{eqnarray}
Then, we have
\begin{eqnarray}
  {}_{\ScrX}\!h_{ab}
  &=&
      \left(
      2 (\partial_{R}r_{1}) (1-f)^{1/2}
      -  f_{1}(R) (1-f)
      + \frac{1-f}{r} r_{1}
      \right)
      (dt)_{a} (dt)_{b}
      \nonumber\\
  &&
     +
     \frac{1}{f}
     \left(
     2 (\partial_{R}r_{1})
     -  f_{1}(R) (1-f)^{1/2}
     - \partial_{t}r_{1}
     \right)
     2 (dt)_{(a} (dr)_{b)}
      \nonumber\\
  &&
     +
     \left(
     2 (\partial_{R}r_{1}) (1-f)^{-1/2}
     -  f_{1}(R)
     + \frac{1-f}{r} r_{1}
     -  2 f (\partial_{r}r_{1})
     \right)
     f^{-2} (dr)_{a} (dr)_{b}
     \nonumber\\
  &&
     + {\pounds}_{V_{(1)}}g_{ab}
     .
     \label{eq:LTB-linear-Schw-BG-sc-LieV1gab-pre}
\end{eqnarray}


Here, we note the inverse relation of
Eqs.~(\ref{eq:dRa-f=0-tau0=R-with-Killing-time-sum-R}) and
(\ref{eq:dRa-f=0-tau0=R-with-Killing-time-sum-tau}) as follows:
\begin{eqnarray}
  (dt)_{a}
  &=&
      \frac{1}{f} (d\tau)_{a} - \frac{1-f}{f} (dR)_{a}
      ,
      \label{eq:dt-dtau-dR-reation}
  \\
  (dr)_{a}
  &=&
      -  (1-f)^{1/2} (d\tau)_{a} + (1-f)^{1/2} (dR)_{a}
      .
      \label{eq:dr-dtau-dR-reation}
\end{eqnarray}
From Eqs.~(\ref{eq:dt-dtau-dR-reation}) and
(\ref{eq:dr-dtau-dR-reation}), we obtain
\begin{eqnarray}
  (\partial_{R}r_{1})
  &=&
      \frac{\partial t}{\partial R} (\partial_{t}r_{1})
      +
      \frac{\partial r}{\partial R} (\partial_{r}r_{1})
      \nonumber\\
  &=&
      - \frac{1-f}{f} (\partial_{t}r_{1})
      + (1-f)^{1/2} (\partial_{r}r_{1})
      .
      \label{eq:LTB-partialRr1-is-partialtr1-and-partialrr1}
\end{eqnarray}
Then, we obtain
\begin{eqnarray}
  (\partial_{t}r_{1})
  =
  - \frac{f}{1-f} (\partial_{R}r_{1})
  + \frac{f}{1-f} (1-f)^{1/2} (\partial_{r}r_{1})
  .
  \label{eq:LTB-partialtr1-is-partialRr1-and-partialrr1}
\end{eqnarray}
Substituting
Eq.~(\ref{eq:LTB-partialtr1-is-partialRr1-and-partialrr1}) into
Eq.~(\ref{eq:LTB-linear-Schw-BG-sc-LieV1gab-pre}), we obtain
\begin{eqnarray}
  {}_{\ScrX}\!h_{ab}
  \!\!\!\!&=&\!\!\!\!
      \left(
      2 (\partial_{R}r_{1}) (1-f)^{1/2}
      -  f_{1}(R) (1-f)
      + \frac{1-f}{r} r_{1}
      \right)
      (dt)_{a} (dt)_{b}
      \nonumber\\
  \!\!\!\!&&\!\!\!\!
     +
     \frac{1}{f} (1-f)^{-1/2}
     \left(
     + (2-f) (1-f)^{-1/2} (\partial_{R}r_{1})
     -  f_{1}(R) (1-f)
     -  f (\partial_{r}r_{1})
     \right)
     2 (dt)_{(a} (dr)_{b)}
      \nonumber\\
  \!\!\!\!&&\!\!\!\!
     +
     \left(
     2 (\partial_{R}r_{1}) (1-f)^{-1/2}
     -  f_{1}(R)
     + \frac{1-f}{r} r_{1}
     -  2 f (\partial_{r}r_{1})
     \right)
     f^{-2} (dr)_{a} (dr)_{b}
     \nonumber\\
  \!\!\!\!&&\!\!\!\!
     + {\pounds}_{V_{(1)}}g_{ab}
     .
     \label{eq:LTB-linear-Schw-BG-sc-LieV1gab}
\end{eqnarray}


Here, we also note that, apart from the term ${\pounds}_{V}g_{ab}$,
the solution (\ref{eq:calFab+poundsVg-l=0-non-vac-final}) is
traceless.
Therefore, the trace part of $(t,r)$ components in
Eq.~(\ref{eq:LTB-linear-Schw-BG-sc-LieV1gab}) should be
included in the term of the Lie derivative of the background metric
$g_{ab}$.
To see this, we consider the components ${\pounds}_{V_{(2)}}g_{ab}$
with the generator $V_{(2)a} = V_{(2)t}(dt)_{a}$ as follows:
\begin{eqnarray}
  \label{eq:poundsWgtt-tr-l=0-LTB-V2tneq0}
  {\pounds}_{V_{(2)}}g_{tt}
  =
  2 \partial_{t}V_{(2)t}
  ,
  \quad
  {\pounds}_{V_{(2)}}g_{tr}
  =
  \partial_{r}V_{(2)t} - \frac{f'}{f} V_{(2)t}
  .
\end{eqnarray}
Substituting Eq.~(\ref{eq:poundsWgtt-tr-l=0-LTB-V2tneq0}) into
Eq.~(\ref{eq:LTB-linear-Schw-BG-sc-LieV1gab}), we obtain
\begin{eqnarray}
  {}_{\ScrX}\!h_{ab}
  &=&
      \left(
      2 (\partial_{R}r_{1}) (1-f)^{1/2}
      -  f_{1}(R) (1-f)
      + \frac{1-f}{r} r_{1}
      - 2 \partial_{t}V_{(2)t}
      \right)
      (dt)_{a} (dt)_{b}
      \nonumber\\
  &&
     +
     \frac{(1-f)^{-1/2}}{f}
     \left(
     (2-f) (1-f)^{-1/2} (\partial_{R}r_{1})
     -  f_{1}(R) (1-f)
     -  f (\partial_{r}r_{1})
     \right.
     \nonumber\\
  && \quad\quad\quad\quad\quad\quad
     \left.
     - f (1-f)^{1/2} \partial_{r}V_{(2)t}
     + \frac{1-f}{r} (1-f)^{1/2} V_{(2)t}
     \right)
     2 (dt)_{(a} (dr)_{b)}
      \nonumber\\
  &&
     +
     \left(
     2 (\partial_{R}r_{1}) (1-f)^{-1/2}
     -  f_{1}(R)
     + \frac{1-f}{r} r_{1}
     -  2 f (\partial_{r}r_{1})
     \right)
     f^{-2} (dr)_{a} (dr)_{b}
     \nonumber\\
  &&
     + {\pounds}_{V_{(1)}+V_{(2)}}g_{ab}
     .
     \label{eq:LTB-linear-Schw-BG-sc-LieV1gab-LieV2gab}
\end{eqnarray}


Apart from the term ${\pounds}_{V_{(1)}+V_{(2)}}g_{ab}$, the remaining
term in ${}_{\ScrX}\!h_{ab}$ should be traceless.
Then, we obtain
\begin{eqnarray}
  0
  &=&
      g^{ab} \left[
      {}_{\ScrX}\!h_{ab}
      - {\pounds}_{V_{(1)}+V_{(2)}}g_{ab}
      \right]
      \nonumber\\
  &=&
      \frac{1}{f} \left(
      + 2 f (\partial_{R}r_{1}) (1-f)^{-1/2}
      -  2 f (\partial_{r}r_{1})
      -  f f_{1}(R)
      + 2 \partial_{t}V_{(2)t}
      \right)
      .
      \label{eq:LTB-linear-pert-with-Schw-BG-sc-trace-vansh-cond}
\end{eqnarray}
Here, we choose $V_{(2)t}$ so that
\begin{eqnarray}
  \partial_{t}V_{(2)t}
  =
  -  f (\partial_{R}r_{1}) (1-f)^{-1/2}
  + f (\partial_{r}r_{1})
  + \frac{1}{2} f f_{1}(R)
  .
  \label{eq:V2t-choice}
\end{eqnarray}
Through this expression of $(\partial_{R}r_{1})$ given by
Eq.~(\ref{eq:LTB-partialRr1-is-partialtr1-and-partialrr1}),
Eq.~(\ref{eq:V2t-choice}) is given by
\begin{eqnarray}
  \partial_{t}V_{(2)t}
  =
  \partial_{t}((1-f)^{1/2} r_{1})
  + \frac{1}{2} f f_{1}(R)
  \label{eq:V2t-choice-2}
\end{eqnarray}
and
\begin{eqnarray}
  V_{(2)t}
  =
  (1-f)^{1/2} r_{1}
  + \frac{1}{2} f \int dt f_{1}(R)
  ,
  \label{eq:V2t-choice-2-sol}
\end{eqnarray}
where we choose the arbitrary function $r$ to be zero.
From Eq.~(\ref{eq:V2t-choice-2-sol}), we obtain
\begin{eqnarray}
  &&
     - f (1-f)^{1/2} \partial_{r}V_{(2)t}
     + \frac{1-f}{r} (1-f)^{1/2} V_{(2)t}
     \nonumber\\
  &=&
      \frac{1}{2r} (1-f) (2-f) r_{1}
      -  f (1-f) \partial_{r}r_{1}
      - \frac{1}{2} f^{2} (1-f)^{1/2}  \int dt \partial_{r}f_{1}(R)
  .
  \label{eq:-fpartialrV2t+fprimeV2t}
\end{eqnarray}
Here, we note that $f_{1}=f_{1}(R)$ and its derivative with respect to
$r$ is given by
\begin{eqnarray}
  \label{eq:partialrf1R}
  \partial_{r}f_{1}(R)
  =
  \frac{\partial R}{\partial r} \frac{d}{dR}f_{1}(R)
  =
  \frac{1}{f(1-f)^{1/2}} \frac{d}{dR}f_{1}(R)
  .
\end{eqnarray}
On the other hand, the derivative of $f_{1}(R)$ with respect to $t$ is
given by
\begin{eqnarray}
  \label{eq:partialtf1R}
  \partial_{t}f_{1}(R)
  =
  \frac{\partial R}{\partial t} \frac{d}{dR}f_{1}(R)
  =
  \frac{d}{dR}f_{1}(R)
  .
\end{eqnarray}
Then, we obtain
\begin{eqnarray}
  \label{eq:partialrf1R-is-partialtf1R}
  \partial_{r}f_{1}(R)
  =
  \frac{1}{f(1-f)^{1/2}} \partial_{t}f_{1}(R)
  .
\end{eqnarray}
Substituting Eq.~(\ref{eq:partialrf1R-is-partialtf1R}) into
Eq.~(\ref{eq:-fpartialrV2t+fprimeV2t}), we obtain
\begin{eqnarray}
  &&
  - f (1-f)^{1/2} \partial_{r}V_{(2)t}
  + \frac{1-f}{r} (1-f)^{1/2} V_{(2)t}
     \nonumber\\
  &=&
  \frac{1}{2r} (1-f) (2-f) r_{1}
  -  f (1-f) \partial_{r}r_{1}
  -  \frac{1}{2} f f_{1}
  .
  \label{eq:-fpartialrV2t+fprimeV2t-2}
\end{eqnarray}
Furthermore, the substitution of Eqs.~(\ref{eq:V2t-choice}) and
(\ref{eq:-fpartialrV2t+fprimeV2t-2}) into
Eq.~(\ref{eq:LTB-linear-Schw-BG-sc-LieV1gab-LieV2gab}), we obtain
\begin{eqnarray}
  {}_{\ScrX}\!h_{ab}
  \!\!\!\!&=&\!\!\!\!
     2\left(
     (\partial_{R}r_{1}) (1-f)^{-\frac{1}{2}}
     + \frac{1-f}{2r} r_{1}
     -  f (\partial_{r}r_{1})
     -  \frac{1}{2} f_{1}(R)
     \right)
              \nonumber\\
  && \quad\quad\quad
     \times
      \left(
      (dt)_{a} (dt)_{b} + \frac{1}{f^{2}} (dr)_{a} (dr)_{b}
      \right)
      \nonumber\\
  \!\!\!\!&&\!\!\!\!
     +
     \frac{(2-f) (1-f)^{-\frac{1}{2}}}{f}
     \left(
     (\partial_{R}r_{1}) (1-f)^{-1/2}
     + \frac{1-f}{2r} r_{1}
     -  f (\partial_{r}r_{1})
     -  \frac{1}{2} f_{1}(R)
     \right)
              \nonumber\\
  && \quad\quad\quad
     \times
     2 (dt)_{(a} (dr)_{b)}
      \nonumber\\
  \!\!\!\!&&\!\!\!\!
     + {\pounds}_{V_{(1)}+V_{(2)}}g_{ab}
     .
     \label{eq:LTB-linear-Schw-BG-sc-LieV1gab-LieV2gab-2}
\end{eqnarray}


Here, we consider the information from the solution
(\ref{eq:LTB-Hubble-equation-linear-pert-sol.-sum}) of the linearized
LTB solution.
To consider the necessary information from the solution
(\ref{eq:LTB-Hubble-equation-linear-pert-sol.-sum}), we consider the
derivative $\partial_{r}r_{1}$ in
Eq.~(\ref{eq:LTB-linear-Schw-BG-sc-LieV1gab-LieV2gab-2}).
Using Eqs.~(\ref{eq:dRa-f=0-tau0=R-with-Killing-time-sum-R}) and
(\ref{eq:dRa-f=0-tau0=R-with-Killing-time-sum-tau}), we obtain
\begin{eqnarray}
  \label{eq:partialrr1-solutions}
  \partial_{r}r_{1}
  &=&
      \frac{\partial R}{\partial r} (\partial_{R}r_{1})
      +
      \frac{\partial \tau}{\partial r} (\partial_{\tau}r_{1})
      \nonumber\\
  &=&
      \frac{1}{f} (1-f)^{-1/2} (\partial_{R}r_{1})
      + \frac{1}{f} (1-f)^{1/2} (\partial_{\tau}r_{1})
      .
\end{eqnarray}
From Eqs.~(\ref{eq:partialrr1-solutions}) and
(\ref{eq:LTB-Hubble-equation-linear-pert}), we obtain
\begin{eqnarray}
  2 (\partial_{R}r_{1}) (1-f)^{-1/2}
  -  2 f \partial_{r}r_{1}
  +  \frac{1-f}{r} r_{1}
  - f_{1}(R)
  =
  \frac{2m_{1}(R)}{r}
  .
  \label{eq:2m1overr-expression}
\end{eqnarray}
Through Eq.~(\ref{eq:2m1overr-expression}), we obtain
\begin{eqnarray}
  {}_{\ScrX}\!h_{ab}
  &=&
      \frac{2m_{1}(R)}{r}
      \left[
      (dt)_{a} (dt)_{b} + \frac{1}{f^{2}} (dr)_{a} (dr)_{b}
      \right]
      +
      \frac{2-f}{f (1-f)^{1/2}}
      \frac{m_{1}(R)}{r}
      2 (dt)_{(a} (dr)_{b)}
      \nonumber\\
  &&
     + {\pounds}_{V_{(LTB)}}g_{ab}
     ,
     \label{eq:LTB-linear-Schw-BG-sc-LieV1gab-LieV2gab-fnal}
\end{eqnarray}
where $V_{(LTB)a}$ are given by
\begin{eqnarray}
  \label{eq:V1a-V2a-summary}
  V_{(LTB)a}
  :=
  V_{(1)a}
  +
  V_{(2)a}
  =
  \left[(1-f)^{1/2} r_{1} + \frac{1}{2} f \int dt f_{1}(R)\right] (dt)_{a}
  +
  \frac{r_{1}}{f} (dr)_{a}
  .
\end{eqnarray}


Now, we check whether the linear-order perturbative solution
(\ref{eq:LTB-linear-Schw-BG-sc-LieV1gab-LieV2gab-fnal}) have the form
of the general solution (\ref{eq:calFab+poundsVg-l=0-non-vac-final})
for the $l=0$ mode perturbations, or not.
Here, we only consider the case $M_{1}=0$ in
Eq.~(\ref{eq:calFab+poundsVg-l=0-non-vac-final}), because we can add
the term $M_{1}$ with an appropriate term of the Lie derivative of the
background metric..
First, we consider the first term in
Eq.~(\ref{eq:LTB-linear-Schw-BG-sc-LieV1gab-LieV2gab-fnal}).
From Eq.~(\ref{eq:calFab+poundsVg-l=0-non-vac-final}), the expression
\begin{eqnarray}
  \label{eq:m1=M1+4piintdrr2overfTrr}
  m_{1}(t,r) = 4 \pi \int dr \left[\frac{r^{2}}{f} \tilde{T}_{tt}\right]
\end{eqnarray}
should appear in
Eq.~(\ref{eq:LTB-linear-Schw-BG-sc-LieV1gab-LieV2gab-fnal}).
From Eqs.~(\ref{eq:LTB-dust-Ttt-Ttr-Trr-def}) and
(\ref{eq:LTB-energy-density-perturbation-3}), we obtain
\begin{eqnarray}
  \label{eq:4piintdrr2overfTrr}
  4 \pi \int dr \left[\frac{r^{2}}{f} \tilde{T}_{tt}\right]
  =
  4 \pi \int dr \left[\frac{r^{2}}{f} \rho \right]
  =
  \int dr \left[
  f^{-1}(1-f)^{-1/2} \partial_{R}m_{1}(R)
  \right]
  .
\end{eqnarray}
Here, we note that
\begin{eqnarray}
  \partial_{r}m_{1}(R)
  =
  \frac{\partial R}{\partial r} \partial_{R}m_{1}(R)
  +
  \frac{\partial \tau}{\partial r} \partial_{\tau}m_{1}(R)
  =
  f^{-1} (1-f)^{-1/2}
  \partial_{R}m_{1}(R)
  .
  \label{eq:partialrmR-partialRmR}
\end{eqnarray}
Substituting Eq.~(\ref{eq:partialrmR-partialRmR}) into
Eq.~(\ref{eq:4piintdrr2overfTrr}), we obtain
\begin{eqnarray}
  \label{eq:4piintdrr2overfTrr-2}
  4 \pi \int dr \left[\frac{r^{2}}{f} \tilde{T}_{tt}\right]
  =
  \int dr \partial_{r}m_{1}(R)
  =
  m_{1}(R)
  .
\end{eqnarray}
Thus, we may regard that the first term in
Eq.~(\ref{eq:calFab+poundsVg-l=0-non-vac-final}) realizes the first
term in Eq.~(\ref{eq:LTB-linear-Schw-BG-sc-LieV1gab-LieV2gab-fnal}) of
the linearized LTB solution.


Next, we consider the second term in
Eq.~(\ref{eq:calFab+poundsVg-l=0-non-vac-final}).
In this case, we evaluate the integration
\begin{eqnarray}
  \label{eq:4pirintdtfinvTtt+fTrr}
  4 \pi r \int dt \left(
  \frac{1}{f}\tilde{T}_{tt} + f \tilde{T}_{rr}
  \right).
\end{eqnarray}
From Eq.~(\ref{eq:LTB-dust-Ttt-Ttr-Trr-def}), we obtain
\begin{eqnarray}
  \label{eq:4pirintdtfinvTtt+fTrr-2}
  \frac{1}{f}\tilde{T}_{tt} + f \tilde{T}_{rr}
  =
  \frac{1}{f} \rho + f \frac{1-f}{f^{2}} \rho
  =
  \frac{2-f}{f} \rho
  =
  \frac{1}{4\pi r^{2}f} (2-f) (1-f)^{-1/2} \partial_{R}m_{1}(R)
      .
\end{eqnarray}
Here, we consider $\partial_{t}m_{1}(R)$ as
\begin{eqnarray}
  \partial_{t}m_{1}(R)
  =
  \frac{\partial R}{\partial t} \partial_{R}m_{1}(R)
  +
  \frac{\partial \tau}{\partial t} \partial_{\tau}m_{1}(R)
  =
  \partial_{R}m_{1}(R)
  .
  \label{eq:partialtmR-partialRmR}
\end{eqnarray}
Then, we obtain
\begin{eqnarray}
  4 \pi r \int dt \left(
  \frac{1}{f}\tilde{T}_{tt} + f \tilde{T}_{rr}
  \right)
  &=&
      4 \pi r \int dt \left(
      \frac{1}{4\pi r^{2}f} (2-f) (1-f)^{-1/2} \partial_{t}m_{1}(R)
      \right)
      \nonumber\\
  &=&
      \frac{2-f}{rf(1-f)^{1/2}} \int dt \partial_{t}m_{1}(R)
      =
      \frac{2-f}{f(1-f)^{1/2}} \frac{m_{1}(R)}{r}
      .
      \label{eq:4pirintdtfinvTtt+fTrr-3}
\end{eqnarray}
Thus, we confirmed that the second term in
Eq.~(\ref{eq:calFab+poundsVg-l=0-non-vac-final}) realizes the second
term in Eq.~(\ref{eq:LTB-linear-Schw-BG-sc-LieV1gab-LieV2gab-fnal}) in
the linearized LTB solution.


The remaining term in
Eq.~(\ref{eq:LTB-linear-Schw-BG-sc-LieV1gab-LieV2gab-fnal}) is the Lie
derivative of the background spacetime.
Here, we note that there is always ambiguity of the gauge-choice
${\pounds}_{V}g_{ab}$ with an arbitrary vector field $V^{a}$ in the
linear perturbation ${}_{\ScrX}\!h_{ab}$.
Through this degree of freedom, we can always adjust the solution
${}_{\ScrX}\!h_{ab}$ so that the last term in
Eq.~(\ref{eq:LTB-linear-Schw-BG-sc-LieV1gab-LieV2gab-fnal}) is
identical with the last term in
Eq.~(\ref{eq:calFab+poundsVg-l=0-non-vac-final}).


Thus, the linear perturbation version
(\ref{eq:LTB-linear-Schw-BG-sc-LieV1gab-LieV2gab-fnal}) of the LTB
exact solution with Schwarzschild background spacetime is realized
from the solution (\ref{eq:calFab+poundsVg-l=0-non-vac-final}) of the
$l=0$ mode perturbations.
In this sense, the solutions
(\ref{eq:calFab+poundsVg-l=0-non-vac-final}) of the $l=0$ mode
perturbations are justified by the LTB solutions.
It is important to note that the arbitrary functions $f_{1}(R)$ and
$\tau_{1}(R)$ of the perturbative LTB solution is included only in
the vector field $V_{(LTB)a}$ in
Eq.~(\ref{eq:LTB-linear-Schw-BG-sc-LieV1gab-LieV2gab-fnal}).
Therefore, we may say that the term ${\pounds}_{V_{(LTB)}}g_{ab}$
includes physical information of the LTB solution, i.e., this term has its
physical meaning.


\section{Realization of the linearized non-rotating C-metric}
\label{sec:Reconsideration_of_the_linearized_C-metric_again}


\subsection{The linearized non-rotating C-metric}
\label{sec:the_linearized_C-metric_again}


Here, we consider the non-rotating vacuum
C-metric~\cite{J.B.Griffiths-P.Krtous-J.Podolsky-2006}, in which
conical singularities may occur both in the axis $\theta=0$ and
$\theta=\pi$.
The C-metric is well-known as the solution describing uniformly
accelerating black holes which are pulled or pushed by the straight string at
$\theta=0$ or $\theta=\pi$.
The C-metric is described by the metric
\begin{eqnarray}
  \label{eq:J.B.Griffiths-P.Krtous-J.Podolsky-2006-6-2}
  g_{ab}
  &=&
  \frac{1}{(1+\alpha r\cos\theta)^{2}}
  \left(
  - Q (dt)_{a}(dt)_{b}
  + \frac{1}{Q} (dr)_{a}(dr)_{b}
  \right.
  \nonumber\\
  && \quad\quad\quad\quad\quad\quad\quad
  \left.
  + \frac{r^{2}}{P} (d\theta)_{a}(d\theta)_{b}
  + P r^{2}\sin^{2}\theta (d\varphi)_{a}(d\varphi_{b})
  \right)
  ,
\end{eqnarray}
where
\begin{eqnarray}
  \label{eq:J.B.Griffiths-P.Krtous-J.Podolsky-2006-7-2}
  P = 1 + 2 \alpha m \cos\theta, \quad
  Q = (1 - \alpha^{2}r^{2}) \left(1 - \frac{2m}{r}\right), \quad
  \varphi\in(-C\pi,+C\pi)
\end{eqnarray}
includes the singularities both in the axis $\theta=0$ and $\theta=\pi$.
To see this, we note that the metric given by
Eqs.~(\ref{eq:J.B.Griffiths-P.Krtous-J.Podolsky-2006-6-2}) and
(\ref{eq:J.B.Griffiths-P.Krtous-J.Podolsky-2006-7-2}) includes three
positive real parameters $m$, $\alpha$ (satisfying $2\alpha m<1$), and
$C$ (which is hidden in the range of the rotational coordinate
$\varphi\in(-C\pi,+C\pi)$).


Here, we consider the two-dimensional section of the spacetime with
the metric (\ref{eq:J.B.Griffiths-P.Krtous-J.Podolsky-2006-6-2}) as
\begin{eqnarray}
  \label{eq:acceleration_perturbation_again_of_SCH-00300}
  (1+\alpha r \cos\theta)^{2} \left.g_{ab}\right|_{r=const.,t=const.}
  =:
  \frac{r^{2}}{P} \bar{\gamma}_{ab}
  =
  \frac{r^{2}}{P}
  \left(
  (d\theta)_{a}(d\theta)_{b}
  +
  P^{2}\sin^{2}\theta
  (d\varphi)_{a}(d\varphi)_{b}
  \right)
  .
\end{eqnarray}
Besides the conformal factor $r^{2}/P$, the ``radius'' which is the
proper distance along the $(\partial/\partial\theta)^{a}$ is given by
\begin{eqnarray}
  &&
  \int_{0}^{\theta}
  \sqrt{
  \bar{\gamma}_{ab}
  (\partial/\partial\theta)^{a}
  (\partial/\partial\theta)^{a}
  }
  d\theta
  =
  \theta
  ,
  \nonumber\\
  &&
  \int_{\theta}^{\pi}
  \sqrt{
  \bar{\gamma}_{ab}
  (\partial/\partial\theta)^{a}
  (\partial/\partial\theta)^{a}
  }
  d\theta
  =
  \pi - \theta
  .
  \label{eq:acceleration_perturbation_again_of_SCH-00400}
\end{eqnarray}
On the other hand, the ``circumference'', which is the proper distance
along $(\partial/\partial\varphi)^{a}$ from $\varphi=-C\pi$ to
$\varphi=C\pi$, for any $\theta$ is given by
\begin{eqnarray}
  &&
     \int_{-C\pi}^{+C\pi}
     \sqrt{
     \bar{\gamma}_{ab}
     (\partial/\partial\varphi)^{a}
     (\partial/\partial\varphi)^{a}
     }
     d\varphi
     \nonumber\\
  &=&
      \int_{-C\pi}^{+C\pi}
      (1 + 2 \alpha m \cos\theta)\sin\theta
      d\varphi
      =
      2 \pi C
      (1 + 2 \alpha m \cos\theta)\sin\theta
      .
  \label{eq:acceleration_perturbation_again_of_SCH-00500}
\end{eqnarray}
From Eqs.~(\ref{eq:acceleration_perturbation_again_of_SCH-00400}) and
(\ref{eq:acceleration_perturbation_again_of_SCH-00500}), we obtain the
following results: in the neighborhood of $\theta=0$,
\begin{eqnarray}
  \label{eq:acceleration_perturbation_again_of_SCH-00600}
  \frac{
  \mbox{circumference at $\theta$}
  }{
  \mbox{radius from $\theta=0$}
  }
  =
  \frac{
  2\pi C (1 + 2 \alpha m \cos\theta)\sin\theta
  }{
  \theta
  }
  ;
\end{eqnarray}
and in the neighborhood of $\theta=\pi$,
\begin{eqnarray}
  \frac{
  \mbox{circumference at $\theta$}
  }{
  \mbox{radius from $\theta=\pi$}
  }
  &=&
      \frac{
      2\pi C (1 + 2 \alpha m \cos\theta)\sin\theta
      }{
      \pi-\theta
      }
      \nonumber\\
  &=&
      \frac{
      2\pi C (1 - 2 \alpha m \cos(\pi-\theta))\sin(\pi-\theta)
      }{
      \pi-\theta
      }
      .
      \label{eq:acceleration_perturbation_again_of_SCH-00700}
\end{eqnarray}
Then, we obtain
\begin{eqnarray}
  &&
     \lim_{\theta\rightarrow 0}
     \frac{
     \mbox{circumference at $\theta$}
     }{
     \mbox{radius from $\theta=0$}
     }
     =
     2\pi C ( 1+2\alpha m)
     ,
     \nonumber\\
  &&
     \lim_{\theta\rightarrow\pi-0}
     \frac{
     \mbox{circumference at $\theta$}
     }{
     \mbox{radius from $\theta=\pi$}
     }
     =
     2\pi C (1 - 2 \alpha m)
     .
     \label{eq:acceleration_perturbation_again_of_SCH-00800}
\end{eqnarray}
These imply the existence of a conical singularity with a different
conicity (unless $\alpha m=0$).
The deficit or excess angle of either of these two conical singularity
can be removed in an appropriate choice of the constant $C$, but not
both simultaneously.
In general, the constant $C$ can thus be seen to determine the balance
between the deficit/excess angles on the two halves of the axis.
In particular, one natural choice is to remove the conical singularity
at $\theta=0$ by setting $C=(1+2\alpha m)^{-1}$.
In this choice, the deficit angle at the poles $\theta=0,\pi$ are
respectively
\begin{eqnarray}
  \label{eq:acceleration_perturbation_again_of_SCH-00900}
  \delta_{0}=0, \quad
  \delta_{\pi}=2\pi - \frac{2\pi(1-2\alpha m)}{1+2\alpha m} = 2\pi
  \frac{4\alpha m}{1+2\alpha m}.
\end{eqnarray}


To compare the Schwarzschild spacetime, it is convenient to rescale
the range of the rotational coordinate is $2\pi$.
This can be achieved by the simple rescaling
\begin{eqnarray}
  \label{eq:acceleration_perturbation_again_of_SCH-01000}
  \varphi = C \phi,
\end{eqnarray}
where $\phi\in(-\pi,\pi)$.
For this choice, the metric
(\ref{eq:J.B.Griffiths-P.Krtous-J.Podolsky-2006-6-2}) is given by
\begin{eqnarray}
  g_{ab}
  &=&
      \frac{1}{(1+\alpha r\cos\theta)^{2}}
      \left(
      - Q (dt)_{a}(dt)_{b}
      + \frac{1}{Q} (dr)_{a}(dr)_{b}
      \right.
      \nonumber\\
  && \quad\quad\quad\quad\quad\quad\quad
     \left.
     + \frac{r^{2}}{P} (d\theta)_{a}(d\theta)_{b}
     + P C^{2} r^{2}\sin^{2}\theta (d\phi)_{a}(d\phi_{b})
     \right)
     ,
     \label{eq:acceleration_perturbation_again_of_SCH-010010}
\end{eqnarray}
where $P$ and $Q$ are still given by
Eq.~(\ref{eq:J.B.Griffiths-P.Krtous-J.Podolsky-2006-7-2}).


Now, we consider the situation where the black hole mass $m$ is
finite, and the acceleration $\alpha$ is infinitesimally small.
In this case, the Rindler horizon $r=1/\alpha$ is larger than the black
hole horizon $r=2m$.
Therefore, this situation is naturally given by the inequality
\begin{eqnarray}
  \label{eq:acceleration_perturbation_again_of_SCH-01100}
  1/\alpha > 2m, \quad \mbox{i.e.}, \quad 2m\alpha < 1.
\end{eqnarray}
Furthermore, we consider the situation
\begin{eqnarray}
  \label{eq:acceleration_perturbation_again_of_SCH-01200}
  2 m \alpha \ll 1.
\end{eqnarray}
This situation is appropriate for the consideration of the linearized
C-metric spacetime around the Schwarzschild spacetime.
Moreover, we consider the situation where the constant $C$ is finite.
We regard that the metric on the physical spacetime is given by
\begin{eqnarray}
  \label{eq:acceleration_perturbation_again_of_SCH-01300}
  \bar{g}_{ab}(\bar{M},\bar{\alpha},\bar{C};\bar{x})
  =
  \bar{g}_{\mu\nu}(\bar{M},\bar{\alpha},\bar{C};\bar{x})
  (d\bar{x}^{\mu})_{a}(d\bar{x}^{\nu})_{b}
  .
\end{eqnarray}
This metric is given by the replacements $m\rightarrow\bar{M}$,
$\alpha\rightarrow\bar{\alpha}$, $C\rightarrow\bar{C}$, and
$\{t,r,\theta,\phi\}\rightarrow\{\bar{t},\bar{r},\bar{\theta},\bar{\phi}\}$
in Eqs.~(\ref{eq:J.B.Griffiths-P.Krtous-J.Podolsky-2006-6-2}) and
(\ref{eq:J.B.Griffiths-P.Krtous-J.Podolsky-2006-7-2}).


As a gauge choice of the second kind, we consider the point-identification
between the background spacetime with the metric
(\ref{eq:background-metric-2+2})--(\ref{eq:background-metric-2+2-gamma-comp-Schwarzschild})
in terms of the coordinates $\{x^{\mu}\}=\{t,r,\theta,\phi\}$
and the physical spacetime with the metric
(\ref{eq:acceleration_perturbation_again_of_SCH-01300}) by
\begin{eqnarray}
  \label{eq:acceleration_perturbation_again_of_SCH-01600}
  \bar{x}^{\mu} \stackrel{\leftarrow}{=} x^{\mu}.
\end{eqnarray}
We call this gauge choice $\ScrX_{\epsilon}$.
We also consider the situation of perturbation
\begin{eqnarray}
  \label{eq:acceleration_perturbation_again_of_SCH-01700}
  \bar{M} &:=& M + \epsilon M_{1}, \\
  \label{eq:acceleration_perturbation_again_of_SCH-01800}
  \bar{\alpha} &:=& \alpha + \epsilon \alpha_{1}, \\
  \label{eq:acceleration_perturbation_again_of_SCH-01900}
  \bar{C} &:=& C + \epsilon C_{1}.
\end{eqnarray}
Then, the pull-back $\ScrX_{\epsilon} ^{*}$ of the metric
$\bar{g}_{ab}$ on the physical spacetime with the metric
(\ref{eq:acceleration_perturbation_again_of_SCH-01300}) to the
background spacetime with the metric
(\ref{eq:background-metric-2+2})--(\ref{eq:background-metric-2+2-gamma-comp-Schwarzschild})
is given by
\begin{eqnarray}
  \label{eq:acceleration_perturbation_again_of_SCH-02100}
  \ScrX_{\epsilon}^{*}\bar{g}_{ab}
  &=:&
      {}_{\ScrX}\!\bar{g}_{ab}
       \nonumber\\
  &=&
      g_{ab}(M,\alpha,C:x)
      + \epsilon M_{1} \partial_{M} {}_{\ScrX}\!\bar{g}_{ab}(M,\alpha,C;x)
      \nonumber\\
  &&
      + \epsilon \alpha_{1} \partial_{\alpha} {}_{\ScrX}\!\bar{g}_{ab}(M,\alpha,C;x)
      + \epsilon C_{1} \partial_{C} {}_{\ScrX}\!\bar{g}_{ab}(M,\alpha,C;x)
      \nonumber\\
  &&
      +
      O(\epsilon^{2})
\end{eqnarray}
in the second-kind gauge choice $\ScrX_{\epsilon}$.
Since the linear-order perturbations ${}_{\ScrX}\!h_{ab}$ under the
gauge-choice $\ScrX_{\epsilon}$ is defined by
Eq.~(\ref{eq:metric-expansion}), we obtain the representation of the
linear perturbation ${}_{\ScrX}\!h_{ab}$ under the gauge choice
$\ScrX_{\epsilon}$ as
\begin{eqnarray}
  \label{eq:acceleration_perturbation_again_of_SCH-02300}
  {}_{\ScrX}\!h_{ab}
  &=&
  M_{1} \partial_{M} {}_{\ScrX}\bar{g}_{ab}(M,\alpha,C;x)
  +
  \alpha_{1} \partial_{\alpha} {}_{\ScrX}\bar{g}_{ab}(M,\alpha,C;x)
  \nonumber\\
  &&
  +
  C_{1} \partial_{C} {}_{\ScrX}\bar{g}_{ab}(M,\alpha,C;x)
  .
\end{eqnarray}


On the other hand, if we apply the other gauge choice
$\ScrY_{\epsilon}$, we have other representation of the linear-order
perturbation
\begin{eqnarray}
  \label{eq:acceleration_perturbation_again_of_SCH-02400}
  \ScrY_{\epsilon}^{*}\bar{g}_{ab}
  =
  g_{ab}
  +
  \epsilon {}_{\ScrY}\!h_{ab}
  +
  O(\epsilon^{2})
  .
\end{eqnarray}
As the gauge choice $\ScrY_{\epsilon}$, we consider the
point-identification
\begin{eqnarray}
  \label{eq:acceleration_perturbation_again_of_SCH-02500}
  \bar{x}^{\mu} \stackrel{\leftarrow}{=} {x'}^{\mu}
\end{eqnarray}
from the background spacetime with the metric
(\ref{eq:acceleration_perturbation_again_of_SCH-010010}) to the
physical spacetime with the metric
(\ref{eq:acceleration_perturbation_again_of_SCH-01300}).
We assume that the coordinates $\{{x'}^{\mu}\}$ in the gauge choice
$\ScrY_{\epsilon}$ is related to the coordinate $\{x^{\mu}\}$ as
\begin{eqnarray}
  \label{eq:acceleration_perturbation_again_of_SCH-02600}
  {x'}^{\mu} = x^{\mu} + \epsilon \xi^{\mu} + O(\epsilon^{2}).
\end{eqnarray}
This is the coordinate transformation induced by the second-kind
gauge-transformation
$\Phi_{\epsilon}=\ScrX_{\epsilon}\circ\ScrY_{\epsilon}^{-1}$ $:$
$\ScrY_{\epsilon}\rightarrow\ScrX_{\epsilon}$.
The metric on the physical spacetime pulled-back by the second-kind gauge choice
$\ScrY_{\epsilon}$ is given by
\begin{eqnarray}
  \ScrY_{\epsilon}^{*}
  \bar{g}_{ab}(\bar{M},\bar{\alpha},\bar{C};{x'}^{\mu})
  &=&
      g_{ab}
     \nonumber\\
  &&
      + \epsilon \left(
      M_{1}\partial_{M}g_{\mu\nu}(M,\alpha,C;x)
      + \alpha_{1}\partial_{\alpha}\bar{g}_{\mu\nu}(M,\alpha,C;x)
     \right.
     \nonumber\\
  && \quad\quad
     \left.
      + C_{1}\partial_{C}\bar{g}_{\mu\nu}(M,\alpha,C;x)
      \right) (dx^{\mu})_{a} (dx^{\nu})_{b}
     \nonumber\\
  &&
      + \epsilon {\pounds}_{\xi}g_{ab}
     \nonumber\\
  &&
      +
      O(\epsilon^{2})
      .
      \label{eq:acceleration_perturbation_again_of_SCH-02800}
\end{eqnarray}
Comparing
Eqs.~(\ref{eq:background-metric-2+2})--(\ref{eq:background-metric-2+2-gamma-comp-Schwarzschild})
and (\ref{eq:acceleration_perturbation_again_of_SCH-02800}), the
perturbation ${}_{\ScrY}\!h_{ab}$ in the gauge choice $\ScrY$ is
given by
\begin{eqnarray}
  \label{eq:acceleration_perturbation_again_of_SCH-02900}
  {}_{\ScrY}\!h_{ab}
  &=&
  \left(
  M_{1}\partial_{M}\bar{g}_{\mu\nu}(M,\alpha,C;x)
  + \alpha_{1}\partial_{\alpha}\bar{g}_{\mu\nu}(M,\alpha,C;x)
     \right.
     \nonumber\\
  && \quad
     \left.
  + C_{1}\partial_{C}\bar{g}_{\mu\nu}(M,\alpha,C;x)
  \right) (dx^{\mu})_{a} (dx^{\nu})_{b}
  + {\pounds}_{\xi}g_{ab}
  .
\end{eqnarray}
Together with
Eq.~(\ref{eq:acceleration_perturbation_again_of_SCH-02300}), we obtain
the gauge-transformation
\begin{eqnarray}
  \label{eq:acceleration_perturbation_again_of_SCH-03000}
  {}_{\ScrY}\!h_{ab} - {}_{\ScrX}\!h_{ab}
  =
  {\pounds}_{\xi}g_{ab}
  .
\end{eqnarray}


Now, we consider the explicit expression of the perturbation $h_{ab}$.
From the definitions
(\ref{eq:J.B.Griffiths-P.Krtous-J.Podolsky-2006-7-2}) of the
functions $P$ and $Q$, we obtain
\begin{eqnarray}
  \label{eq:acceleration_perturbation_again_of_SCH-03100}
  && \partial_{M}P = 2 \alpha \cos\theta, \quad \partial_{\alpha}P = 2M \cos\theta, \\
  \label{eq:acceleration_perturbation_again_of_SCH-03200}
  && \partial_{M}Q = (1-\alpha^{2}r^{2}) \left(-\frac{2}{r}\right), \quad
  \partial_{\alpha}P = - 2 \alpha r^{2} \left(1 - \frac{2m}{r}\right).
\end{eqnarray}
Through these formulae and
Eq.~(\ref{eq:acceleration_perturbation_again_of_SCH-010010}), we
obtain
\begin{eqnarray}
  \partial_{M}\bar{g}_{ab}
  &=&
      \frac{1}{(1+\alpha r\cos\theta)^{2}}
      \nonumber\\
  &&
     \times
      \left[
      (1-\alpha^{2}r^{2}) \left(\frac{2}{r}\right) \left(
      + (dt)_{a}(dt)_{b}
      + \frac{1}{Q^{2}} (dr)_{a}(dr)_{b}
      \right)
      \right.
      \nonumber\\
  && \quad\quad
     \left.
      + 2 \alpha \cos\theta \frac{r^{2}}{P^{2}}
      \left(
      - (d\theta)_{a}(d\theta)_{b}
      + C^{2} P^{2} \sin^{2}\theta (d\phi)_{a}(d\phi_{b})
      \right)
      \right]
      ,
      \label{eq:acceleration_perturbation_again_of_SCH-03600}
  \\
  \partial_{\alpha}\bar{g}_{ab}
  &=&
      - 2 \frac{r\cos\theta}{(1+\alpha r\cos\theta)}
      g_{ab}
      \nonumber\\
  &&
      + \frac{2r^{2}}{(1+\alpha r\cos\theta)^{2}}
     \nonumber\\
  && \times
      \left[
      + \alpha \left(1-\frac{2M}{r}\right) \left(
      + (dt)_{a}(dt)_{b}
      + \frac{1}{Q^{2}} (dr)_{a}(dr)_{b}
      \right)
      \right.
      \nonumber\\
  && \quad\quad\quad
     \left.
      + M \cos\theta \left(
      - \frac{1}{P^{2}} (d\theta)_{a}(d\theta)_{b}
      + C^{2} \sin^{2}\theta (d\phi)_{a}(d\phi_{b})
      \right)
      \right]
     ,
      \label{eq:acceleration_perturbation_again_of_SCH-03700}
  \\
  \partial_{C}\bar{g}_{ab}
  &=&
      \frac{1}{(1+\alpha r\cos\theta)^{2}}
      2 P C r^{2}\sin^{2}\theta (d\phi)_{a}(d\phi_{b})
      .
      \label{eq:acceleration_perturbation_again_of_SCH-03800}
\end{eqnarray}


In the case of $\alpha=0$ and $C=1$, that are background value of
these parameter for the Schwarzschild spacetime, we obtain
\begin{eqnarray}
  \left.\partial_{M}\bar{g}_{ab}\right|_{\alpha=0,C=1}
  \!\!\!\!&=&\!\!\!\!
      \left(\frac{2}{r}\right) \left(
      (dt)_{a}(dt)_{b}
      + f^{-2} (dr)_{a}(dr)_{b}
      \right)
      ,
      \quad
      f = 1 - \frac{2M}{r}
      ,
      \label{eq:acceleration_perturbation_again_of_SCH-03900}
  \\
  \left.\partial_{\alpha}\bar{g}_{ab}\right|_{\alpha=0,C=1}
  \!\!\!\!&=&\!\!\!\!
      -  2 r \cos\theta g_{ab}
      + 2 M r^{2} \cos\theta \left(
      - (d\theta)_{a}(d\theta)_{b}
      + \sin^{2}\theta (d\phi)_{a}(d\phi_{b})
      \right)
      ,
      \label{eq:acceleration_perturbation_again_of_SCH-04000}
\end{eqnarray}
and
\begin{eqnarray}
  \left.\partial_{C}\bar{g}_{ab}\right|_{\alpha=0,C=1}
  =
  2 r^{2}\sin^{2}\theta (d\phi)_{a}(d\phi)_{b}
  .
  \label{eq:acceleration_perturbation_again_of_SCH-04100}
\end{eqnarray}
Thus, the linear-order perturbation ${}_{\ScrX}\!h_{ab}$ defined by
Eq.~(\ref{eq:acceleration_perturbation_again_of_SCH-03900}) in the
second-kind gauge choice $\ScrX_{\epsilon}$ is given by
\begin{eqnarray}
  {}_{\ScrX}\!h_{ab}
  &=&
      \frac{2M_{1}}{r}
      \left(
      (dt)_{a}(dt)_{b}
      + f^{-2} (dr)_{a}(dr)_{b}
      \right)
      \nonumber\\
  &&
      +
      2 \alpha_{1}
      \left[
      -  r \cos\theta g_{ab}
      + M r^{2} \cos\theta \left(
      - (d\theta)_{a}(d\theta)_{b}
      + \sin^{2}\theta (d\phi)_{a}(d\phi)_{b}
      \right)
      \right]
      \nonumber\\
  &&
      +
      2 C_{1} r^{2} \sin^{2}\theta (d\phi)_{a}(d\phi)_{b}
     \label{eq:acceleration_perturbation_again_of_SCH-04200}
\end{eqnarray}
and that in the gauge $\ScrY$ is given by
\begin{eqnarray}
  {}_{\ScrY}\!h_{ab}
  &=&
      \frac{2M_{1}}{r}
      \left(
      (dt)_{a}(dt)_{b}
      + f^{-2} (dr)_{a}(dr)_{b}
      \right)
      \nonumber\\
  &&
      +
      2 \alpha_{1}
      \left[
      -  r \cos\theta g_{ab}
      + M r^{2} \cos\theta \left(
      - (d\theta)_{a}(d\theta)_{b}
      + \sin^{2}\theta (d\phi)_{a}(d\phi)_{b}
      \right)
      \right]
      \nonumber\\
  &&
      +
      2 C_{1} r^{2}\sin^{2}\theta (d\phi)_{a}(d\phi)_{b}
      \nonumber\\
  &&
     +
     {\pounds}_{\xi}g_{ab}
      .
     \label{eq:acceleration_perturbation_again_of_SCH-04300}
\end{eqnarray}


Here, we note that the first lines in
Eqs.~(\ref{eq:acceleration_perturbation_again_of_SCH-04200}) and
(\ref{eq:acceleration_perturbation_again_of_SCH-04300}), which
corresponds to the mass perturbations, are included  in the
perturbation $\ScrF_{AB}$ in
Eqs.~(\ref{eq:2+2-gauge-invariant-variables-calFAB}).
In the previous papers~\cite{K.Nakamura-2021a,K.Nakamura-2021d}, we
already saw this term in the analyses of the $l=0$ mode vacuum
perturbations.
On the other hand, the second and third lines in
Eqs.~(\ref{eq:acceleration_perturbation_again_of_SCH-04200}) and
(\ref{eq:acceleration_perturbation_again_of_SCH-04300}), which
corresponds to the acceleration perturbations and perturbations of the
deficit/excess angle are not so simple.
We also note that the deficit/excess angle perturbation of the third
line in Eqs.~(\ref{eq:acceleration_perturbation_again_of_SCH-04200})
and (\ref{eq:acceleration_perturbation_again_of_SCH-04300}) may
depends on the mass perturbation and acceleration perturbations.


\subsection{Components of metric perturbation of the linearized
  non-rotating C-metric}
\label{sec:Realization_of_the_linearized_C-metric}


Here, we consider the components ${}_{\ScrX}\!h_{ab}$ which is given
by Eq.~(\ref{eq:acceleration_perturbation_again_of_SCH-04200}).
We omit the gauge-index $\ScrX$ in the notation of
${}_{\ScrX}\!h_{ab}$.
The first term is the additional mass parameter perturbation of the
Schwarzschild spacetime shown in the
papers~\cite{K.Nakamura-2021a,K.Nakamura-2021d}, which is also
described in Eq.~(\ref{eq:calFab+poundsVg-l=0-non-vac-final}) for
$l=0$ mode perturbation.
The $(t-r)$-part $h_{ab}$ is given by
\begin{eqnarray}
  h_{AB}
  =
  \frac{2M_{1}}{r}
  \left(
  (dt)_{A}(dt)_{B}
  + f^{-2} (dr)_{A}(dr)_{B}
  \right)
  - 2 \alpha_{1} r \cos\theta y_{AB}
  .
  \label{eq:calXhAB-Cmetric}
\end{eqnarray}
In the expression of $h_{ab}$ in
Eq.~(\ref{eq:acceleration_perturbation_again_of_SCH-04200}),
there is no component of $h_{Ap}$.
Furthermore, the angular part of $h_{ab}$ is given by
\begin{eqnarray}
  \label{eq:calXhpq-Cmetric}
  h_{pq}
  &=&
      2 \alpha_{1}
      \left[
      -  r \cos\theta r^{2} \gamma_{pq}
      + M r^{2} \cos\theta \left(
      - (d\theta)_{p}(d\theta)_{q}
      + \sin^{2}\theta (d\phi)_{p}(d\phi)_{q}
      \right)
      \right]
      \nonumber\\
  &&
      +
      2 C_{1} r^{2} \sin^{2}\theta (d\phi)_{a}(d\phi)_{b}
     .
\end{eqnarray}
This component $h_{pq}$ is decomposed as Eq.~(\ref{eq:hpq-fourier}).
The trace of $h_{pq}$ is given by
\begin{eqnarray}
  \sum_{l,m} \tilde{h}_{(e0)} S
  =
  \frac{1}{r^{2}} \gamma^{pq} h_{pq}
  =
  -  4 \alpha_{1} r \cos\theta
  + 2 C_{1}
  .
  \label{eq:pq-part-of-Xhab-C-metric-trace}
\end{eqnarray}
On the other hand, the traceless part of $h_{pq}$ is given by
\begin{eqnarray}
  \frac{1}{r^{2}} h_{pq}
  -
  \frac{1}{2} \gamma_{pq} \frac{1}{r^{2}} \gamma^{rs} h_{rs}
  =
  \left(
  2 \alpha_{1} M \cos\theta + C_{1}
  \right)
  \left(
  - (d\theta)_{p}(d\theta)_{q}
  + \sin^{2}\theta (d\phi)_{p}(d\phi)_{q}
  \right)
  .
  \label{eq:pq-part-of-Xhab-C-metric-traceless}
\end{eqnarray}


\subsection{Harmonic decomposition of the perturbative non-rotating C-metric}
\label{sec:Realization_of_the_linearized_C-metric-harmonic-decomp}


Here, we choose the mode function $S$ as the Legendre function
$P_{l}(\cos\theta)$ as
\begin{eqnarray}
  \label{eq:harmonics-S-is-Pl}
  S_{l} = P_{l}(\cos\theta).
\end{eqnarray}
This choice corresponds to the fact that we concentrate only on the
$m=0$ modes for arbitrary $l$.
Then, the vector harmonics is defined by
\begin{eqnarray}
  \label{eq:vector-harmonics-when-S-is-Pl-even}
  \hat{D}_{p}S_{l}
  &=&
      - \sqrt{1-z^{2}} \frac{d}{dz}P_{l}(z) (d\theta)_{p}, \quad z := \cos\theta
      ,
  \\
  \label{eq:vector-harmonics-when-S-is-Pl-odd}
  \epsilon_{pq}\hat{D}^{q}S_{l}
  &=&
      (1-z^{2}) \frac{d}{dz}P_{l}(z) (d\phi)_{p}
      .
\end{eqnarray}
Furthermore, the tensor harmonics consists of the trace part
\begin{eqnarray}
  \label{eq:tensor-harmonics-when-S-is-Pl-trace}
  \frac{1}{2} \gamma_{pq} S_{l} = \frac{1}{2} \gamma_{pq} P_{l}(z),
\end{eqnarray}
the traceless even part
\begin{eqnarray}
  \left(
  \hat{D}_{p}\hat{D}_{q}
  -
  \frac{1}{2} \gamma_{pq} \hat{D}^{r}\hat{D}_{r}
  \right)S_{l}
  =
  P_{l}^{2}(z)
  \frac{1}{2} \left( \theta_{p} \theta_{q} - \phi_{p}\phi_{q}\right)
  , \quad
  (l\geq 2)
  ,
  \label{eq:tensor-harmonics-when-S-is-Pl-traceless-even}
\end{eqnarray}
and the traceless odd part
\begin{eqnarray}
  2 \epsilon_{r(p}\hat{D}_{q)}\hat{D}^{r} S_{l}
  &=&
      P_{l}^{2}(z)
      2 \theta_{(p}\phi_{q)}
      , \quad
      (l\geq 2).
  \label{eq:tensor-harmonics-when-S-is-Pl-traceless-odd}
\end{eqnarray}
These are derived from the formula
\begin{eqnarray}
  \label{eq:hatDqhatDrS-when-S-is-Pl}
  \hat{D}_{q}\hat{D}_{r}S_{l}
  =
  \left[
  (1-z^{2}) \frac{d^{2}}{dz^{2}}P_{l}(z) - z \frac{d}{dz}P_{l}(z)
  \right]
  \theta_{q}\theta_{r}
  +
  \left[
  -
  z \frac{d}{dz}P_{l}(z)
  \right]
  \phi_{r}\phi_{q}
  .
\end{eqnarray}
For $l=0,1$ modes, tensor harmonics which correspond to
Eqs.~(\ref{eq:tensor-harmonics-when-S-is-Pl-traceless-even}) and
(\ref{eq:tensor-harmonics-when-S-is-Pl-traceless-odd}) vanish.
These correspond to the fact that we have already impose the
regularity $\delta=0$ in
Proposal~\ref{proposal:treatment-proposal-on-pert-on-spherical-BG}.


Here we check the formulae for the orthogonality of the harmonics
which are necessary later.
First, we point out that the orthogonality of the scalar harmonics
(\ref{eq:harmonics-S-is-Pl}) for $l,l'\geq 0$ modes:
\begin{eqnarray}
  \label{eq:orthogonality-of-harmonics-S-is-Pl}
  \int d^{2}\Omega S_{l'}S_{l}
  =
  2 \pi \int_{0}^{\pi} \sin\theta d\theta P_{l'}(\cos\theta)P_{l}(\cos\theta)
  =
  2 \pi \int_{-1}^{-1} dz P_{l'}(z)P_{l}(z)
  =
  \frac{4\pi}{2l+1} \delta_{ll'}
  .
\end{eqnarray}
Next, we consider the orthogonality of the even tensor harmonics
(\ref{eq:tensor-harmonics-when-S-is-Pl-traceless-even}) for
$l,l'\geq 2$ modes:
\begin{eqnarray}
  &&
  \int d^{2}\Omega
  \left(
  \hat{D}_{p}\hat{D}_{q}
  -
  \frac{1}{2} \gamma_{pq} \hat{D}^{r}\hat{D}_{r}
  \right)S_{l}
  \left(
  \hat{D}^{p}\hat{D}^{q}
  -
  \frac{1}{2} \gamma^{pq} \hat{D}^{r}\hat{D}_{r}
  \right)S_{l'}
     \nonumber\\
  &=&
      \frac{2\pi(l-1) l (l+1) (l+2)}{(2l+1)}\delta_{ll'}
      .
  \label{eq:orthogonality-of-even-tensor-harmonics-S-is-Pl}
\end{eqnarray}


Now, we consider the perturbative C-metric given by
Eqs.~(\ref{eq:calXhAB-Cmetric}), (\ref{eq:calXhpq-Cmetric}), and
$h_{Ap}=0$.
The angular part $h_{pq}$ given by Eq.~(\ref{eq:calXhpq-Cmetric}) is
also decomposed into the trace and the traceless part as
Eqs.~(\ref{eq:pq-part-of-Xhab-C-metric-trace}) and
(\ref{eq:pq-part-of-Xhab-C-metric-traceless}).
Since there is no $h_{Ap}$ component, the perturbative C-metric does
not have any vector part.
Furthermore, the traceless part of the angular components
(\ref{eq:pq-part-of-Xhab-C-metric-traceless}) does not have
$(\theta-\phi)$ component.
Therefore, the perturbative C-metric does not have any tensor odd
mode.
Furthermore, we do not have any vector- and tensor-modes in $l=0,1$
modes.


For our convention, we introduce the notation $|K_{pq}^{l}\rangle$
to consider the orthogonality
Eq.~(\ref{eq:orthogonality-of-even-tensor-harmonics-S-is-Pl}) of the
even-mode tensor harmonics by
\begin{eqnarray}
  |K_{pq}^{l}\rangle
  :=
  \left[
  \hat{D}_{p}\hat{D}_{q}
  -
  \frac{1}{2} \gamma_{pq} \hat{D}_{r}\hat{D}^{r}
  \right]S_{l}
  .
  \label{eq:Spqelrangle-def}
\end{eqnarray}
The orthogonality condition
(\ref{eq:orthogonality-of-even-tensor-harmonics-S-is-Pl}) is
denoted as
\begin{eqnarray}
  \label{eq:orthogonality-of-even-tensor-harmonics-S-is-Pl-braket}
  \langle K^{pq}_{l'}|K_{pq}^{l}\rangle
  &:=&
       \int d^{2}\Omega
       \left[
       \hat{D}^{p}\hat{D}^{q}
       -
       \frac{1}{2} \gamma^{pq} \hat{D}_{r}\hat{D}^{r}
       \right]P_{l'}
       \left[
       \hat{D}_{p}\hat{D}_{q}
       -
       \frac{1}{2} \gamma_{pq} \hat{D}_{s}\hat{D}^{s}
       \right]P_{l}
       \nonumber\\
  &=&
      \frac{2\pi(l-1)l(l+1)(l+2)}{2l+1} \delta_{ll'}
      .
\end{eqnarray}
When an arbitrary traceless tensor $f_{pq}$, which represent the
vector $|f_{pq}\rangle$ of the function space, is given by
\begin{eqnarray}
  |f_{pq}\rangle
  =
  \sum_{l}
  g_{l}
  |K_{pq}^{l}\rangle
  ,
\end{eqnarray}
applying $\langle K^{pq}_{l'}|$ from the left, we obtain
\begin{eqnarray}
  \langle K^{pq}_{l'}|f_{pq}\rangle
  &=&
      \sum_{l\geq 2}^{\infty}
      g_{l}
      \langle K^{pq}_{(e)l'}|K_{pq}^{(e)l}\rangle
      \nonumber\\
  &=&
      \sum_{l\geq 2}^{\infty}
      g_{l}
      \frac{2\pi(l-1)l(l+1)(l+2)}{2l+1}
      \delta_{ll'}
      \nonumber\\
  &=&
      g_{l}
      \frac{2\pi(l-1)l(l+1)(l+2)}{2l+1}
      \quad
      (l\geq 2)
      .
\end{eqnarray}
Then, we have
\begin{eqnarray}
  g_{l}
  =
  \frac{2l+1}{2\pi(l-1)l(l+1)(l+2)}
  \langle K^{pq}_{l}|f_{pq}\rangle
  \quad
  (l\geq 2)
  .
\end{eqnarray}


Since we note that the traceless part
(\ref{eq:pq-part-of-Xhab-C-metric-traceless}) in $h_{pq}$ does not
have the odd-mode part, we obtain
\begin{eqnarray}
  \frac{1}{r^{2}} {}_{\ScrX}\!h_{pq}
  -
  \frac{1}{2} \gamma_{pq} \frac{1}{r^{2}} \gamma^{rs} {}_{\ScrX}\!h_{rs}
  &=&
      \sum_{l\geq 2} \tilde{h}_{(e2)} |K_{pq}^{l}\rangle
      \nonumber\\
  &=&
      \left(
      2 \alpha_{1} M \cos\theta + C_{1}
      \right)
      \left(
      - \theta_{p} \theta_{q}
      + \phi_{p} \phi_{q}
      \right)
      .
  \label{eq:pq-part-of-Xhab-C-metric-traceless-4}
\end{eqnarray}
Then, we obtain
\begin{eqnarray}
  \tilde{h}_{(e2)}
  &=&
      -
      \frac{2l+1}{2\pi(l-1)l(l+1)(l+2)}
      2 \pi \int_{-1}^{1} dx
      \left(
      2 \alpha_{1} M x + C_{1}
      \right)
      P_{l}^{2}(x)
      \nonumber\\
  &=&
      -
      \frac{2l+1}{2\pi(l-1)l(l+1)(l+2)}
      2 \pi \int_{-1}^{1} dx
      \left(
      2 \alpha_{1} M x + C_{1}
      \right)
      (1-x^{2})
      \frac{d^{2}}{dx^{2}}P_{l}(x)
      \nonumber\\
  &=&
      -
      \frac{2(2l+1)}{(l-1)l(l+1)(l+2)}
      \left[
      2 \alpha_{1} M + C_{1}
      +
      ( - 2 \alpha_{1} M + C_{1} )
      (-1)^{l}
      \right]
      .
      \label{eq:pq-part-of-Xhab-C-metric-traceless-each-mode-result}
\end{eqnarray}


In summary, the perturbative C-metric $h_{ab}$ given by
Eqs.~(\ref{eq:calXhAB-Cmetric})--(\ref{eq:calXhpq-Cmetric}) is
decomposed based on the scalar harmonic function $S_{l}=P_{l}(\cos\theta)$.
Together with corresponding gauge-invariant and gauge-variant variables,
these are summarized as follows:
For $l=0$ mode, the mode coefficients of the harmonic decomposition are
given by
\begin{eqnarray}
  \label{eq:calXtildehAB-Cmetric-l=0}
  \tilde{h}_{AB}
  &=&
      \frac{2M_{1}}{r}
      \left(
      (dt)_{A}(dt)_{B}
      + f^{-2} (dr)_{A}(dr)_{B}
      \right)
      ,
  \\
  \label{eq:calXtildehAp-even-odd-Cmetric-l=0}
  \tilde{h}_{(e1)A}
  &=&
      0,
      \quad
      \tilde{h}_{(o1)A} = 0
      ,
  \\
  \label{eq:calXtildehe0-calXtildeho1-calXtildehe1-pq-Cmetric-l=0}
  \tilde{h}_{(e0)}
  &=&
      2 C_{1}
      ,
      \quad
      \tilde{h}_{(e2)}
      =
      0
      ,
      \quad
      \tilde{h}_{(o2)}
      =
      0
      .
\end{eqnarray}
Components of gauge-variant part $\tilde{Y}_{A}$, $\tilde{Y}_{(o)}$,
and $\tilde{Y_{(e)}}$ of the metric perturbation for $l=0$ mode are
given by Eqs.~(\ref{eq:2+2-gauge-trans-tildeYA-def-sum}),
(\ref{eq:tildeYo-def}), and (\ref{eq:tildeYe-def}).
Substituting
Eqs.~(\ref{eq:calXtildehAB-Cmetric-l=0})--(\ref{eq:calXtildehe0-calXtildeho1-calXtildehe1-pq-Cmetric-l=0})
into these equations, we obtain
\begin{eqnarray}
  \label{eq:gauge-variant-l=0-def-C-metric}
  \tilde{Y}_{A}
  :=
       r \tilde{h}_{(e1)A}
       - \frac{r^{2}}{2} \bar{D}_{A}\tilde{h}_{(e2)}
       =
       0
       ,
  \quad
  \tilde{Y}_{(o1)}
  :=
       - r^{2} \tilde{h}_{(o2)}
       =
       0
       ,
  \quad
  \tilde{Y}_{(e1)}
  :=
       \frac{r^{2}}{2} \tilde{h}_{(e2)}
       =
       0
       .
\end{eqnarray}
Components of the gauge-invariant variables $\tilde{F}_{A}$,
$\tilde{F}$, and $\tilde{F}_{AB}$ defined by
Eqs.~(\ref{eq:2+2-gauge-inv-def-tildeFA-sum})--(\ref{eq:gauge-inv-tildeFAB-def-sum})
are given by
\begin{eqnarray}
  \label{eq:2+2-gauge-inv-kernel-Delta-def-tildeFA-C-metric}
  \tilde{F}_{A}
  &:=&
       \tilde{h}_{(k_{\hat{\Delta}},o)A}
       + r \bar{D}_{A}\tilde{h}_{(k_{\hat{\Delta}},o1)}
       =
       0
       ,
  \\
  \label{eq:2+2-gauge-inv-kernel-Delta-tildeF-def-C-metric}
  \tilde{F}
  &:=&
       \tilde{h}_{(k_{\hat{\Delta}},e0)}
       - \frac{4}{r} \tilde{Y}_{A} \bar{D}^{A}r
       =
       2 C_{1}
       , \\
  \tilde{F}_{AB}
  &:=&
       \tilde{h}_{(k_{\hat{\Delta}})AB}
       - 2 \bar{D}_{(A}\tilde{Y}_{(k_{\hat{\Delta}})B)}
       =
       \frac{2M_{1}}{r}
       \left(
       (dt)_{A}(dt)_{B}
       + f^{-2} (dr)_{A}(dr)_{B}
       \right)
       .
       \label{eq:gauge-inv-kernel-Delta-tildeFAB-def-C-metric}
\end{eqnarray}


For $l=1$ mode, the mode coefficients of harmonic decomposition are
given by
\begin{eqnarray}
  \label{eq:calXtildehAB-Cmetric-l=1}
  h_{AB}
  &=&
      - 2 \alpha_{1} r (- f (dt)_{A} (dt)_{B} + f^{-1} (dr)_{A} (dr)_{B} )
      ,
  \\
  \label{eq:calXtildehAp-even-odd-Cmetric-l=1}
  \tilde{h}_{(e1)A}
  &=&
      0,
      \quad
      \tilde{h}_{(o1)A} = 0
      ,
  \\
  \label{eq:calXtildehe0-calXtildeho1-calXtildehe1-pq-Cmetric-l=1}
  \tilde{h}_{(e0)}
  &=&
      -
      4 \alpha_{1} r
      ,
      \quad
      \tilde{h}_{(e2)}
      =
      0
      ,
      \quad
      \tilde{h}_{(o2)}
      =
      0
      .
\end{eqnarray}
Components of gauge-variant part $\tilde{Y}_{A}$, $\tilde{Y}_{(o)}$,
and $\tilde{Y_{(e)}}$ of the metric perturbation for $l=1$ modes are
given by Eqs.~(\ref{eq:2+2-gauge-trans-tildeYA-def-sum}),
(\ref{eq:tildeYo-def}), and (\ref{eq:tildeYe-def}).
Substituting
Eqs.~(\ref{eq:calXtildehAB-Cmetric-l=1})--(\ref{eq:calXtildehe0-calXtildeho1-calXtildehe1-pq-Cmetric-l=1})
into these equations, we obtain
\begin{eqnarray}
  \label{eq:gauge-variant-l=1-def-C-metric}
  \tilde{Y}_{A}
  :=
  r \tilde{h}_{(e1)A}
  - \frac{r^{2}}{2} \bar{D}_{A}\tilde{h}_{(e2)}
  =
  0
  ,
  \quad
  \tilde{Y}_{(o)}
  :=
  - r^{2} \tilde{h}_{(o2)}
  =
  0
  ,
  \quad
  \tilde{Y}_{(e)}
  :=
  \frac{r^{2}}{2} \tilde{h}_{(e2)}
  =
  0
  .
\end{eqnarray}
Components of the gauge-invariant variables $\tilde{F}_{A}$,
$\tilde{F}$, and $\tilde{F}_{AB}$ defined by
Eqs.~(\ref{eq:2+2-gauge-inv-def-tildeFA-sum})--(\ref{eq:gauge-inv-tildeFAB-def-sum})
are given by
\begin{eqnarray}
  \label{eq:2+2-gauge-inv-kernel-Delta+2-def-tildeFA-sum-C-metric}
  \tilde{F}_{A}
  &:=&
       \tilde{h}_{(o1)A}
       + r \bar{D}_{A}\tilde{h}_{(o2)}
       =
       0
       ,
  \\
  \label{eq:2+2-gauge-inv-kernel-Delta+2-tildeF-def-sum-C-metric}
  \tilde{F}
  &:=&
       - 4 \alpha_{1} r
       , \\
  \label{eq:gauge-inv-kernel-Delta+2-tildeFAB-def-sum-C-metric}
  \tilde{F}_{AB}
  &:=&
       - 2 \alpha_{1} r ( - f (dt)_{A} (dt)_{B} + f^{-1} (dr)_{A} (dr)_{B} )
       .
\end{eqnarray}


For $l\geq 2$ modes, the mode coefficients of harmonic decomposition
are given by
\begin{eqnarray}
  \label{eq:calXtildehAB-calXtildehAp-even-odd-Cmetric-lgeq2}
  h_{AB}
  &=&
      0
      ,
      \quad
      \tilde{h}_{(e1)A}
      =
      0
      ,
      \quad
      \tilde{h}_{(o1)A} = 0
      ,
  \\
  \label{eq:calXtildehe0-pq-Cmetric-lgeq2}
  \tilde{h}_{(e0)}
  &=&
      0
      ,
      \quad
      \tilde{h}_{(o2)}
      =
      0
      ,
  \\
  \label{eq:calXtildehe2-pq-Cmetric-lgeq2}
  \tilde{h}_{(e2)}
  &=&
      -
      \frac{2(2l+1)}{(l-1)l(l+1)(l+2)}
      \left[
      (+2 \alpha_{1} M + C_{1})
      +
      ( - 2 \alpha_{1} M + C_{1} )
      (-1)^{l}
      \right]
      .
\end{eqnarray}
Components of gauge-variant part $\tilde{Y}_{A}$, $\tilde{Y}_{(o)}$,
and $\tilde{Y_{(e)}}$ of the metric perturbation for $l\geq 2$ mode
given by Eqs.~(\ref{eq:2+2-gauge-trans-tildeYA-def-sum}),
(\ref{eq:tildeYo-def}), and (\ref{eq:tildeYe-def}).
Substituting
Eqs.~(\ref{eq:calXtildehAB-calXtildehAp-even-odd-Cmetric-lgeq2})--(\ref{eq:calXtildehe2-pq-Cmetric-lgeq2})
into these equations, we obtain
\begin{eqnarray}
  \label{eq:2+2-gauge-trans-tildeYA-lgeq2-def-C-metric}
  \!\!\!\!\!
  \tilde{Y}_{A}
  \!\!\!\!&:=&\!\!\!\!
       r \tilde{h}_{(e1)A}
       - \frac{r^{2}}{2} \bar{D}_{A}\tilde{h}_{(e2)}
       =
       0
       ,
       \quad
       \tilde{Y}_{(o)}
      :=
       - r^{2} \tilde{h}_{(o2)}
       =
       0
       ,
      \\
  \label{eq:tildeYe-def-lgeq2-C-metric}
  \!\!\!\!\!
  \tilde{Y}_{(e)}
  \!\!\!\!&:=&\!\!\!\!
       \frac{r^{2}}{2} \tilde{h}_{(e2)}
       =
       -
       \frac{r^{2}(2l+1)}{(l-1)l(l+1)(l+2)}
       \left[
       (+ 2 \alpha_{1} M + C_{1})
       +
       ( - 2 \alpha_{1} M + C_{1} )
       (-1)^{l}
       \right]
       .
\end{eqnarray}
Components of the gauge-invariant variables $\tilde{F}_{A}$,
$\tilde{F}$, and $\tilde{F}_{AB}$ defined by
Eqs.~(\ref{eq:2+2-gauge-inv-def-tildeFA-sum})--(\ref{eq:gauge-inv-tildeFAB-def-sum})
are given by
\begin{eqnarray}
  \label{eq:2+2-gauge-inv-def-tildeFA-lgeq2-C-metric}
  \tilde{F}_{A}
  &:=&
       \tilde{h}_{(o1)A}
       + r \bar{D}_{A}\tilde{h}_{(o2)}
       =
       0
       ,
       \quad
       \tilde{F}_{AB}
       :=
       \tilde{h}_{AB}
       - 2 \bar{D}_{(A}\tilde{Y}_{B)}
       =
       0
       ,
  \\
  \tilde{F}
  &:=&
       \tilde{h}_{(e0)}
       - \frac{4}{r} \tilde{Y}_{A} \bar{D}^{A}r
       + \tilde{h}_{(e2)} l(l+1)
       \nonumber\\
  &=&
      -
      \frac{2(2l+1)}{(l-1)(l+2)}
      \left[
      ( + 2 \alpha_{1} M + C_{1} )
      +
      ( - 2 \alpha_{1} M + C_{1} )
      (-1)^{l}
      \right]
      .
      \label{eq:2+2-gauge-inv-tildeF-def-lgeq2-C-metric}
\end{eqnarray}


\subsection{Realization of $l\geq 2$ mode perturbations}
\label{sec:lgeq2-mode-for-the-C-metric}


As shown in above, the perturbative expression of the C-metric does
not include odd-mode perturbation as in the first equation
in Eq.~(\ref{eq:2+2-gauge-inv-def-tildeFA-lgeq2-C-metric}).
Furthermore, we also obtain $\tilde{F}_{AB}=0$ from the second
equation in Eq.~(\ref{eq:2+2-gauge-inv-def-tildeFA-lgeq2-C-metric})
and the gauge-invariant variable $\tilde{F}$ is constant given by
Eq.~(\ref{eq:2+2-gauge-inv-tildeF-def-lgeq2-C-metric}).
Therefore, we obtain
\begin{eqnarray}
  \label{eq:gauge-inv-metric-variable-lgeq2-C-metric-1}
  \tilde{F}_{AB} = 0, \quad \partial_{r}\tilde{F} = \partial_{t}\tilde{F} = 0,
\end{eqnarray}
for $l\geq 2$ mode perturbations.


For $l\geq 2$ mode, the linearized Einstein equations for even-mode
perturbations are given in Sec.~\ref{sec:Even_Einstein_equations}.
From Eq.~(\ref{eq:linearized-Einstein-pq-traceless-even}) and the
first condition (\ref{eq:gauge-inv-metric-variable-lgeq2-C-metric-1}),
we obtain
\begin{eqnarray}
  \label{eq:eventildeFtrace-matter-reduce-C-metric}
  \tilde{T}_{(e2)} = 0.
\end{eqnarray}
Since Eqs.~(\ref{eq:gauge-inv-metric-variable-lgeq2-C-metric-1})
implies $X_{(e)}=Y_{(e)}=0$ and $\tilde{F}$ is constant, we obtain
\begin{eqnarray}
  \label{eq:even-FAB-divergence-3-A-C-metric}
  \tilde{T}_{(e1)A} = 0
\end{eqnarray}
from Eqs.~(\ref{eq:even-FAB-divergence-3}) and
(\ref{eq:eventildeFtrace-matter-reduce-C-metric}).
Furthermore, from
Eq.~(\ref{eq:gauge-inv-metric-variable-lgeq2-C-metric-1}) and
Eq.~(\ref{eq:1st-pert-Einstein-non-vac-AB-traceless-final-3}), we
obtain
\begin{eqnarray}
  \label{eq:SFFAB-lgeq2-vanishes}
  S_{(\FF)AB} = 0.
\end{eqnarray}
Together with Eqs.~(\ref{eq:eventildeFtrace-matter-reduce-C-metric})
and (\ref{eq:even-FAB-divergence-3-A-C-metric}),
Eq.~(\ref{eq:SFFAB-lgeq2-vanishes}) yields
\begin{eqnarray}
  \label{eq:even-mode-Ye-evol-eq-reduce-C-metric-sum}
  && \tilde{T}_{tr} = 0, \\
  \label{eq:even-mode-Xe-evol-eq-reduce-C-metric-sum}
  && \tilde{T}_{tt} + f^{2} \tilde{T}_{rr} = 0.
\end{eqnarray}
Moreover, from Eqs.~(\ref{eq:Moncrief-master-variable-final}) and
(\ref{eq:gauge-inv-metric-variable-lgeq2-C-metric-1}),
Eq.~(\ref{eq:even-mode-tildeF-eq-Phie-sum}) with the source term
(\ref{eq:source(F)-def-sum-component}) is given by
\begin{eqnarray}
  \tilde{F}
  =
  16 \pi \frac{r^{2}}{(l-1)(l+2)}
  \left(
  \frac{1}{f} \tilde{T}_{tt}
  -  f \tilde{T}_{rr}
  \right)
  =
  \frac{32 \pi r^{2}}{(l-1)(l+2) f} \tilde{T}_{tt} = \mbox{constant}
  .
  \label{eq:even-mode-tildeF-evol-eq-reduce-C-metric-sum}
\end{eqnarray}
Finally, from Eqs.~(\ref{eq:eventildeFtrace-matter-reduce-C-metric})
and (\ref{eq:even-FAB-divergence-3-A-C-metric}) the component
(\ref{eq:div-barTab-linear-p-even-mode}) yields
\begin{eqnarray}
  \tilde{T}_{(e0)} = 0
  \label{eq:1st-pert-Einstein-pq-trace-Fourier-2-C-metric-sum}
\end{eqnarray}
for $l\geq 2$ modes.
Thus, from the definition of these components
(\ref{eq:1st-pert-calTab-dd-decomp}), for $l\geq 2$ mode, we obtain
\begin{eqnarray}
  {}^{(1)}\!\ScrT_{ac}
  &=&
      \sum_{l=2}^{\infty} P_{l}(\cos\theta)
      \left(
      \tilde{T}_{tt}
      (dt)_{a} (dt)_{c}
      +
      \tilde{T}_{rr}
      (dr)_{a} (dr)_{c}
      \right)
      \nonumber\\
  &=:&
      -
      \frac{1}{r^{2}}
      y_{ab}
      \sum_{l=2}^{\infty}
      \lambda_{l}
      P_{l}(\cos\theta)
      ,
     \label{eq:1st-pert-calTab-dd-decomp-2-C-metric}
\end{eqnarray}
where we defined
\begin{eqnarray}
  \label{eq:constant-tension-from-Einstein}
  16 \pi \lambda_{l} := 16 \pi \frac{r^{2}}{f} \tilde{T}_{tt} = -
  (2l+1) \left[
  (+2\alpha_{1}M + C_{1}) + (-2\alpha_{1}M+C_{1}) (-1)^{l}
  \right]
\end{eqnarray}
from Eqs.~(\ref{eq:even-mode-tildeF-evol-eq-reduce-C-metric-sum}) and
(\ref{eq:2+2-gauge-inv-tildeF-def-lgeq2-C-metric}).


We check the other components of the linearized Einstein equation.
To carry out this, we see that the Moncrief variable $\Phi_{(e)}$ in
the case of Eq.~(\ref{eq:gauge-inv-metric-variable-lgeq2-C-metric-1})
is given by
\begin{eqnarray}
  \label{eq:Moncrief-master-variable-final-C-metric}
  \Phi_{(e)}
  =
  - \frac{r}{4} \tilde{F}
  ,
\end{eqnarray}
through the definition (\ref{eq:Moncrief-master-variable-final}).
Through Eqs.~(\ref{eq:eventildeFtrace-matter-reduce-C-metric}) and
(\ref{eq:even-mode-Ye-evol-eq-reduce-C-metric-sum}), and from
Eqs.~(\ref{eq:SLambdaF-def-explicit}) and
(\ref{eq:sourceYe-SPsie-explicit-sum}), the source terms
$S_{(\Lambda\tilde{F})}$ and $S_{(Y_{2})}$ are given by
\begin{eqnarray}
  S_{(\Lambda\tilde{F})}
  =
  \tilde{T}_{tt}
  ,
  \quad
  S_{(Y_{(e)})}
  =
  0
  .
  \label{eq:SLambdaF-sourceYe-explicit-C-metric}
\end{eqnarray}
From $S_{(Y_{(e)})}=0$ and the vanishing components
$(X_{(e)},Y_{(e)})$ of $\tilde{\FF}_{AB}$ and the
derivative of $\tilde{F}$ yields that
Eq.~(\ref{eq:ll+1fYe-1st-pert-Ein-non-vac-tr-FAB-div-t-r-sum-3}) is
trivial.
On the other hand, substituting
Eqs.~(\ref{eq:Moncrief-master-variable-final}) and
$S_{(\Lambda\tilde{F})}=\tilde{T}_{tt}$ into
Eq.~(\ref{eq:-Ein-non-vac-XF-constraint-apha-Phie-with-SLambdaF-sum}),
we obtain the same result as
Eq.~(\ref{eq:even-mode-tildeF-evol-eq-reduce-C-metric-sum}).


Finally, we check the master equation
(\ref{eq:Zerilli-Moncrief-eq-final-sum}) with the source term
$S_{(\Phi_{(e)})}$ given by Eq.~(\ref{eq:SPhie-def-explicit-sum}).
Substituting Eqs.~(\ref{eq:eventildeFtrace-matter-reduce-C-metric}),
(\ref{eq:even-FAB-divergence-3-A-C-metric}),
(\ref{eq:even-mode-Xe-evol-eq-reduce-C-metric-sum}),
(\ref{eq:even-mode-tildeF-evol-eq-reduce-C-metric-sum}), and (\ref{eq:1st-pert-Einstein-pq-trace-Fourier-2-C-metric-sum}),
\begin{eqnarray}
  S_{(\Phi_{(e)})}
  &=&
      \frac{1}{2f} \Lambda \tilde{T}_{tt}
      + \frac{2f-1}{f} \tilde{T}_{tt}
      - \frac{3(1-f)}{\Lambda} \tilde{T}_{tt}
     .
      \label{eq:SPhie-def-explicit-sum-C-metric}
\end{eqnarray}
Substituting Eqs.~(\ref{eq:Moncrief-master-variable-final-C-metric}),
(\ref{eq:SPhie-def-explicit-sum-C-metric}),
(\ref{eq:Zerilli-Moncrief-master-potential-final-sum}), and
(\ref{eq:even-mode-tildeF-evol-eq-reduce-C-metric-sum}) into
Eq.~(\ref{eq:Zerilli-Moncrief-eq-final-sum}), and using
Eq.~(\ref{eq:mu-Lambda-defs}), we can confirm that
Eq.~(\ref{eq:Zerilli-Moncrief-eq-final-sum}) is trivial.


\subsection{Realization of $l=0$ mode perturbations}
\label{sec:l=0-mode-for-the-C-metric}


The solutions to the Einstein equation for $l=0$ mode with a generic
matter field are extensively discussed in the Part II
paper~\cite{K.Nakamura-2021d}.
Following
Proposal~\ref{proposal:treatment-proposal-on-pert-on-spherical-BG},
we impose the regularity to the harmonic $S_{\delta}$ and the
components $\tilde{T}_{(e1)A}$ and $\tilde{T}_{(e2)}$ do not appear
in the expression of the linear perturbations of the energy-momentum
tensor.
Therefore, we may safely choose,
\begin{eqnarray}
  \label{eq:l=0-constraint-for-ene-mon}
  \tilde{T}_{(e1)A} = 0, \quad \tilde{T}_{(e2)} = 0.
\end{eqnarray}
From Eqs.~(\ref{eq:l=0-constraint-for-ene-mon}) and
(\ref{eq:div-barTab-linear-p-even-mode}), we have
\begin{eqnarray}
  \tilde{T}_{(e0)}
  =
  0
  ,
  \label{eq:div-barTab-linear-p-l=0-C-metric}
\end{eqnarray}
and Eqs.~(\ref{eq:div-barTab-linear-AB-t})
and (\ref{eq:div-barTab-linear-AB-r}) are given by
\begin{eqnarray}
  &&
     -  \partial_{t}\tilde{T}_{tt}
     + f^{2} \partial_{r}\tilde{T}_{rt}
     + \frac{(1+f)f}{r} \tilde{T}_{rt}
     =
     0
     ,
     \label{eq:div-barTab-linear-AB-t-l=0-C-metric}
  \\
  &&
     -  \partial_{t}\tilde{T}_{tr}
     + \frac{1-f}{2rf} \tilde{T}_{tt}
     + f^{2} \partial_{r}\tilde{T}_{rr}
     + \frac{(3+f)f}{2r} \tilde{T}_{rr}
     =
     0
     .
     \label{eq:div-barTab-linear-AB-r-l=0-C-metric}
\end{eqnarray}
On the other hand, the metric perturbation of $l=0$ mode summarized in
Eq.~(\ref{eq:calXtildehAB-Cmetric-l=0})--(\ref{eq:calXtildehe0-calXtildeho1-calXtildehe1-pq-Cmetric-l=0})
is given by
\begin{eqnarray}
  h_{ab}(l=0)
  &=&
      \frac{2M_{1}}{r}
      \left(
      (dt)_{a}(dt)_{b}
      + f^{-2} (dr)_{a}(dr)_{b}
      \right)
     + r^{2} C_{1} \gamma_{ab}
     .
     \label{eq:acceleration_perturbation_of_SCH-07000-00500}
\end{eqnarray}
The mass perturbation $M_{1}$ is discussed in the Part II
paper~\cite{K.Nakamura-2021d}.
To include this mass perturbation parameter $M_{1}$ as the solution to
the Einstein equation in our formulation, we have to introduce the
term
\begin{eqnarray}
  \label{eq:mass-gauge-added}
  {}_{\ScrZ}\!h_{ab}(l=0)
  =
  {}_{\ScrX}\!h_{ab}(l=0)
  +
  {\pounds}_{V}g_{ab}
  ,
\end{eqnarray}
where the generator $V_{a}$ is given
\begin{eqnarray}
  \label{eq:l=0-vac-generator-sum-C-metric}
  V_{a}
  =
  \left(
  \frac{f}{4} \Upsilon
  + \frac{rf}{4} \partial_{r}\Upsilon
  -  \frac{r}{1-3f} \Xi(r)
  + f \int dr \frac{2}{f(1-3f)^{2}} \Xi(r)
  \right)
  (dt)_{a}
  +
  \frac{r}{4f} \partial_{t}\Upsilon
  (dr)_{a}
  .
\end{eqnarray}
Here, the function $\Upsilon(t,r)$ is the solution to the equation
\begin{eqnarray}
  -  \frac{1}{f} \partial_{t}^{2}\Upsilon
  + \partial_{r}( f \partial_{r}\Upsilon )
  + \frac{1}{r^{2}} 3(1-f) \Upsilon
  -  \frac{4}{r^{3}} 2 M_{1} t
  =
  0
  ,
  \label{eq:Upsilon-eq-int-dt-2-for-mass-perturbation-C-metric}
\end{eqnarray}
and $\Xi(r)$ in Eq.~(\ref{eq:l=0-vac-generator-sum-C-metric}) is an
arbitrary function of $r$.


Since we always introduce the mass parameter perturbation $M_{1}$
if we introduce the last term in Eq.~(\ref{eq:mass-gauge-added}), we
ignore this mass perturbation at this moment.
Then, we may concentrate on the perturbation $C_{1}$ in the metric
perturbation
\begin{eqnarray}
  h_{ab}(l=0)
  &=&
     r^{2} C_{1} \gamma_{ab}
     .
     \label{eq:acceleration_perturbation_of_SCH-07000-00500-2}
\end{eqnarray}
Since the $\theta\theta$- and $\phi\phi$-components in the solution
(\ref{eq:calFab+poundsVg-l=0-non-vac-final}) is described by the term
${\pounds}_{V}g_{ab}$, we consider the components
${\pounds}_{V_{(1)}}g_{ab}$ with the generator
$V_{(1)a}=V_{(1)r}(r)(dr)_{a}$.
In this case, the non-vanishing components of
${\pounds}_{V_{(1)}}g_{ab}$ are summarized as
\begin{eqnarray}
  &&
     {\pounds}_{V_{(1)}}g_{tt}
     =
     - f f' V_{(1)r}
     ,
     \quad
     {\pounds}_{V_{(1)}}g_{rr}
     =
     2 \partial_{r}V_{(1)r} + \frac{f'}{f} V_{(1)r}
     ,
     \nonumber
  \\
  &&
     {\pounds}_{V_{(1)}}g_{\theta\theta}
     =
     2 rf V_{(1)r}
     ,
     \quad
     {\pounds}_{V_{(1)}}g_{\phi\phi}
     =
     2 rf \sin^{2}\theta V_{(1)r}
     .
     \label{eq:poundsVgab-tt-rr-thetatheta-phiphiCmetric-l=0-Vtheta=Vphi=0}
\end{eqnarray}
Here, we choose
\begin{eqnarray}
  V_{(1)r} = \frac{r^{2}}{2rf} C_{1}
\end{eqnarray}
so that $\theta\theta$- and $\phi\phi$-components are described by the
term ${\pounds}_{V_{(1)}}g_{ab}$.
Then, we have
\begin{eqnarray}
  h_{ab}(l=0)
  &=&
      \frac{1-f}{2} C_{1} (dt)_{a}(dt)_{b}
      + \frac{1-3f}{2f^{2}} C_{1} (dr)_{a}(dr)_{b}
      + {\pounds}_{V_{(1)}}g_{ab}
      .
     \label{eq:acceleration_perturbation_of_SCH-07000-00500-3}
\end{eqnarray}


As in the case of LTB solution, we make $(t,r)$ part in
Eq.~(\ref{eq:acceleration_perturbation_of_SCH-07000-00500-3}) to be
traceless through the introduction of the term
${\pounds}_{V_{(2)}}g_{ab}$ with $V_{(2)a}=V_{(2)t}(dt)_{a}$ with the
condition $\partial_{\theta}V_{(2)t}=\partial_{\phi}V_{(2)t}=0$.
In this choice of $V_{(2)a}$, the nonvanishing components of
${\pounds}_{V_{(2)}}g_{ab}$ are given by
\begin{eqnarray}
  \label{eq:poundsVgab-tt-tr-Cmetric-l=0}
  {\pounds}_{V_{(2)}}g_{tt}
  =
  2 \partial_{t}V_{(2)t}
  ,
  \quad
  {\pounds}_{V}g_{tr}
  =
  \partial_{r}V_{(2)t} - \frac{f'}{f} V_{(2)t}
  .
\end{eqnarray}
Through this expression (\ref{eq:poundsVgab-tt-tr-Cmetric-l=0}),
$h_{ab}(l=0)$ is given by
\begin{eqnarray}
  h_{ab}(l=0)
  &=&
      \left( \frac{1-f}{2} C_{1} - 2 \partial_{t}V_{(2)t} \right) (dt)_{a}(dt)_{b}
      + \frac{1-3f}{2} C_{1} f^{-2} (dr)_{a}(dr)_{b}
      \nonumber\\
  &&
      - \left( \partial_{r}V_{(2)t} - \frac{f'}{f} V_{(2)t} \right) 2 (dt)_{(a} (dr)_{b)}
      + {\pounds}_{V_{(1)}+V_{(2)}}g_{ab}
      \nonumber\\
  &=&
      \left( f C_{1} - 2 \partial_{t}V_{(2)t} \right) (dt)_{a}(dt)_{b}
      + \frac{1-3f}{2} C_{1} \left(
      (dt)_{a}(dt)_{b} + f^{-2} (dr)_{a}(dr)_{b}
      \right)
      \nonumber\\
  &&
      - \left( \partial_{r}V_{(2)t} - \frac{f'}{f} V_{(2)t} \right) 2 (dt)_{(a} (dr)_{b)}
      + {\pounds}_{V_{(1)}+V_{(2)}}g_{ab}
     .
     \label{eq:l=0-mode-perturbation-sol-C-meric-tmp}
\end{eqnarray}
Here, we choose $V_{(2)t}$ so that
\begin{eqnarray}
  \label{eq:Wt-determination-by-C1}
  V_{(2)t} = \frac{1}{2} f t C_{1}.
\end{eqnarray}
This choice (\ref{eq:Wt-determination-by-C1}) yields
\begin{eqnarray}
  \partial_{r}V_{(2)t} - \frac{f'}{f} V_{(2)t} = 0,
  \label{eq:t-r-comp-confirm}
\end{eqnarray}
and
\begin{eqnarray}
  {}_{\ScrX}\!h_{ab}(l=0)
  &=&
      \frac{1-3f}{2} C_{1} \left(
      (dt)_{a}(dt)_{b}
      + f^{-2} (dr)_{a}(dr)_{b}
      \right)
      + {\pounds}_{V_{(C_{1})}}g_{ab}
      ,
     \label{eq:l=0-mode-perturbation-sol-C-meric}
\end{eqnarray}
where
\begin{eqnarray}
  \label{eq:accel_pert_of_SCH-07000-01200-generators-2}
  V_{(C_{1})a} = \frac{1}{2} f t C_{1} (dt)_{a} + \frac{r}{2f} C_{1} (dr)_{a}.
\end{eqnarray}


Now, we compare the metric perturbation
(\ref{eq:l=0-mode-perturbation-sol-C-meric}) with the derived $l=0$
solution (\ref{eq:calFab+poundsVg-l=0-non-vac-final}) with the
generator (\ref{eq:Va-result-non-vac-final}).
Since we ignore the mass parameter perturbation $M_{1}$, we obtain the
relations
\begin{eqnarray}
  &&
     4 \pi \int dr \left[\frac{r^{2}}{f} \tilde{T}_{tt}\right]
     =
     \frac{1-3f}{4} r C_{1}
     ,
     \label{eq:l=0-general-sol-C-metric-Compare-1}
  \\
  &&
     4 \pi r \int dt \left(
     \frac{1}{f} \tilde{T}_{tt}
     + f \tilde{T}_{rr}
     \right)
     =
     0
     .
     \label{eq:l=0-general-sol-C-metric-Compare-2}
\end{eqnarray}
If the condition (\ref{eq:l=0-general-sol-C-metric-Compare-2}) is
satisfied for an arbitrary time $t$, we obtain
\begin{eqnarray}
  \label{eq:l=0-general-sol-C-metric-Compare-2-sol}
  \tilde{T}_{tt} + f^{2} \tilde{T}_{rr} = 0.
\end{eqnarray}
Since we ignore the integration constant $M_{1}$, the condition
(\ref{eq:l=0-general-sol-C-metric-Compare-1}) gives
\begin{eqnarray}
  \frac{r^{2}}{f} \tilde{T}_{tt}
  =
  \partial_{r}\left(\frac{1-3f}{16\pi} r\right) C_{1}
  =
  - \frac{1}{8\pi} C_{1}
  .
  \label{eq:l=0-general-sol-C-metric-Compare-1-sol}
\end{eqnarray}
Furthermore, we may choose $\tilde{T}_{tr}=0$ without contradiction to
the linear perturbation of the continuity equations
(\ref{eq:div-barTab-linear-AB-t}) and
(\ref{eq:div-barTab-linear-AB-r}) with $l=0$.
From the definition (\ref{eq:1st-pert-calTab-dd-decomp}) of the
components $\tilde{T}_{tt}$ and $\tilde{T}_{rr}$, for $l=0$ mode, we
obtain
\begin{eqnarray}
  {}^{(1)}\!\ScrT_{ac}
  &=&
      -
      \frac{1}{r^{2}}
      y_{ab}
      \lambda_{l=0}
     \label{eq:1st-pert-calTab-dd-decomp-2-C-metric-2-l=0}
\end{eqnarray}
from Eq.~(\ref{eq:l=0-general-sol-C-metric-Compare-2-sol}) and
(\ref{eq:l=0-general-sol-C-metric-Compare-1-sol}).
Here, we defined
\begin{eqnarray}
  \label{eq:constant-tension-from-Einstein-l=0}
  \lambda_{l=0} := \frac{r^{2}}{f} \tilde{T}_{tt} = - \frac{C_{1}}{8\pi}.
\end{eqnarray}


Next, we compare the generator $V_{(C_{1})a}$ defined by
Eq.~(\ref{eq:accel_pert_of_SCH-07000-01200-generators-2})
and the generator $V_{a}$ given by
Eq.~(\ref{eq:Va-result-non-vac-final}) in the $l=0$ mode solution
(\ref{eq:calFab+poundsVg-l=0-non-vac-final}).
Comparing the $r$-component of
Eq.~(\ref{eq:accel_pert_of_SCH-07000-01200-generators-2}) with
Eq.~(\ref{eq:Va-result-non-vac-final}), we choose
\begin{eqnarray}
  \label{eq:Upsilon-choice-C-metric}
  \frac{1}{4f} r \partial_{t}\Upsilon = \frac{r}{2f} C_{1},
\end{eqnarray}
and obtain
\begin{eqnarray}
  \label{eq:Upsilon-choice-C-metric-2}
  \Upsilon = 2 C_{1} t.
\end{eqnarray}
Here, we ignore the integration constant in the integration of
Eq.~(\ref{eq:Upsilon-choice-C-metric}).
Substituting this result (\ref{eq:Upsilon-choice-C-metric-2}) into
Eq.~(\ref{eq:Va-result-non-vac-final})
\begin{eqnarray}
  \label{eq:Va-result-non-vac-final-C-metric}
  V_{a}
  =
  \left(
  \frac{f}{4} 2 C_{1} t
  -  \frac{r \Xi(r)}{(1-3f)}
  + f \int dr \frac{2 \Xi(r)}{f(1-3f)^{2}}
  \right) (dt)_{a}
  +
  \frac{r}{2f} C_{1} (dr)_{a}
  .
\end{eqnarray}
Choosing $\Xi(r)=0$, the generator
(\ref{eq:Va-result-non-vac-final-C-metric}) coincides with the
generator $V_{(C_{1})a}$ given by
Eq.~(\ref{eq:accel_pert_of_SCH-07000-01200-generators-2}).
Thus, the $l=0$ mode solution
(\ref{eq:calFab+poundsVg-l=0-non-vac-final}) realizes the $l=0$ mode
part (\ref{eq:l=0-mode-perturbation-sol-C-meric}) of the C-metric.


In summary, we have obtained the $l=0$ metric perturbation
\begin{eqnarray}
  {}_{\ScrZ}\!h_{ab}(l=0)
  &=&
      {}_{\ScrX}\!h_{ab}(l=0)
      +
      {\pounds}_{V_{M_{1}}}g_{ab}
      \nonumber\\
  &=&
      \left(
      \frac{2M_{1}}{r}
      +
      \frac{1-3f}{2} C_{1}
      \right)
      \left(
      (dt)_{a}(dt)_{b}
      + f^{-2} (dr)_{a}(dr)_{b}
      \right)
      \nonumber\\
  &&
      +
      {\pounds}_{V_{(M_{1})}+V_{(C_{1})}}g_{ab}
      ,
      \label{eq:C-metric-l=0-mode-perturbation-sum}
\end{eqnarray}
where
\begin{eqnarray}
  \label{eq:Va-generator-even-l=0-components-mass-perturbation-C-metric-sum}
  V_{(M_{1})a}
  &=&
      \frac{1}{4} f \left(
      \Upsilon_{M_{1}}
      + r \partial_{r}\Upsilon_{M_{1}}
      \right)
      (dt)_{a}
      +
      \frac{1}{4f} r \partial_{t}\Upsilon
      (dr)_{a}
      ,
  \\
  \label{eq:acceleration_perturbation_of_SCH-07000-01200-generators-2-sum}
  V_{(C_{1})a} &=& \frac{1}{2} f t C_{1} (dt)_{a} + \frac{r}{2f} C_{1} (dr)_{a}.
\end{eqnarray}
Here, the function $\Upsilon_{M_{1}}(t,r)$ is the solution to the
equation (\ref{eq:even-mode-tildeF-eq-Phie-sum}) with
$\tilde{F}=:\partial_{t}\Upsilon$, i.e.,
\begin{eqnarray}
  -  \frac{1}{f} \partial_{t}^{2}\Upsilon_{M_{1}}
  + \partial_{r}( f \partial_{r}\Upsilon_{M_{1}} )
  + \frac{1}{r^{2}} 3(1-f) \Upsilon_{M_{1}}
  -  \frac{4}{r^{3}} 2 M_{1} t
  =
  0
  .
  \label{eq:Upsilon-eq-int-dt-2-for-mass-perturbation-C-metric-sum}
\end{eqnarray}
The $l=0$ mode energy-momentum tensor for C-metric is given by
\begin{eqnarray}
  {}^{(1)}\!\ScrT_{ac}
  &=&
      -
      \frac{1}{r^{2}}
      y_{ab}
      \lambda_{l=0}
     \label{eq:1st-pert-calTab-dd-decomp-2-C-metric-2-l=0-sum}
\end{eqnarray}
with
\begin{eqnarray}
  \label{eq:constant-tension-from-Einstein-C-metric-l=0-sum}
  \lambda_{l=0}
  =
  - \frac{C_{1}}{8\pi}
  .
\end{eqnarray}
Here, we note that the result
(\ref{eq:constant-tension-from-Einstein-C-metric-l=0-sum}) is also
realized by the substitution $l=0$ into
Eq.~(\ref{eq:constant-tension-from-Einstein}), although
Eq.~(\ref{eq:constant-tension-from-Einstein}) is derived only in the
case of $l\geq 2$ modes.
This indicates that the formula
(\ref{eq:constant-tension-from-Einstein}) is also valid even for $l=0$
mode perturbations.


\subsection{Realization of $l=1$ mode perturbations}
\label{sec:l=1-mode-for-the-C-metric}


The $l=1$ mode of the C-metric is summarized as
Eqs.~(\ref{eq:calXtildehAB-Cmetric-l=1})--(\ref{eq:gauge-inv-kernel-Delta+2-tildeFAB-def-sum-C-metric}).
We consider the continuity equation of the energy-momentum tensor
(\ref{eq:div-barTab-linear-AB-t})--(\ref{eq:div-barTab-linear-p-even-mode})
with $l=1$.
As in the $l\geq 2$ and $l=0$ cases, we choose
$\tilde{T}_{(e1)A}=0=\tilde{T}_{(e2)}$.
These and Eq.~(\ref{eq:div-barTab-linear-p-even-mode}) with $l=1$
yield
\begin{eqnarray}
  \label{eq:Te0-condition-C-metric}
  \tilde{T}_{(e0)} = 0.
\end{eqnarray}
Furthermore, we also assume that
\begin{eqnarray}
  \label{eq:tildeTrt-C-metric-assumption}
  \tilde{T}_{rt} = 0,
\end{eqnarray}
inspecting the $l\geq 2$ and $l=0$ cases.
Then,
Eqs.~(\ref{eq:div-barTab-linear-AB-t}) and
(\ref{eq:div-barTab-linear-AB-r}) with $l=1$ are given by
\begin{eqnarray}
  &&
     \partial_{t}\tilde{T}_{tt} = 0 ,
     \label{eq:div-barTab-linear-AB-t-comp-reduce-l=1-C-metric}
  \\
  &&
     \partial_{r}(\tilde{T}_{tt}+f^{2}\tilde{T}_{rr})
     -  \frac{f}{r^{2}} \partial_{r}(\frac{r^{2}}{f}\tilde{T}_{tt})
     + \frac{5f-1}{2rf} \left(\tilde{T}_{tt} + f^{2} \tilde{T}_{rr}\right)
     =
     0
     .
     \label{eq:div-barTab-linear-AB-r-comp-reduce-l=1-C-metric}
\end{eqnarray}
Equation (\ref{eq:div-barTab-linear-AB-t-comp-reduce-l=1-C-metric})
indicates that we have the static energy density $T_{tt}$.
As in the case of $l\geq 2$ and $l=0$ modes, we define $\lambda_{l=1}$
by
\begin{eqnarray}
  \label{eq:f2overfTtt=constant-l=1}
  \lambda_{l=1} := \frac{r^{2}}{f} \tilde{T}_{tt}
\end{eqnarray}
and we assume that $\lambda_{l=1}$ is constant and
\begin{eqnarray}
  \label{eq:tildeTtt+f2tildeTrr=0-l=1}
  \tilde{T}_{tt} + f^{2} \tilde{T}_{rr} = 0.
\end{eqnarray}
Due to these assumptions,
Eq.~(\ref{eq:div-barTab-linear-AB-r-comp-reduce-l=1-C-metric}) is
trivial.
Through the above components of the energy-momentum tensor,
Eq.~(\ref{eq:tildeFab-l=1-m=0-nonvacsum-2-cov}) is given by
\begin{eqnarray}
  \ScrF_{ab}
  &=&
      {\pounds}_{V_{(vac)}}g_{ab}
      \nonumber\\
  &&
      - \frac{16 \pi \lambda_{l=1} f}{3(1-f)}\left[
      \frac{1+f}{2} (dt)_{a}(dt)_{b}
      -  \frac{1-3f}{2f^{2}} (dr)_{a}(dr)_{b}
      + r^{2}\gamma_{ab}
     \right] \cos\theta
     .
     \label{eq:tildeFab-l=1-m=0-nonvacsum-2-cov-C-metric}
\end{eqnarray}


Here, we note that the $l=1$ mode solution
(\ref{eq:2+2-gauge-inv-kernel-Delta+2-def-tildeFA-sum-C-metric})--(\ref{eq:gauge-inv-kernel-Delta+2-tildeFAB-def-sum-C-metric})
is summarized as
\begin{eqnarray}
  \ScrF_{ab}
  &=&
      \tilde{F}_{AB} \cos\theta (dx^{A})_{a} (dx^{B})_{b}
      + \frac{1}{2} \gamma_{pq} \tilde{F} \cos\theta (dx^{p})_{a} (dx^{q})_{b}
      \nonumber\\
  &=&
      - 2 \alpha_{1} r  \cos\theta g_{ab}
      .
      \label{eq:calFab-linearized-C-meric-l=1}
\end{eqnarray}
Comparing (\ref{eq:tildeFab-l=1-m=0-nonvacsum-2-cov-C-metric}) and
(\ref{eq:calFab-linearized-C-meric-l=1}), we rewrite
(\ref{eq:tildeFab-l=1-m=0-nonvacsum-2-cov-C-metric}) as
\begin{eqnarray}
  \ScrF_{ab}
  &=&
      {\pounds}_{V_{(vac)}}g_{ab}
      \nonumber\\
  &&
     - \frac{16 \pi \lambda_{l=1} f}{3(1-f)}\left[
     - \frac{1-f}{2} (dt)_{a}(dt)_{b}
     + \frac{1+3f}{2f^{2}} (dr)_{a}(dr)_{b}
     + \frac{1+f}{f} r^{2} \gamma_{ab}
     \right] \cos\theta
      \nonumber\\
  &&
     + \frac{16 \pi \lambda_{l=1}}{3(1-f)} \cos\theta g_{ab}
     .
     \label{eq:tildeFab-l=1-m=0-nonvacsum-2-cov-C-metric-2}
\end{eqnarray}


Here, we explain the choice of the coefficients of the last term in
Eq.~(\ref{eq:tildeFab-l=1-m=0-nonvacsum-2-cov-C-metric-2}).
If the expression (\ref{eq:constant-tension-from-Einstein}) of the
$\lambda_{l}$ for $l\geq 2$ is also valid even for $l=1$,
Eq.~(\ref{eq:constant-tension-from-Einstein}) is given by
\begin{eqnarray}
  \label{eq:constant-tension-from-Einstein-l=1}
  16 \pi \lambda_{l=1} =  - 12 \alpha_{1}M
\end{eqnarray}
and the last term in
Eq.~(\ref{eq:tildeFab-l=1-m=0-nonvacsum-2-cov-C-metric-2}) is
given by
\begin{eqnarray}
  \label{eq:tildeFab-l=1-m=0-nonvacsum-2-cov-C-metric-last}
  \frac{16 \pi \lambda_{l=1}}{3(1-f)} \cos\theta g_{ab}
  =
  - 2 \alpha_{1} r  \cos\theta g_{ab}
  .
\end{eqnarray}
This is the $l=1$ mode solution described by
Eqs.~(\ref{eq:calXtildehAB-Cmetric-l=1})--(\ref{eq:gauge-inv-kernel-Delta+2-tildeFAB-def-sum-C-metric}).
As the remaining problem, we have to consider the problem whether the
middle term in
Eq.~(\ref{eq:tildeFab-l=1-m=0-nonvacsum-2-cov-C-metric-2}) has the
form ${\pounds}_{W}g_{ab}$, or not.
If  the middle term in
Eq.~(\ref{eq:tildeFab-l=1-m=0-nonvacsum-2-cov-C-metric-2}) does have the
form ${\pounds}_{W}g_{ab}$, we may say that our $l=1$ mode solution
(\ref{eq:tildeFab-l=1-m=0-nonvacsum-2-cov}) does describe the
linearized C-metric apart from the term of the Lie derivative of the
background metric $g_{ab}$.


Now, we concentrate on the problem whether the middle term in
Eq.~(\ref{eq:tildeFab-l=1-m=0-nonvacsum-2-cov-C-metric-2}) has the form
${\pounds}_{W}g_{ab}$, or not.
To show this, we consider the components of ${\pounds}_{W}g_{ab}$ for
an appropriate vector field $W_{a}$.
We consider the generator $W_{a}$ which satisfies $W_{\phi}=0$,
$\partial_{\phi}W_{\theta}=\partial_{\phi}W_{r}=\partial_{\phi}W_{t}=0$.
Furthermore, we assume that $W_{t}=:w_{t}\cos\theta$,
$W_{r}=:w_{r}\cos\theta$, and $W_{\theta}=:w_{\theta}\sin\theta$ using
${\pounds}_{W}g_{t\theta}=0$.
Then, the non-trivial components of ${\pounds}_{W}g_{ab}$ are
summarized as follows:
\begin{eqnarray}
  \label{eq:poundsWgab-components-tt-Cmetric-00300}
  &&
     {\pounds}_{W}g_{tt}
     =
     \left(
     2 \partial_{t}w_{t} - f f' w_{r}
     \right)
     \cos\theta
     ,
  \\
  \label{eq:poundsWgab-components-tr-Cmetric-00300}
  &&
     {\pounds}_{W}g_{tr}
     =
     \left(
     \partial_{t}w_{r} + \partial_{r}w_{t} - \frac{f'}{f} w_{t}
     \right)
     \cos\theta
     ,
  \\
  \label{eq:poundsWgab-components-ttheta-Cmetric-00300}
  &&
     {\pounds}_{W}g_{t\theta}
     =
     \left(
     \partial_{t}w_{\theta} - w_{t}
     \right)
     \sin\theta
     ,
  \\
  \label{eq:poundsWgab-components-rr-Cmetric-00300}
  &&
     {\pounds}_{W}g_{rr}
     =
     \left(
     2 \partial_{r}w_{r} + \frac{f'}{f} w_{r}
     \right)
     \cos\theta
     ,
  \\
  \label{eq:poundsWgab-components-rtheta-Cmetric-00300}
  &&
     {\pounds}_{W}g_{r\theta}
     =
     \left(
     \partial_{r}w_{\theta}
     - w_{r}
     - \frac{2}{r} w_{\theta}
     \right)
     \sin\theta
     ,
  \\
  \label{eq:poundsWgab-components-thetatheta-Cmetric-00300}
  &&
     {\pounds}_{W}g_{\theta\theta}
     =
     2
     \left(
     w_{\theta} + rf w_{r}
     \right)
     \cos\theta
     ,
  \\
  \label{eq:poundsWgab-components-phiphi-Cmetric-00300}
  &&
     {\pounds}_{W}g_{\phi\phi}
     =
     2
     \left(
     w_{\theta}
     + rf w_{r}
     \right)
     \sin^{2}\theta \cos\theta
     .
\end{eqnarray}
The middle term in
Eq.~(\ref{eq:tildeFab-l=1-m=0-nonvacsum-2-cov-C-metric-2}) has only its
diagonal components, we may choose
${\pounds}_{W}g_{t\theta}={\pounds}_{W}g_{r\theta}=0$.
From these equations,
Eqs.~(\ref{eq:poundsWgab-components-ttheta-Cmetric-00300}) and
(\ref{eq:poundsWgab-components-rtheta-Cmetric-00300}) yield
\begin{eqnarray}
  \label{eq:C-metric-PoundsWgab-wt-wr-results}
  w_{t} = \partial_{t}w_{\theta}, \quad
  w_{r} = \partial_{r}w_{\theta} - \frac{2}{r} w_{\theta}.
\end{eqnarray}


Furthermore, from
Eqs.~(\ref{eq:poundsWgab-components-thetatheta-Cmetric-00300}), (\ref{eq:poundsWgab-components-phiphi-Cmetric-00300}), the second
equation in Eq.~(\ref{eq:C-metric-PoundsWgab-wt-wr-results}), and the
term proportional to $\gamma_{ab}$ in the second line of
Eq.~(\ref{eq:tildeFab-l=1-m=0-nonvacsum-2-cov-C-metric-2}), we have
\begin{eqnarray}
  \label{eq:calFthetatheta-middle-is-poundWgthetatheta}
  rf \partial_{r}w_{\theta} + (1- 2f) w_{\theta}
  =
  - \frac{r^{2}(1+f)}{6(1-f)} 16 \pi \lambda_{l=1}
  .
\end{eqnarray}
A solution to
Eq.~(\ref{eq:calFthetatheta-middle-is-poundWgthetatheta}) is given by
\begin{eqnarray}
  \label{eq:calFthetatheta-middle-is-poundWgthetatheta-sol}
  w_{\theta}
  =
  - \frac{r^{2}}{6(1-f)} 16 \pi \lambda_{l=1}
  .
\end{eqnarray}
Then, from Eqs.~(\ref{eq:C-metric-PoundsWgab-wt-wr-results}), we
obtain
\begin{eqnarray}
  \label{eq:C-metric-PoundsWgab-wt-results-sol}
  w_{t} = 0, \quad
  w_{r} = - \frac{r}{6(1-f)} 16 \pi \lambda_{l=1}.
\end{eqnarray}
Substituting
Eqs.~(\ref{eq:calFthetatheta-middle-is-poundWgthetatheta-sol}) and
(\ref{eq:C-metric-PoundsWgab-wt-results-sol}) into
Eqs.~(\ref{eq:poundsWgab-components-tt-Cmetric-00300})--(\ref{eq:poundsWgab-components-phiphi-Cmetric-00300}),
the non-vanishing components of ${\pounds}_{W}g_{ab}$ is given by
\begin{eqnarray}
  \label{eq:poundsWgab-C-metricsol}
     {\pounds}_{W}g_{ab}
     =
     \frac{16 \pi \lambda_{l=1} f}{3(1-f)}
     \left[
     \frac{1-f}{2} (dt)_{a} (dt)_{b}
     - \frac{1+3f}{2f^{2}} (dr)_{a} (dr)_{b}
     - \frac{1+f}{f} r^{2} \gamma_{ab}
     \right] \cos\theta
     .
\end{eqnarray}
Through Eq.~(\ref{eq:poundsWgab-C-metricsol}),
Eq.~(\ref{eq:tildeFab-l=1-m=0-nonvacsum-2-cov-C-metric-2}) is given by
\begin{eqnarray}
  \ScrF_{ab}
  &=&
      {\pounds}_{V_{(l=1)}}g_{ab}
      + \frac{16 \pi \lambda_{l=1}}{3(1-f)} \cos\theta g_{ab}
      ,
      \label{eq:tildeFab-l=1-m=0-nonvacsum-2-cov-C-metric-final}
\end{eqnarray}
where $V_{(l=1)a}:=V_{(vac)a}+W_{a}$.
Comparing with
Eqs.~(\ref{eq:2+2-gauge-inv-kernel-Delta+2-tildeF-def-sum-C-metric})
and (\ref{eq:gauge-inv-kernel-Delta+2-tildeFAB-def-sum-C-metric}), we
obtain Eqs.~(\ref{eq:constant-tension-from-Einstein-l=1}) and
(\ref{eq:tildeFab-l=1-m=0-nonvacsum-2-cov-C-metric-last}) as
expected.
This also indicates that the coefficient
(\ref{eq:constant-tension-from-Einstein}) for $l\geq 2$ is also valid
not only for $l=0$ mode but also $l=1$ mode, i.e., the coefficient
(\ref{eq:constant-tension-from-Einstein}) is valid for any $l\geq 0$.
Furthermore, we have seen above the equation of state
(\ref{eq:tildeTtt+f2tildeTrr=0-l=1}) with
(\ref{eq:f2overfTtt=constant-l=1}) yields
\begin{eqnarray}
  {}^{(1)}\!\ScrT_{ac}(l=1)
  =
  -
  \frac{1}{r^{2}}
  y_{ab}
  \lambda_{l=1}
  P_{1}(\cos\theta)
  ,
  \label{eq:1st-pert-calTab-dd-decomp-l=1}
\end{eqnarray}
where $\lambda_{l=1}$ is constant which is given by
Eq.~(\ref{eq:constant-tension-from-Einstein-l=1}).
At the same time, we may say that our $l=1$ mode solution
(\ref{eq:tildeFab-l=1-m=0-nonvacsum-2-cov}) does describe the
linearized C-metric apart from the term of the Lie derivative of the
background metric $g_{ab}$.


\subsection{Source term of the linearized C-metric}
\label{sec:Source_term_of_the_linearized_C-metric}


Here, we summarize the energy-momentum tensor $\ScrT_{ab}$ for the
linearized C-metric as follows
\begin{eqnarray}
  {}^{(1)}\!\ScrT_{ac}
  &=&
      -
      \frac{1}{r^{2}}
      y_{ab}
      \sum_{l=0}^{\infty}
      \lambda_{l}
      P_{l}(\cos\theta)
      ,
     \label{eq:1st-pert-calTab-dd-decomp-all_value_of_l}
\end{eqnarray}
where the constant $\lambda_{l}$ is given by
\begin{eqnarray}
  \label{eq:constant-tension-from-Einstein-C-metric-sum-lgeq0-2}
  16 \pi \lambda_{l}
  =
  - (2l+1) \left[
  (+ 2 \alpha_{1} M + C_{1})
  +
  ( - 2 \alpha_{1} M + C_{1} )
  (-1)^{l}
  \right]
\end{eqnarray}
for $l\geq 0$.
Substituting
Eq.~(\ref{eq:constant-tension-from-Einstein-C-metric-sum-lgeq0-2})
into Eq.~(\ref{eq:1st-pert-calTab-dd-decomp-all_value_of_l}), we obtain
\begin{eqnarray}
  {}^{(1)}\!\ScrT_{ac}
  &=&
      \frac{1}{4r^{2}}
      y_{ab}
      \left[
      \left( + 2 \alpha_{1} M + C_{1} \right)
      \left(
      \frac{1}{4\pi}
      \sum_{l=0}^{\infty}
      (2l+1)
      P_{l}(\cos\theta)
      \right)
      \right.
      \nonumber\\
  && \quad\quad\quad\quad
      \left.
      +
      \left( - 2 \alpha_{1} M + C_{1} \right)
      \left(
      \frac{1}{4\pi}
      \sum_{l=0}^{\infty}
      (2l+1)
      (-1)^{l}
      P_{l}(\cos\theta)
      \right)
      \right]
      .
     \label{eq:1st-pert-calTab-dd-sum-of-all_value_of_l}
\end{eqnarray}


Inspecting the work by Kodama~\cite{H.Kodama-2008-Accelerating-BH}, we
consider the mode decomposition of the $\delta$-function on
$S^{2}$~\cite{R.Szmytkowski-2006a,R.Szmytkowski-2007}
\begin{eqnarray}
  \label{eq:completeness-of-Ylm}
  \delta^{(2)}({\bf n}-{\bf n}')
  =
  \sum_{l=0}^{\infty} \sum_{m=-l}^{m=l}
  Y_{lm}({\bf n}) Y_{lm}^{*}({\bf n}')
  ,
\end{eqnarray}
where ${\bf n}$ and ${\bf n}'$ are the position vectors which point to
the points on $S^{2}$ in embedded in $\RF^{3}$, respectively.
The summation over $m$ is given
by~\cite{R.Szmytkowski-2006a,R.Szmytkowski-2007}
\begin{eqnarray}
  \label{eq:sum-over-m-Ylm-Ylm}
  \sum_{m=-l}^{l} Y_{lm}({\bf n}) Y_{lm}^{*}({\bf n}')
  =
  \frac{2l+1}{4\pi} C_{l}^{1/2}({\bf n}\cdot{\bf n}')
  ,
\end{eqnarray}
where $C_{l}^{1/2}(x)$ is the Gegenbauer polynomial.
In $\RF^{3}$, any point of the unit sphere $S^{2}$ is specified by the
orthogonal coordinates $(x,y,z)$ with the center of $S^{2}$ in
$\RF^{3}$
\begin{eqnarray}
  x = \sin\theta \cos\phi, \quad
  y = \sin\theta \sin\phi, \quad
  z = \cos\theta.
\end{eqnarray}
The north pole is specified as $(x,y,z)=(0,0,1)$ and the south pole is
specified as $(x,y,z)=(0,0,-1)$.
The inner product ${\bf n}\cdot{\bf n}'$ in the case where ${\bf n}'$
is the north pole or the south pole is given by
\begin{eqnarray}
  \label{eq:nn-inner-north-south}
  {\bf n}\cdot{\bf n}_{\rm{north}} = \cos\theta, \quad
  {\bf n}\cdot{\bf n}_{\rm{south}} = - \cos\theta,
\end{eqnarray}
respectively, and we have
\begin{eqnarray}
  \label{eq:sum-over-m-north}
  \sum_{m=-l}^{l} Y_{lm}({\bf n}) Y_{lm}^{*}({\bf n}_{\rm{north}})
  &=&
      \frac{2l+1}{4\pi} C_{l}^{1/2}(\cos\theta)
      =
      \frac{2l+1}{4\pi} P_{l}(\cos\theta)
      ,
  \\
  \label{eq:sum-over-m-south}
  \sum_{m=-l}^{l} Y_{lm}({\bf n}) Y_{lm}^{*}({\bf n}_{\rm{south}})
  &=&
      \frac{2l+1}{4\pi} C_{l}^{1/2}(-\cos\theta)
      =
      \frac{2l+1}{4\pi} P_{l}(-\cos\theta)
      \nonumber\\
  &=&
      \frac{2l+1}{4\pi} (-1)^{l} P_{l}(\cos\theta)
      .
\end{eqnarray}
Then, from Eq.~(\ref{eq:completeness-of-Ylm}), we obtain
\begin{eqnarray}
  \label{eq:delta-func-north-pole}
  \delta^{(2)}({\bf n}-{\bf n}_{\rm{north}})
  \!\!\!\!&=&\!\!\!\!
      \sum_{l=0}^{\infty} \sum_{m=-l}^{m=l}
      Y_{lm}({\bf n}) Y_{lm}^{*}({\bf n}_{\rm{north}})
      =
      \frac{1}{4\pi} \sum_{l=0}^{\infty} (2l+1) P_{l}(\cos\theta)
      ,
  \\
  \label{eq:delta-func-south-pole}
  \delta^{(2)}({\bf n}-{\bf n}_{\rm{south}})
  \!\!\!\!&=&\!\!\!\!
      \sum_{l=0}^{\infty} \sum_{m=-l}^{m=l}
      Y_{lm}({\bf n}) Y_{lm}^{*}({\bf n}_{\rm{south}})
      =
      \frac{1}{4\pi} \sum_{l=0}^{\infty} (2l+1) (-1)^{l} P_{l}(\cos\theta)
      .
\end{eqnarray}


Through these expressions of the $\delta$-functions, the first-order
perturbation of the energy-momentum tensor
(\ref{eq:1st-pert-calTab-dd-sum-of-all_value_of_l}) yields
\begin{eqnarray}
  {}^{(1)}\!\ScrT_{ac}
  &=&
      -
      \frac{1}{4r^{2}}
      y_{ab}
      \left[
      -
      \left( 2 \alpha_{1} M + C_{1} \right)
      \delta^{(2)}({\bf n}-{\bf n}_{\rm{north}})
      \right.
      \nonumber\\
  && \quad\quad\quad\quad\quad\quad
      \left.
      +
      \left( 2 \alpha_{1} M - C_{1} \right)
      \delta^{(2)}({\bf n}-{\bf n}_{\rm{south}})
      \right]
      \nonumber\\
  &=&
      -
      \frac{1}{r^{2}}
      y_{ab}
      \left[
      \mu_{n}
      \delta^{(2)}({\bf n}-{\bf n}_{\rm{north}})
      +
      \mu_{s}
      \delta^{(2)}({\bf n}-{\bf n}_{\rm{south}})
      \right]
      ,
     \label{eq:1st-pert-calTab-delta-func-rep}
\end{eqnarray}
where
\begin{eqnarray}
  \label{eq:string-tension-north-south}
  \mu_{n}
  =
  -
  \frac{1}{4}
  \left( 2 \alpha_{1} M + C_{1} \right)
  ,
  \quad
  \mu_{s}
  =
  \frac{1}{4}
  \left( 2 \alpha_{1} M - C_{1} \right)
  .
\end{eqnarray}
The first-order perturbation of the energy-momentum tensor
(\ref{eq:1st-pert-calTab-delta-func-rep}) coincides with the
energy-momentum tensor for a half-infinite string with constant line
densities $\mu_{n}$ on the north half of the symmetry axis and
$\mu_{s}$ on the south half of the symmetry axis.
If we impose the regularity at the north pole, i.e., $\mu_{n}=0$,
we have $\mu_{s}=\alpha_{1}M>0$ which corresponds to the positive
energy density of string.
On the other hand, if we impose the regularity at the south pole,
i.e., $\mu_{s}=0$, we have $\mu_{n}=- \alpha_{1} M <0$ which
corresponds to the negative energy density of string.
These results are consistent with the stringy interpretation of the
singularity of the C-metric~\cite{H.Kodama-2008-Accelerating-BH}.
Finally, we have to emphasize that the treatment of $l=1$ mode
perturbations in our derivation of the energy density of the C-metric
is essentially different from those in
Ref.~\cite{H.Kodama-2008-Accelerating-BH}.


\section{Summary and Discussions}
\label{sec:Summary_and_Discussions}


In summary, after reviewing our general framework of the
gauge-invariant perturbation theory and its application to the
perturbation theory on the Schwarzschild background spacetime
developed in
Refs.~\cite{K.Nakamura-2021a,K.Nakamura-2021c,K.Nakamura-2021d}, we
checked the fact that our linearized solutions derived in
Refs.~\cite{K.Nakamura-2021a,K.Nakamura-2021c,K.Nakamura-2021d}
realizes the linearized LTB solution and the linearized C-metric
around the Schwarzschild background spacetime.
These facts yield that our derived $l=0,1$ solutions to the linearized
Einstein equation following
Proposal~\ref{proposal:treatment-proposal-on-pert-on-spherical-BG}
are physically reasonable.
Then, we may say that
Proposal~\ref{proposal:treatment-proposal-on-pert-on-spherical-BG}
itself is also physically reasonable.
Our general framework of the gauge-invariant perturbation theory
developed in
Refs.~\cite{K.Nakamura-2003,K.Nakamura-2005,K.Nakamura-2011,K.Nakamura-IJMPD-2012,K.Nakamura-2013,K.Nakamura-2014}
was applied to the cosmological perturbation theory in
Refs.~\cite{K.Nakamura-2006,K.Nakamura-2007,K.Nakamura-2009a,K.Nakamura-2009b,K.Nakamura-LTVII-2008,A.J.Christopherson-K.A.Malik-D.R.Matravers-K.Nakamura-2011,K.Nakamura-2020}.
On the other hand, in this series of our
papers~\cite{K.Nakamura-2021a,K.Nakamura-2021b,K.Nakamura-2021c,K.Nakamura-2021d},
we apply our general framework of the gauge-invariant perturbation
theory to the perturbations on the Schwarzschild background spacetime.
Thus, we may say that the applicability of our general framework of the
gauge-invariant perturbation theory developed in
Refs.~\cite{K.Nakamura-2003,K.Nakamura-2005,K.Nakamura-2011,K.Nakamura-IJMPD-2012,K.Nakamura-2013,K.Nakamura-2014}
is very wide.


Our general-framework is based on the single non-trivial
Conjecture~\ref{conjecture:decomposition-conjecture}.
This conjecture is almost proved in Ref.~\cite{K.Nakamura-2013} except
for the ``zero-mode problem.''
In the proof in Ref.~\cite{K.Nakamura-2013}, we introduced the Green
functions for some elliptic derivative operators.
This means that the kernel modes of these elliptic derivative
operators were out of our considerations.
We call these kernel modes as ``zero modes'' and the problem to find
gauge-invariant treatments for these kernel modes as ``zero-mode
problem.''
To carry out the application to the perturbations on the Schwarzschild
background spacetime, we have to propose a gauge-invariant treatment
of $l=0,1$ mode perturbation on the Schwarzschild background
spacetime, because these modes correspond to the above ``zero modes''
and the gauge-invariant treatments was unclear until our proposal in
Refs.~\cite{K.Nakamura-2021a,K.Nakamura-2021c,K.Nakamura-2021d}.
We should also note that such ``zero-mode problem'' exists even in the
cosmological perturbation theory which are developed in
Refs.~\cite{K.Nakamura-2006,K.Nakamura-2007,K.Nakamura-2009a,K.Nakamura-2009b,K.Nakamura-LTVII-2008,A.J.Christopherson-K.A.Malik-D.R.Matravers-K.Nakamura-2011,K.Nakamura-2020}.


In conventional perturbation theory on spherically symmetric
background spacetimes, we use the spherical harmonics
$S=Y_{lm}$ as the scalar-harmonics and construct vector and tensor
harmonics from the derivative of this scalar harmonics.
However, in this construction of tensor harmonics, the set
(\ref{eq:harmonic-fucntions-set}) of the scalar-, vector- and
tensor-harmonics loses its linear-independence as the basis of the
tangent space on $S^{2}$ in $l=0,1$ mode.
To recover this linear-independence of the set
(\ref{eq:harmonic-fucntions-set}), we introduced the singular
harmonics for $l=0,1$ modes at once and proposed the strategy to
construct gauge-invariant variables and derive the Einstein equation
as
Proposal~\ref{proposal:treatment-proposal-on-pert-on-spherical-BG}.
The conventional expansion using the spherical harmonic functions
$Y_{lm}$ is the restriction of the function space to the $L^{2}$
space on $S^{2}$.
This restriction corresponds to the imposition of the regular boundary
condition for the functions on $S^{2}$ at the starting point.
On the other hand, our introduction of the singular harmonic functions
at once and
Proposal~\ref{proposal:treatment-proposal-on-pert-on-spherical-BG}
state that the boundary condition on $S^{2}$ should be imposed when we
solve the linearized Einstein equations.
Owing to
Proposal~\ref{proposal:treatment-proposal-on-pert-on-spherical-BG}, we
could prove Conjecture~\ref{conjecture:decomposition-conjecture} for
perturbations on the spherically symmetric background spacetime.
Then, we reached to the statement
Theorem~\ref{theorem:decomposition-theorem-with-spherical-symmetry}.


Actually, following
Proposal~\ref{proposal:treatment-proposal-on-pert-on-spherical-BG}, we
could construct gauge-invariant variables not only for $l\geq 2$ modes
but also for $l=0,1$ modes.
Furthermore, in Ref.~\cite{K.Nakamura-2021c}, we derive the solution
of $l=1$ odd-mode perturbations and, in Ref.~\cite{K.Nakamura-2021d},
we derive the solutions of $l=0,1$ even-mode perturbations.


In this paper, we also reviewed the strategy to solve even-mode
perturbation on the Schwarzschild background spacetime including
$l=0,1$ modes which was discussed in the Part II
paper~\cite{K.Nakamura-2021d}, and then, we showed that it is possible
to confirm the realizations of the LTB solutions and non-rotating
C-metric through the even-mode solutions derived in the Part II
paper~\cite{K.Nakamura-2021d}.
Because the LTB solution is a spherically symmetric solution, its
linearized version should be realized $l=0$ even-mode perturbations.
On the other hand, non-rotating C-metric includes all $l\geq 0$
even-mode perturbations.
This implies that the realization of the C-metric perturbation
supports the fact that our derived solutions in the Part II
paper~\cite{K.Nakamura-2021d} are reasonable, and then, we may say
that
Proposal~\ref{proposal:treatment-proposal-on-pert-on-spherical-BG} is
also physically reasonable.


The LTB solutions is a spherically symmetric exact solution which
describes the expanding universe with the dust matter or the dust
matter collapse to a black hole.
It is well-known that the LTB solutions include the Schwarzschild
spacetime as a special case.
For this reason, we can regard this LTB solution as a black hole
solution with the perturbative collapsing dust matter.
After reviewing the LTB exact solution, we considered the vacuum black
hole solution, i.e., the Schwarzschild spacetime with the perturbative
dust matter and examine our $l=0$ even-mode solution derived in the
Part II paper~\cite{K.Nakamura-2021d} describes this perturbative
solution at linear level.
From the perturbative treatment, we confirmed the linearized
continuity equations of the dust matter in terms of the static
Schwarzschild coordinate.
We also considered the linear metric perturbation on the Schwarzschild
background spacetime of the LTB exact solution.
In this realization, the perturbative arbitrary functions $f(R)$ and
$\tau_{0}(R)$ in the LTB solutions, which have their physical
meanings, included in the term of the Lie derivative of the background
metric $g_{ab}$.
Therefore, we should regard that such terms of the Lie derivative of
the background metric $g_{ab}$ are physical.
Thus, we confirmed that the $l=0$ even-mode solution derived in the
Part II paper~\cite{K.Nakamura-2021d} does describe this perturbative
LTB solution.


Next, we considered the linearized C-metric with the Schwarzschild
background spacetime, which may have $l=1$ even-mode perturbations.
Since this linearized C-metric actually includes $l=1$ even-mode
perturbations, this solution is appropriate for check whether our
derived $l=1$ even-mode perturbation physically reasonable, or not.
After reviewing the non-rotating vacuum
C-metric~\cite{J.B.Griffiths-P.Krtous-J.Podolsky-2006}, in which
conical singularities may occur both in the axis $\theta=0$ and
$\theta=\pi$, we considered the perturbative form of this solution on
the Schwarzschild background spacetime.
To consider the perturbative expression of the C-metric on the
Schwarzschild background spacetime, we considered the situation where
the acceleration parameter $\alpha_{1}$ is sufficiently small.
Furthermore, we have to consider the deficit/excess angle perturbation
$C_{1}$.
We have to keep in our mind that the fact that we may always change
the point-identification between the physical C-metric spacetime and
the background Schwarzschild spacetime, i.e., we may change the
second-kind gauge at any time as in the LTB case.


Although we follow
Proposal~\ref{proposal:treatment-proposal-on-pert-on-spherical-BG}, we
compare the result after imposing the regularity $\delta=0$.
We only consider the even $m=0$-mode perturbations, because the
non-rotating C-metric does not have the $\phi$-dependence nor the
odd-mode components of perturbations.
We consider the mode-decomposition of the linearized C-metric with the
Schwarzschild background spacetime through the use of Legendre
function $P_{l}(\cos\theta)$ as the scalar harmonics.
Then, we could identify the linearized C-metric with the $l=0,1$- and
$l\geq 2$-mode perturbations on the Schwarzschild background
spacetime.


From the $l\geq 2$ metric perturbation of the linearized C-metric, we
identify the components of the linear perturbations of the
energy-momentum tensor for $l\geq 2$ modes.
Then, we have obtained the linear perturbation of the energy-momentum
tensor (\ref{eq:1st-pert-calTab-dd-decomp-2-C-metric}) with the
constant $\lambda_{l}$ which given by
(\ref{eq:constant-tension-from-Einstein}).
Furthermore, we also checked the consistency of all components of the
Einstein equation for $l\geq 2$ modes.


For $l=0$ modes of the linearized C-metric, we have the additional
mass parameter perturbation of the Schwarzschild spacetime and the
deficit/excess angle perturbation.
The additional mass parameter perturbation can be always added as the
vacuum solution with the term of the Lie derivative of the background
metric $g_{ab}$.
In this case, the term of the Lie derivative of the background metric
$g_{ab}$ play important roles.
On the other hand, the deficit/excess perturbation is proportional to
the metric on $S^{2}$.
This perturbation is also expressed as the traceless part of
$(t,r)$-component and the term of the Lie derivative of the background
metric $g_{ab}$.
From this expression, we obtain the equation of state of the linear
perturbations of the energy momentum tensor and the connection of the
deficit/excess angle perturbation and the energy density.
Then, we showed the formula of the deficit/excess angle perturbation
and the energy density for $l\geq 2$ mode perturbations is also valid
for $l=0$ mode perturbations.


For $l=1$ modes of the linearized C-metric, the equation of state for
the linear perturbations of the energy-momentum tensor and the formula
which gives the relation between the energy density and the
deficit/excess perturbation in the cases $l\geq 2$ modes and $l=0$
modes is also consistent even in the $l=1$-mode case.
Furthermore, we showed that the linearized C-metric perturbation is
given from the $l=1$-mode solution obtained in the Part II
paper~\cite{K.Nakamura-2021d} apart from the Lie derivative of the
background metric $g_{ab}$.
Note that the term of the Lie derivative of the background metric
$g_{ab}$ play quite important roles even in $l=1$ mode solution.


As the results of the above $l\geq 2$-mode, $l=0$-mode, and $l=1$-mode
analyses, we have obtained the expression
(\ref{eq:1st-pert-calTab-dd-decomp-all_value_of_l}) of the
linear-perturbation of the energy-momentum tensor with the relation
(\ref{eq:constant-tension-from-Einstein-C-metric-sum-lgeq0-2}) between
the energy density and the acceleration parameter perturbation and the
deficit/excess angle perturbation.
Substituting
Eq.~(\ref{eq:constant-tension-from-Einstein-C-metric-sum-lgeq0-2})
into Eq.~(\ref{eq:1st-pert-calTab-dd-decomp-all_value_of_l}), we
obtain the $\delta$-function expression at the north and south pole of
$S^{2}$ which is given by
Eq.~(\ref{eq:1st-pert-calTab-delta-func-rep}) with the string tension
formulae (\ref{eq:string-tension-north-south}).
Thus, we have confirmed that the linearized C-metric is realized by
the linear perturbative solutions obtained in our Part II
paper~\cite{K.Nakamura-2021d}.


As the summary of the three papers,
Refs.~\cite{K.Nakamura-2021c,K.Nakamura-2021d} and this paper, we have
formulated a gauge-invariant formulation of the linear perturbations
on the Schwarzschild background spacetime.
Our formulation includes gauge-invariant treatments not only for
$l\geq 2$ modes but also for $l=0,1$ mode perturbations.
To construct gauge-invariant formulation for $l=0,1$-mode
perturbations, we introduce the singular harmonic function at once and
propose
Proposal~\ref{proposal:treatment-proposal-on-pert-on-spherical-BG} as
the strategy to solve the linearized Einstein equations on the
Schwarzschild background spacetime, which state that we eliminate the
singular behavior of introduced singular harmonics when we solve the
linearized Einstein equations.
Following
Proposal~\ref{proposal:treatment-proposal-on-pert-on-spherical-BG}, we
showed the strategy to solve the odd-mode perturbation in the Part I
paper~\cite{K.Nakamura-2021c}, that of the even-mode perturbation in
the Part II paper~\cite{K.Nakamura-2021d}.


We also derived the $l=0,1$ mode solution of the linearized Einstein
equations for the odd and even-mode perturbations in the Part
I~\cite{K.Nakamura-2021c} and in the Part II~\cite{K.Nakamura-2021d}
papers, respectively.
Furthermore, in this paper, we showed that the solutions for
$l=0,1$-mode perturbations derived in the Part II
paper~\cite{K.Nakamura-2021d} realize two exact solutions.
One is the LTB solutions and the other is the non-rotating C-metric.
Thus, the results in this paper support our solutions derived in the
Part II paper~\cite{K.Nakamura-2021d} and our proposal in the Part I
paper~\cite{K.Nakamura-2021c}.
In this sense, we may say that our strategy to solve the linearized
Einstein equations on the Schwarzschild background spacetime proposed
as Proposal~\ref{proposal:treatment-proposal-on-pert-on-spherical-BG}
is physically reasonable.
We also note that the gauge-invariant solutions derived in the Part
I~\cite{K.Nakamura-2021c} and the Part II~\cite{K.Nakamura-2021d}
papers includes the terms of the Lie derivative of the background
metric $g_{ab}$.
In many literature, it is well-known that we have ``residual
gauge-degree of freedom'' if we employ the Regge-Wheeler gauge.
The terms of the Lie derivative of the background metric $g_{ab}$
seems to correspond to this ``residual gauge.''
On the other hand, in this series of paper, we distinct the notion of
the gauge of the first kind and the notion of the gauge of the second
kind.
Furthermore, we declare that the purpose of our gauge-invariant
perturbation theory is to exclude not the gauge of the first kind but
the gauge of the second kind.
Moreover, our formulation completely excludes the gauge of the second
kind.
Therefore, the term of the Lie derivative of the background metric
$g_{ab}$ in our derived solution should be regarded as the gauge
degree of freedom of the first kind.
Even in this Part III paper, these term of Lie derivative played
crucial role.
Therefore, we may say that to take into account of these term of Lie
derivative is also important when we compare the two metric
perturbations.


Owing to
Proposal~\ref{proposal:treatment-proposal-on-pert-on-spherical-BG}, we
could treat $l=0,1$-mode perturbations on the Schwarzschild background
spacetime in a gauge-invariant manner.
This implies that the ``zero mode problem'' on our general framework
of the gauge-invariant perturbation theory was resolved at least in
the perturbations on the spherically symmetric background spacetime.
This also implies that we can apply our general framework of
higher-order gauge-invariant perturbation theory to any-order
perturbations on the spherically symmetric background spacetime.
This extension to any-order perturbations was briefly discussed in our
companion paper~\cite{K.Nakamura-2021b}.
We leave further detailed discussions as future works.







\section*{Acknowledgements}


The author deeply acknowledged to Professor Hiroyuki Nakano for
various discussions and suggestions.



\end{document}